\documentclass[10pt,twocolumn,twoside]{IEEEtran}

\usepackage{amsmath}
\usepackage{amssymb}
\usepackage{latexsym}
\usepackage{verbatim}
\usepackage{subfigure}
\usepackage{graphicx}
\usepackage{psfrag}
\usepackage{hyperref}

\usepackage{tcolorbox}
\definecolor{darkred}{RGB}{250,0,0}
\definecolor{darkgreen}{RGB}{0,150,0}
\definecolor{myblue}{RGB}{0,0,250}
\definecolor{darkblue}{RGB}{0,0,200}
\hypersetup{colorlinks=true, linkcolor=darkred, citecolor=myblue, urlcolor=darkblue}

\newtheorem{theorem}{Theorem}
\newtheorem{proposition}{Proposition}
\newtheorem{lemma}{Lemma}

{\begin{list}               %    with flush left bullets.
    {$\bullet$ \hfill}{
        \setlength{\leftmargin}{\parindent}
        \setlength{\parsep}{0.04\baselineskip}
        \setlength{\itemsep}{0.5\parsep}
        \setlength{\labelwidth}{\leftmargin}
        \setlength{\labelsep}{0em}}
    }
{\end{list}}

\providecommand{\eref}[1]{{\eqref{eq:#1}}}  % call \eqref from amstex
\providecommand{\cref}[1]{Chapter~\ref{chap:#1}}
\providecommand{\sref}[1]{Section~\ref{sec:#1}}
\providecommand{\fref}[1]{Figure~\ref{fig:#1}}

\providecommand{\R}{\ensuremath{\mathbb{R}}}

\providecommand{\abs}[1]{\lvert#1\rvert}
\providecommand{\norm}[1]{\lVert#1\rVert}

\providecommand{\set}[1]{\left\{#1\right\}}

\providecommand{\bydef}{\overset{\text{def}}{=}}

\renewcommand{\vec}[1]{\ensuremath{\boldsymbol{#1}}}
\providecommand{\mat}[1]{\ensuremath{\boldsymbol{#1}}}

\providecommand{\mA}{\mat{A}} 
\providecommand{\mC}{\mat{C}} 
\providecommand{\mD}{\mat{D}}

\providecommand{\mI}{\mat{I}}

 \providecommand{\mG}{\mat{G}}

\providecommand{\va}{\vec{a}} 
\providecommand{\vc}{\vec{c}} \providecommand{\ve}{\vec{e}}
\providecommand{\vh}{\vec{h}}

\providecommand{\vq}{\vec{q}} 
 
\providecommand{\vg}{\vec{g}}
\providecommand{\vu}{\vec{u}} \providecommand{\vw}{\vec{w}}
\providecommand{\vx}{\vec{x}} \providecommand{\vy}{\vec{y}}
\providecommand{\vz}{\vec{z}}

\providecommand{\veta}{\vec{\eta}}

\newcommand{\vsp}{\vspace{3pt}}

\newcommand{\vxt}{\widetilde{\vx}}

\usepackage{lipsum}    
\usepackage{ifthen}
\usepackage{tcolorbox}
\newboolean{showcomments}
\setboolean{showcomments}{true}

\usepackage{mathtools}  
\mathtoolsset{showonlyrefs}

\providecommand{\abs}[1]{\lvert#1\rvert}
\providecommand{\norm}[1]{\lVert#1\rVert}

\providecommand{\set}[1]{\left\{#1\right\}}

\providecommand{\bydef}{\overset{\text{def}}{=}}

\renewcommand{\vec}[1]{\ensuremath{\boldsymbol{#1}}}
\providecommand{\mat}[1]{\ensuremath{\boldsymbol{#1}}}

\providecommand{\mA}{\mat{A}} 
\providecommand{\mC}{\mat{C}} 
\providecommand{\mD}{\mat{D}}

\providecommand{\mI}{\mat{I}}

 \providecommand{\mG}{\mat{G}}

\providecommand{\va}{\vec{a}} 
\providecommand{\vc}{\vec{c}} \providecommand{\ve}{\vec{e}}
\providecommand{\vh}{\vec{h}}

\providecommand{\vq}{\vec{q}} 
 
\providecommand{\vf}{\vec{f}} 
\providecommand{\vg}{\vec{g}}
\providecommand{\vu}{\vec{u}} \providecommand{\vw}{\vec{w}}
\providecommand{\vx}{\vec{x}} \providecommand{\vy}{\vec{y}}
\providecommand{\vz}{\vec{z}}

\providecommand{\vphi}{\vec{\phi}}
\providecommand{\veta}{\vec{\eta}}

\newcommand{\eps}{\epsilon}
\newcommand{\Pro}{\mathbb{P}}

\usepackage{cite}

\usepackage{color}

\usepackage{enumitem}
\usepackage{commath}

\usepackage{esdiff}

\usepackage{pgf,tikz,psfrag}

\usepackage{dsfont}

\DeclareMathOperator{\atan}{atan}

\providecommand{\vxi}{\vec{\xi}}

\begin{document}
%
% Paper Title
\title{Phase Retrieval via Polytope Optimization: Geometry, Phase Transitions, and New Algorithms}
\author{Oussama Dhifallah, Christos Thrampoulidis, and Yue M. Lu
\thanks{O. Dhifallah is with the John A. Paulson School of Engineering and Applied Sciences, Harvard University, Cambridge, MA 02138, USA (e-mail: oussama$\_$dhifallah@g.harvard.edu).}
\thanks{C. Thrampoulidis is with the Research Laboratory of Electronics (RLE) at Massachusetts Institute of Technology, Cambridge, MA 02139, USA (e-mail: cthrampo@mit.edu).}
\thanks{Y. M. Lu is with the John A. Paulson School of Engineering and Applied Sciences, Harvard University, Cambridge, MA 02138, USA (e-mail: yuelu@seas.harvard.edu).}
\thanks{This work was supported in part by the US National Science Foundation under grants CCF-1319140 and CCF-1718698. Preliminary and partial results of this work have been presented at the 55th Annual Allerton Conference on Communication, Control, and Computing in 2017 \cite{Lampouss17}.}
}

\maketitle

\begin{abstract}
We study algorithms for solving quadratic systems of equations based on optimization methods over polytopes. Our work is inspired by a recently proposed convex formulation of the phase retrieval problem, which estimates the unknown signal by solving a simple linear program over a polytope constructed from the measurements. We present a sharp characterization of the high-dimensional geometry of the aforementioned polytope under Gaussian measurements. This characterization allows us to derive asymptotically exact performance guarantees for PhaseMax, which also reveal a phase transition phenomenon with respect to its sample complexity. Moreover, the geometric insights gained from our analysis lead to a new nonconvex formulation of the phase retrieval problem and an accompanying iterative algorithm, which we call PhaseLamp. We show that this new algorithm has superior recovery performance over the original PhaseMax method. Finally, as yet another variation on the theme of performing phase retrieval via polytope optimization, we propose a weighted version of PhaseLamp and  demonstrate, through numerical simulations, that it outperforms several state-of-the-art algorithms under both generic Gaussian measurements as well as more realistic Fourier-type measurements that arise in phase retrieval applications.
\end{abstract}


%!TEX root = phaselamp.tex

\section{Introduction}\label{sec:intro}
\subsection{Background}\label{sec:back}

We study the problem of recovering an unknown vector $\vxi\in\R^n$, up to a global sign change, from  $m$ magnitude measurements of the form:
%$\lbrace y_i, 1\leq i \leq m \rbrace$ of the form:
\begin{equation}\label{eq:abs}
y_i = \abs{\va_i^T \vxi},\quad i=1,\ldots,m,
\end{equation}
where  $\lbrace \va_i\in\R^n, 1 \leq i \leq m \rbrace$ is a set of (known) sensing vectors. This is the real-valued version of the well-known \emph{phase retrieval} problem, which has found numerous applications in science and engineering, including X-ray crystallography, Fourier ptychography, astronomy, radar and wireless communications, to name a few. Despite the problem's long history, developing methods for solving \eqref{eq:abs} remains an active research topic. In particular, the problem has attracted significant attention in the optimization and signal processing communities over the past decade; see, \emph{e.g.}, \cite{Candes:2013xy, jaganathan2015phase,  Waldspurger:2015rz, Netrapalli:2013qv, Candes:2015fv, WangGY:2016, She15} and references therein.

%solving $m$ real-valued quadratic equations in $n$ variables, which is equivalent to   In applied science and engineering, this classical problem is commonly referred to as \emph{phase retrieval problem}, owing to the fact that in order to solve the system of equations in \eqref{eq:abs}, we must resolve the uncertainty due to the missing phase (here: sign) information. Phase retrieval has a rich history  with numerous and diverse applications
%
%In typical modern applications of  phase retrieval the number $n$ of variables is (very) large. Thus, researchers seek methods whose computational efficiency scales with the increasing problem dimensions. At the same time, it is essential that these algorithms come with  \emph{rigorous performance guarantees}. Such guarantees, not only do they serve to assess the validity of the  algorithm's solution, but they also inform the appropriate choice of the system's parameters (e.g., number of measurements) that are required to achieve desired specifications.

Among the most well-established methods are those based on semidefinite relaxation (\emph{e.g.}, \cite{balan2009painless,Candes:2013xy}), which operate by lifting the original $n$-dimensional natural parameter space to a higher dimensional matrix space. Despite the strong theoretical performance guarantees enjoyed by these convex-relaxation methods, the aforementioned lifting step significantly increases the computational complexity and memory requirement for the resulting algorithms. To address these challenges, recent work studies algorithms that directly solve the nonconvex formulations of the phase retrieval problem. Typically, such nonconvex methods follow a two-step approach, combining a careful initialization step \cite{Netrapalli:2013qv, Chen:2015eu, LuL:17} with further local refinement such as iterative gradient descent \cite{Netrapalli:2013qv,Candes:2015fv, Chen:2015eu, WangGY:2016}. 

Taking a different approach, two groups of authors \cite{phmax2, phmax} independently proposed a simple yet highly effective scheme that is based on \emph{convex programming} in the original $n$-dimensional signal space.  The resulting method, referred to as PhaseMax in \cite{phmax}, relaxes the \emph{nonconvex equality} constraints in \eref{abs} to \emph{convex inequality} constraints, and solves the following linear program:
\begin{equation}\label{eq:lp_form}
\begin{aligned}
\widehat{\vx}&=\underset{{\vx}\in\mathbb{R}^n}{\arg\,\max}~~~ {\vx}_\text{init}^{T}\,{\vx}\\	
&~~~~~~~~\text{s.t.}~~~~  \abs{\va_i^T \vx} \leq y_i, \text{ for }  1 \le i \le m.
\end{aligned}
\end{equation} 
Here, $\vx_\text{init}$ represents an initial guess (or ``anchor vector'') that is correlated with the target vector $\vxi$. In practice, $\vx_\text{init}$ can be obtained if we have additional prior knowledge about $\vxi$ (\emph{e.g.}, nonnegativity) or by using a simple spectral method \cite{Netrapalli:2013qv, Chen:2015eu, LuL:17}. The relaxation performed by the PhaseMax method is clearly appealing since it leads to a computationally efficient  convex optimization program over a simple polytope in $\R^n$.

%But, how successful is the relaxation with respect to its ability to solve the system of magnitude equations in \eqref{eq:abs}? More precisely: for given quality of the initialization step, how many measurements $m$ are needed so that the outcome $\widehat{\vx}$ of PhaseMax is the true solution $\vxi$ of \eqref{eq:abs}? Moreover, if the number of measurements is not sufficient for perfect solution, then how far is $\widehat{\vx}$ from the true $\vxi$? 

\subsection{Contributions}

In this paper, we present an \emph{exact performance analysis} of the PhaseMax method in the high-dimensional ($n \to \infty$) limit. In particular, we show that a phase transition phenomenon takes place, with a simple analytical formula characterizing the phase transition boundary. Moreover, we extend the idea of PhaseMax by proposing a new nonconvex formulation of the phase retrieval problem and an accompanying iterative algorithm. We show that this new algorithm, which we call \emph{PhaseLamp}, has provably superior recovery guarantees over the original PhaseMax method. In what follows, we highlight our main results with more technical details.

\emph{1. Exact performance analysis of PhaseMax.} We quantify the performance of PhaseMax in terms of the normalized mean squared error (NMSE), defined as 
\[
\text{NMSE}_n \bydef {{\min\{\norm{\vxi - \widehat\vx}_2^2, \norm{\vxi + \widehat\vx}_2^2\}}}/\,{{\norm{\vxi}_2^2}}.
\]
The NMSE depends on two parameters: the \emph{oversampling ratio}
\[
\alpha \bydef m/n,
\]
and the quality of the initial guess $\vx_\text{init}$, measured via the input \emph{cosine similarity}
\begin{equation}\label{eq:cosin}
\rho_{\text{init}} \bydef \frac{\abs{\vx_\text{init}^T \vxi}}{\norm{\vx_\text{init}}_2 \norm{\vxi}_2}.
\end{equation}
Note that the parameter $\rho_\text{init}$ quantifies the degree of alignment between the target vector $\vxi$ and the initial guess $\vx_\text{init}$. 

As one of the main contributions of our work, we derive the following asymptotically exact characterization of PhaseMax, under the assumption that the sensing vectors are drawn from the normal distribution: as $m,n\rightarrow\infty$ with their ratio $\alpha$ fixed, 
\begin{equation}\label{eq:pt_in}
\mathrm{NMSE}_n \xrightarrow[]{n\to\infty} 
\begin{cases}
0, &\text{if}~\rho_\text{init} > \rho_c(\alpha),\\
%\frac{\frac{\pi}{\alpha}}{\tan(\frac{\pi}{\alpha} )} > 1 - \rho_\text{init}^2,\\
f(\rho_{\text{init}},\alpha)>0, &\text{otherwise},
\end{cases} 
\end{equation} 
where
\begin{equation}\label{eq:ptt}
\rho_c(\alpha)\bydef \bigg({1- \frac{{\pi}/{\alpha}}{\tan({\pi}/{\alpha} )}}\bigg)^{1/2},
\end{equation}
and $f(\rho_{\text{init}},\alpha)$ is a positive function that can be explicitly determined by solving a one-dimensional deterministic fixed point equation (see Theorem~\ref{thm:the1}). We note that the asymptotic characterization in \eref{pt_in} establishes an exact phase transition boundary on the minimum required number of measurements for PhaseMax to be successful: for any fixed sampling ratio $\alpha$, there is a critical threshold $\rho_\text{c}(\alpha)$ such that PhaseMax perfectly recovers $\vxi$ if and only if the input cosine similarity $\rho_\text{init} > \rho_c(\alpha)$. 

\begin{figure}[t]
    \centering
    \subfigure[]{\label{fig:intro_nmse_1}
    \includegraphics[width=0.47\linewidth]{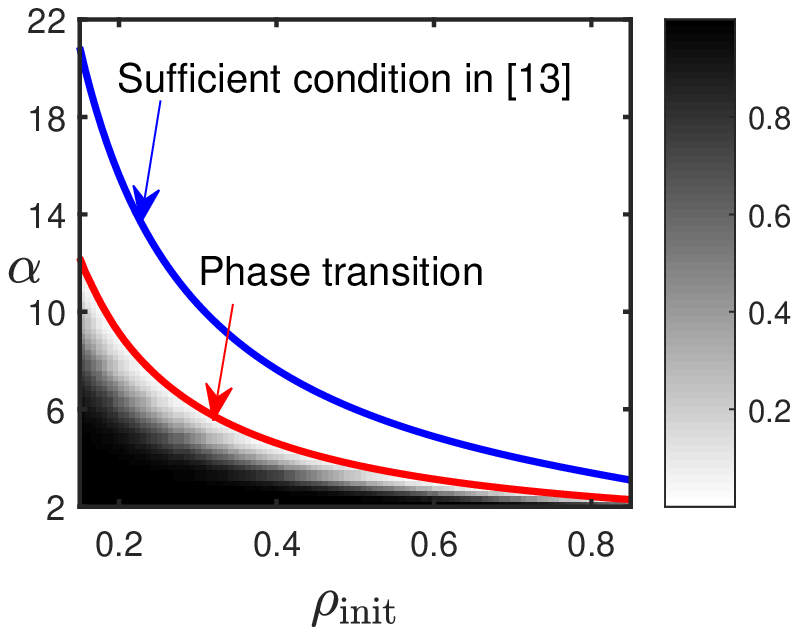}
    }
    \subfigure[]{\label{fig:intro_nmse_2}
        \includegraphics[width=0.47\linewidth]{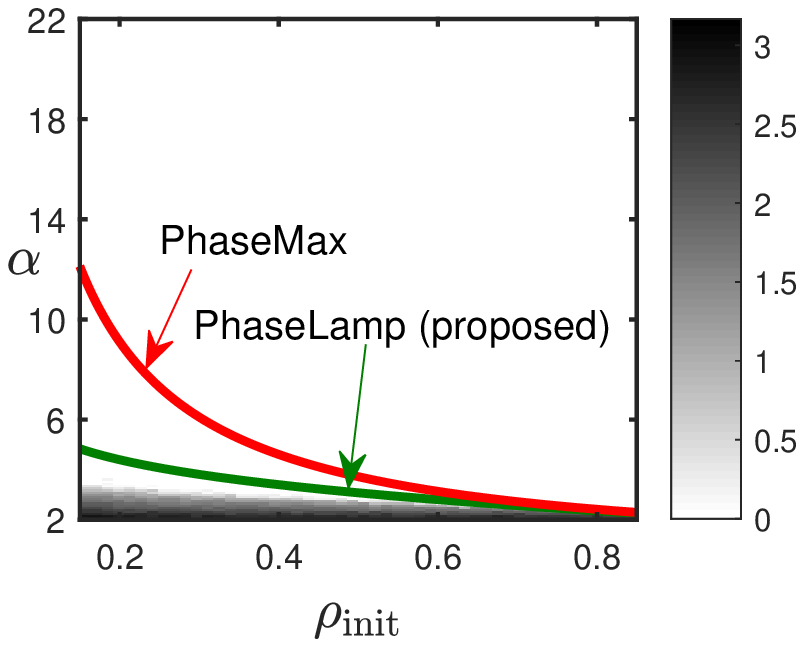}}
    \caption{
    The NMSE: theory versus simulations. (a) The NMSE of the PhaseMax method. (b) The NMSE of the PhaseLamp method. The signal dimension is set to $n=1000$, and the results are averaged over $10$ independent trials. The green curve shows the sufficient condition, as given in \eref{slam_meth_in}, for PhaseLamp to successfully recover the target signal. This is compared against the red curve, which shows the phase transition boundary of the original PhaseMax method as given in \eref{ptt}. The blue curve shows the sufficient condition derived in \cite{phmax}.
    }
        \label{fig:phase}
\end{figure}

Figure \ref{fig:intro_nmse_1} illustrates our asymptotic characterization and compares it with results from numerical simulations. Specifically, the red curve in the figure shows the phase transition boundary $\rho_c(\alpha)$, which can be seen to have excellent agreement with the actual performance of the algorithm. In \cite{phmax}, the authors show that PhaseMax is successful with high probability if 
\begin{equation}\label{eq:pmax_sufficient}
\alpha > \frac{2 \pi}{\pi - \arccos(\rho_\text{init})},
\end{equation}
which is plotted as the blue curve in \fref{intro_nmse_1}. We can see that our theoretical prediction serves to tighten the sufficient condition given in \eref{pmax_sufficient}.

\emph{2. Nonconvex formulation and new algorithms.} The insights gained from the exact analysis of PhaseMax lead us to 
%
%As the second contribution of this paper, we propose a new nonconvex formulation and an efficient iterative algorithm for the phase retrieval problem. Our new formulation is inspired by PhaseMax, the key idea of which is to relax the nonconvex equality constraints in \eref{abs} to convex \emph{inequality} constraints. The intersections of all these inequality constraints form a high-dimensional (random) polytope. Our analysis of the PhaseMax method provides useful insights on the exact high-dimensional geometry of that random polytope. These insights then lead us to 
a new nonconvex formulation of the phase retrieval problem:
\begin{equation}\label{eq:qb_form}
\begin{aligned}
\widehat{\vx}&=\underset{{\vx}\in\mathbb{R}^n}{\arg\,\max}~~~ \norm{\vx}^2_2\\	
&~~~~~~~~\text{s.t.}~~~~  \abs{\va_i^T \vx} \leq y_i, \text{ for }  1 \le i \le m.
\end{aligned}
\end{equation}
Note that \eref{qb_form} is indeed a nonconvex problem, as we aim to \emph{maximize} a convex function over a convex domain. We propose an efficient iterative method, which we call \emph{PhaseLamp}, to solve \eqref{eq:qb_form}. The name comes from the fact that the algorithm is based on the idea of successive \emph{linearization and maximization} over a \emph{polytope}, where in each step we solve a PhaseMax problem with the initialization given by the estimate from the previous iteration.

%
%The PhaseMax method has an appealing simple and computationally efficient formulation. Moreover, existing analysis \cite{phmax2, phmax, Hand:2016cs} shows that it achieves exact signal recovery from a nearly optimal number of \emph{random measurements}. Specifically, in the generic case when the sensing vectors are drawn from the Gaussian distribution, the required number of measurements for perfect reconstruction is shown to be linear with respect to the underlying dimension, \emph{i.e.}, $m = c\,n$ for some constant $c$ that depends on the quality of the initial vector $\vx_\text{init}$. 

We complement PhaseLamp with performance guarantees. Due to the iterative nature of PhaseLamp, the analysis here is more challenging than that of PhaseMax. By carefully characterizing the stationary points of \eqref{eq:qb_form}, we prove that a \emph{sufficient condition} for PhaseLamp to perfectly recover the target signal $\vxi$ (or, $-\vxi$) is
\begin{equation}\label{eq:slam_meth_in}
\rho_\text{init} > \rho_s(\alpha),
\end{equation}
 where $\rho_s(\alpha)$ is determined explicitly by solving a one-dimensional deterministic equation (see \eref{slam_fp} and Theorem~\ref{the2b}.) Importantly, $\rho_s(\alpha)$ is strictly smaller than $\rho_c(\alpha)$ as defined in \eqref{eq:ptt}. Therefore, the proposed PhaseLamp method has (strictly) \emph{superior recovery performance} over PhaseMax with respect to the minimum number of measurements needed to guarantee perfect solution of \eqref{eq:abs}. 
 
We illustrate this improvement in Figure \ref{fig:intro_nmse_2}, where it is shown that PhaseLamp has significantly better recovery performance, especially in the more challenging, and arguably the more practically relevant regime of small input cosine similarities $\rho_\text{init}$. Moreover, the numerical simulations shown at the same figure suggest that, although \eref{slam_meth_in} is only a sufficient condition,  it nevertheless provides a good estimate of the actual performance of the algorithm. Finally, as yet another variation on the theme of performing phase retrieval via polytope optimization, we propose in \sref{WPL} a weighted version of PhaseLamp. This new version is empirically shown to further outperform PhaseLamp. 

\begin{figure}[t!]
    \centering
    \subfigure[Original]{%\label{fig:rp1}
    \includegraphics[width=0.25\linewidth]{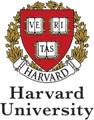}
    }
    \hspace{1ex}
    \subfigure[PhaseLamp]{%\label{fig:rp2}
        \includegraphics[width=0.25\linewidth]{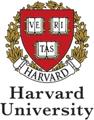}
    }
    \hspace{1ex}
    \subfigure[PhaseMax]{%\label{fig:rp2}
        \includegraphics[width=0.25\linewidth]{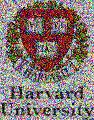}
    }\\    
    \subfigure[TAF]{%\label{fig:rp3}
    \includegraphics[width=0.25\linewidth]{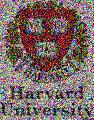}
    }
    \hspace{1ex}
    \subfigure[WF]{%\label{fig:rp4}
        \includegraphics[width=0.25\linewidth]{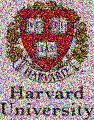}
    }
    \hspace{1ex}
    \subfigure[Fineup]{%\label{fig:rp4}
        \includegraphics[width=0.25\linewidth]{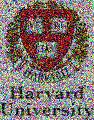}
    }
    \caption{Performance comparison between the proposed PhaseLamp method and existing techniques for the recovery of an image from coded diffraction patterns. The sampling ratio is $\alpha=3$. (a) The original image ($120\times 94\times 3$), (b) PhaseLamp: $\text{NMSE}=1.33e-05$, (c) PhaseMax \cite{phmax2, phmax}: $\text{NMSE}=0.5954$, (d) Truncated amplitude flow (TAF) \cite{WangGY:2016}: $\text{NMSE}=0.6535$, (e) Wirtinger Flow (WF) \cite{Candes:2015fv}: $\text{NMSE}=0.5338$, (f) Fienup \cite{Fienup:82}: $\text{NMSE}=0.6141$. The NMSE values are averaged over $10$ independent trials.}
        \label{fig:real_img}
\end{figure}

Although our theoretical analysis is carried out for generic Gaussian measurements, the proposed PhaseLamp algorithm and its weighted version perform well under more realistic measurement models that arise in phase retrieval applications. In \fref{real_img}, we compare the performance of PhaseLamp to PhaseMax and three other leading methods in the literature, where the measurement model corresponds to coded-diffraction patterns \cite{Candes:2015fv}. In this experiment, PhaseLamp successfully recovers the underlying image and outperforms the other competing methods. More details about the setup of this  experiment as well as additional numerical results can be found in \sref{addsim}.
 
\subsection{Related Work}

The performance of PhaseMax has been previously investigated in the literature. Existing analysis \cite{phmax2, phmax, Hand:2016cs} shows that  PhaseMax can achieve exact signal recovery from a nearly optimal number of random linear measurements. Specifically, in the case where the sensing vectors are drawn from the Gaussian distribution, the required number of measurements for perfect reconstruction is shown to be linear with respect to the underlying dimension, \emph{i.e.}, $m = c\,n$, for some constant $c$ that depends on the quality of the initial guess $\vx_\text{init}$. The analysis in \cite{phmax2, phmax, Hand:2016cs} gives various \emph{upper bounds} on the constant $c$. In a more recent work \cite{Ouss:17}, a subset of the authors of the current paper were able to pinpoint the exact value of $c$,  but the analysis in \cite{Ouss:17} uses the \emph{nonrigorous} replica method from statistical physics. Therefore, the precise nature of the results of our paper serves to(a) tighten up the previously known performance bounds of PhaseMax as given in \cite{phmax2, phmax, hand2016elementary}; and (b) rigorously verify the predictions based on the replica method given in \cite{Ouss:17}. Moreover, our novel theoretical analysis builds upon an exact characterization of the geometry of the feasibility set of the PhaseMax problem in \eref{lp_form}. This geometric insight plays a key role in both the formulation and the analysis of the improved PhaseLamp method proposed in this paper. 

Our analysis builds upon the recently developed convex Gaussian min-max theorem (CGMT) \cite{chris:151, chris:152}, which involves a tight version of a classical Gaussian comparison inequality\cite{Gordon:85}. The CGMT framework has been successfully applied to derive precise performance guarantees for structured signal recovery under (noisy) linear Gaussian measurements, \emph{e.g.}, \cite{Sto:13, thrampoulidis2015asymptotically, chris:151, chris:152}. In \cite{thrampoulidis2015lasso}, the CGMT is used to study signal recovery from a class of nonlinear measurements.  However, this excludes magnitude-only or quadratic measurements that are relevant for the phase retrieval problem considered here. 

%As previously mentioned,  the past decade has seen significant research effort towards establishing computationally efficient phase-retrieval algorithms  with rigorous  recovery guarantees. In this line of work \cite{Candes:2013xy, jaganathan2015phase,  Waldspurger:2015rz, Netrapalli:2013qv, Candes:2015fv, WangGY:2016, soltanolkotabi2017structured}, it is common to assume for the purposes of theoretical analysis that the sensing vectors are generic and sampled from a Gaussian distribution. However, even under that simplifying assumption existing performance guarantees are loose. For example, existing analysis of PhaseMax \cite{phmax2, phmax, Hand:2016cs} shows that the required number of measurements for perfect reconstruction is linear with respect to the underlying dimension, \emph{i.e.}, $m = c\,n$ for some constant $c$ that depends on the quality of the initial vector $\vx_\text{init}$. Unfortunately, the analysis in \cite{phmax2, phmax, Hand:2016cs} only gives loose upper bounds on the constant $c$. In a very recent work 

% he exact value of $c$, namely the sharp \emph{phase transition threshold}, has been previously predicted by a subset of the authors of the current paper in \cite{Ouss:17}, but the analysis in \cite{Ouss:17} uses the \emph{non-rigorous} replica method from statistical physics. Moreover, the exact asymptotic analysis of the PhaseMax method in the complex cases is presented in reference \cite{Phmaxcomp}.  

This paper is a significantly extended version of our earlier conference paper \cite{Lampouss17}, which announced our results with proof sketches. A limitation of our work is that we only consider the real-valued version of the phase retrieval problem. Very recently, our analysis techniques have been extended by the authors of \cite{Phmaxcomp} to the complex-valued case. As another limitation, we assume that we have access to noiseless measurements as in \eref{abs}. However, we believe that our technical approaches based on CGMT can be generalized to study the case of noisy measurements as well as robust versions of PhaseMax (see, \emph{e.g.}, \cite{Hand:2016cs}).

%We mention that this paper is a significantly extended version of a preliminary version that appeared in \cite{Lampouss17}.

%
%Finally, we mention two works that both appeared after the publication of \cite{Lampouss17}. First, the authors in  were able to extend our techniques to the analysis of PhaseMax in the case of complex-valued measurements. Second, a different algorithm for phase retrieval that is based on approximate-message passing (AMP) and comes with precise guarantees has been recently studied in \cite{ma2018optimization}.

\subsection{Paper Outline}

The rest of the paper is organized as follows. Central to our work is an exact characterization of the geometric properties of the feasibility polytope of the optimization in \eqref{eq:lp_form}. Thus, we start by presenting in Section \ref{randpoly} a rigorous high dimensional analysis of this polytope. \sref{main_results} focuses on PhaseMax, where we establish accurate performance guarantees for this method in the high-dimensional limit. The new nonconvex formulation \eref{qb_form} and the accompanying PhaseLamp algorithm are introduced in \sref{slam_alg}. We also provide sufficient conditions for PhaseLamp to achieve perfect recovery. Additional simulation results are shown in \sref{addsim}, comparing PhaseLamp (and its weighted variation) with several other existing algorithms for the phase retrieval problem. \sref{tech} collects the proofs and technical details of all the results introduced in the previous sections. We conclude the paper in \sref{conclusion}.

\subsection{Technical Assumptions and Notation}
\label{sec:assp}

The asymptotic predictions derived in this paper are based on the following assumptions.
\begin{enumerate}[label={(A.\arabic*)}]
\item The sensing vectors $\set{\va_i}_{1 \le i \le m}$ are drawn independently from a Gaussian distribution with zero mean and covariance matrix $\mI_n$.

\item $m = m(n)$ with $\alpha_n = m(n) / n \rightarrow \alpha > 0$ as $n \rightarrow \infty$, where $\alpha > 1$.

\item \label{a:pos} The initial guess $\vx_\text{init}$ has a \emph{positive} correlation with the target signal vector $\vxi$, \emph{i.e.}, $\vxi^T \vx_\text{init} > 0$.

\item \label{a:unit} $\vxi=\ve_1$, \emph{i.e.}, the first vector of the canonical basis of $\mathbb{R}^n$, and $\norm{\vx_\text{init}}_2 = 1$.
\end{enumerate}

The assumption in \ref{a:pos} can be made without loss of generality, as both $\vxi$ and $-\vxi$ are valid target signals. Similarly, the assumptions made in \ref{a:unit} only serve to simplify the notation but they are not restrictive either, thanks to the rotational invariance of the Gaussian distribution and since the optimization problem \eref{lp_form} is scale invariant.

Throughout the paper, we use $\mA \in\mathbb{R}^{m\times n}$ to denote the sensing matrix, whose rows consist of the sensing vectors $\lbrace \va_i^T, 1 \leq i \leq m \rbrace$. Since $\vxi = \ve_1$, the first column of $\mA$ has special significance. We use  $\vq$ to denote the first column of $\mA$ and $\mG\in\mathbb{R}^{m\times(n-1)}$ for the remaining part, \emph{i.e.}, $\mA=[\vq~~\mG]$. More generally, for any $\vx \in \R^n$, we partition it as ${{\vx}}=[x_1~~{{\widetilde{\vx}}}^T]^T$, where $x_1 \in \R$ and $\widetilde{\vx} \in \R^{n-1}$.

%Similarly, define $\eta_1$ and $\widetilde{\veta}$ such that $\vx_{\text{init}}=[\eta_1~~\widetilde{\veta}^T]^T$. 
%%\christos{I suggest that we call the initialization vector $\veta$, not $\mathbf{x}_{\text{init}}$. Then its decomposition  $[\eta_1~~\widetilde{\veta}^T]^T$ is consistent.} 
%Moreover, partition a vectors $\vx\in\mathbb{R}^n$ as follows ${{\vx}}=[x_1~~{{\widetilde{\vx}}}^T]^T$.

For any set $\mathcal{A}$ in a finite-dimensional Euclidean space, we define its $L_2$ norm as
$$\norm{\mathcal{A}}=\sup_{\vx\in\mathcal{A}}~\norm{\vx}_2.$$
When $\mathcal{A}$ is a compact set, we denote its boundary by $\mathrm{bd}(\mathcal{A})$. Additionally, the deviation between two sets $\mathcal{A}$ and $\mathcal{B}$ in a common space is given by
$$
\mathbb{D}(\mathcal{A},\mathcal{B})=\sup_{ \vx\in\mathcal{A} }~\inf_{ \vy\in\mathcal{B} }~\norm{\vx-\vy}_2.
$$

For any vector $\vc$, we let  $\abs{\vc}$ and $\text{sign}(\vc)$ to denote its component-wise absolute value and sign, respectively. Moreover, we let $\min(\vc)$ return the minimum value in the vector, and $\vz=\vc\wedge\mathbf{0}$ represents a vector such that $\vz_i=\min(\vc_i,0)$.
Finally, for a sequence of random variables $\lbrace \mathcal{X}^{(n)}\rbrace_{n\in\mathbb{N}}$ and a constant $c\in\mathbb{R}$ independent of $n$, we write $\mathcal{X}^{(n)}\xrightarrow[]{n\to\infty}c$, to denote convergence in probability, \emph{i.e.}, $\forall \epsilon >0, \lim_{n\to \infty}\mathbb{P}(|\mathcal{X}^{(n)}-c|>\epsilon)=0.$

%!TEX root = phaselamp.tex

\section{Polytope geometry}
\label{randpoly}

In this section, we study the geometry of the \emph{feasibility set} of  PhaseMax in \eqref{eq:lp_form}, which is given as follows: 
\begin{equation}\label{eq:cfeas}
{\mathcal C}_{\rm feas} := \bigcap_{i=1}^{m} \{ \vx\in\mathbb{R}^n:~ |\va_i^T \vx|\leq |\va_i^T \vxi| \}.
\end{equation}
Under the assumption of Gaussian sensing vectors, ${\mathcal C}_{\rm feas}$ forms a high-dimensional \emph{random polytope}. It is essential, both for the analysis of PhaseMax and also for motivating PhaseLamp, to understand the exact structure of the above polytope. 

\subsection{How Does ${\mathcal C}_{\rm feas}$ Look like? Some Intuitions}
\label{hitworks}

Before we delve into the details of our analysis, we provide a visualization of  ${\mathcal C}_{\rm feas}$  via a simulation example, which aims to explain intuitively why PhaseMax is expected to succeed at recovering the unknown signal as the number of measurements increases. 
%
% it is instructive to gain some intuition about the behavior of ${\mathcal C}_{\rm feas}$ as the number of measurements increases 
%%
%%It has been shown in \cite{Lampouss17,Ouss:17} the existence of a sharp phase transition phenomenon in the performance of the PhaseMax method. This section 
%provides a non-rigorous explanation of this result. A simulation example is provided to visualize the feasibility set ${\mathcal C}_{\rm feas}$ of the PhaseMax problem. 
Specifically, \fref{rpoly1} shows a projection of the random polytope ${\mathcal C}_{\rm feas}$ as a function of the number of measurements $m$. 
\begin{figure}[t!]
    \centering
    \includegraphics[width=0.9\linewidth]{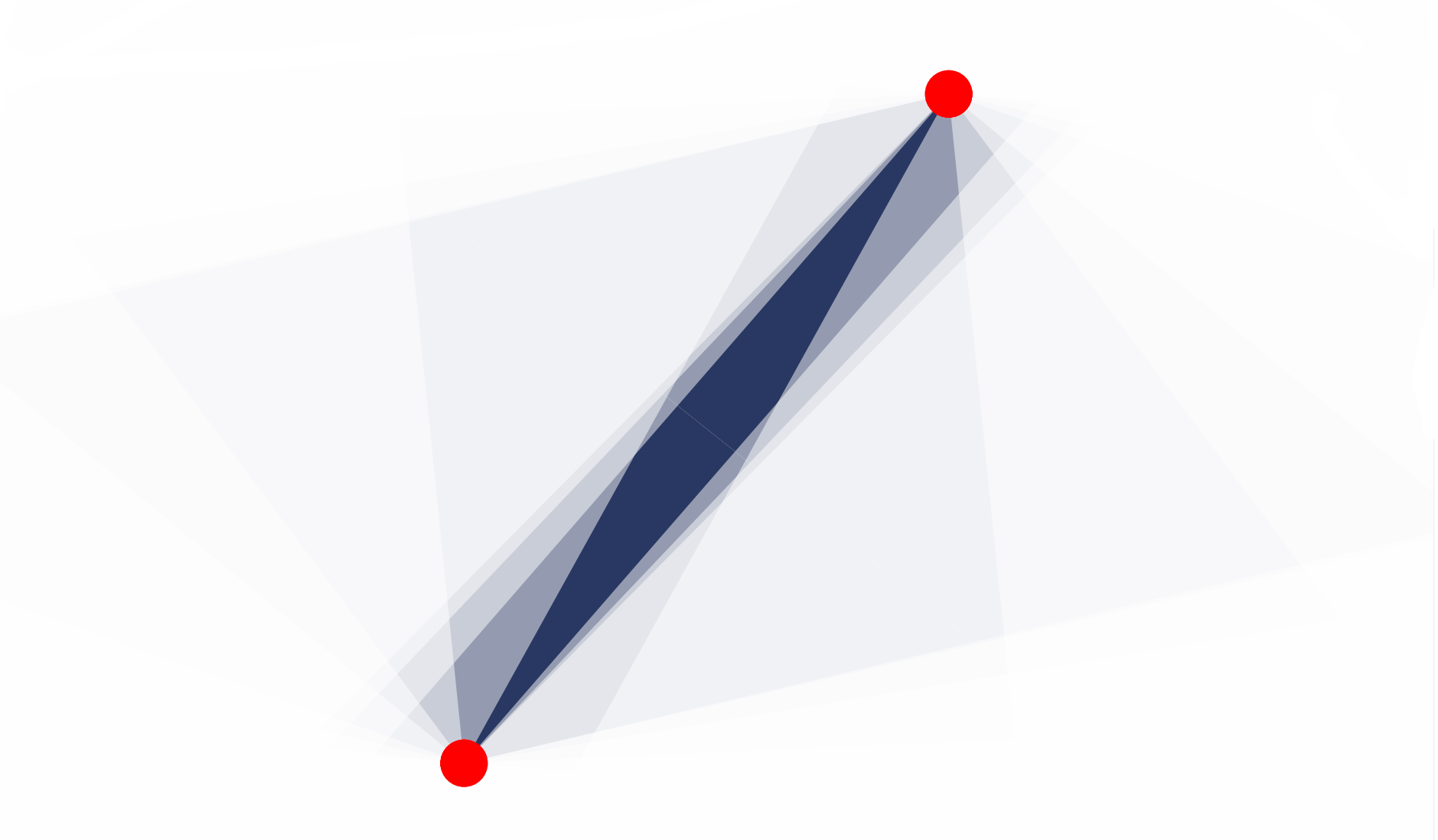}
    \caption{The two dimensional geometry of the random feasibility set of the PhaseMax method, denoted ${\mathcal C}_{\rm feas}$. Red dots: the target signal vectors $\vxi$ and $-\vxi$. Blue region: the feasibility set ${\mathcal C}_{\rm feas}$ of PhaseMax defined in \eref{cfeas}.}
    \label{fig:rpoly1}
\end{figure}
Note that as the number $m$ of constraints increases, the feasibility set ${\mathcal C}_{\rm feas}$ looks more and more like a needle pointed towards the target signal vectors $\vxi$ and $-\vxi$. This observation suggests the existence of a phase transition behavior in the performance of PhaseMax method. In particular, the target signal vector $\vxi$ has the highest correlation with the initial guess vector $\vx_{\text{init}}$ among all the feasible vectors as long as $m$ is sufficiently large (as a function of the correlation of $\vx_{\text{init}}$ with $\vxi$).
% with an appropriate initial guess $\vx_{\text{init}}$ and for sufficiently large $m$, the target signal vector $\vxi$ has the highest correlation with the initial guess vector $\vx_{\text{init}}$ among all the feasible vectors. 
% This means that the solution of the PhaseMax method is always the target signal $\vxi$ for sufficiently large oversampling ratio $\alpha$. 
%While this simple illustration intuitively explains the potential success of PhaseMax with increasing number of measurements, 
Of course, if we hope to make this observation rigorous, we need to develop formal analytic results regarding the properties of the random high-dimensional set ${\mathcal C}_{\rm feas}$. Despite the challenge of the task at hand, we show in the next sections that this is possible.
% The main contribution of this paper is to precisely characterize the set ${\mathcal C}_{\rm feas}$ in the high  limit.
\subsection{The Sufficient Feasible Set}
\label{suffst}

Note that what determines the error in a solution $\vx$ of either \eqref{eq:lp_form} or \eqref{eq:qb_form} are the magnitudes of $|\vx_1 - 1|$ and of $\|\vxt\|_2$. This essentially simplifies our task to that of understanding the geometry of the two dimensional projection of ${\mathcal C}_{\rm feas}$:
\begin{align}\notag%\label{eq:S_feas}
{\mathcal S}_{\rm feas} := \{ (s,r)\in\mathbb{R}^2:~ [x_1~\vxt^T]\in{\mathcal C}_{\rm feas},x_1=s,\|\vxt\|_2=r \}.
\end{align}
%With some abuse of notation we refer to the boundary of this two-dimensional projection ${\mathcal S}_{\rm feas}$ as the \emph{feasibility boundary}. Note that the feasibility boundary can be described as follows
%$$
%{\rm bd}\big({\mathcal S}_{\rm feas}) = \{ (s,r_{\max}(s))~|~s\in\R \},
%$$
%where for fixed $s$, $r_{\max}(s)$ can be expressed as follows
%\begin{align}\label{eq:rmax}
%r_{\max}(s) := \arg\max_r~r\quad\text{s.t.}\quad (s,r)\in{\mathcal S}_{\rm feas}.
%\end{align}
In this section, we describe a high probability upper bound on the feasibility boundary of ${\mathcal S}_{\rm feas}$. Specifically, in Theorem \ref{lem:fb} that follows, we compute a deterministic set ${\mathcal D}^\epsilon_{\rm feas}\subset \R^2$ such that the following holds with high probability:
$
{\mathcal S}_{\rm feas} \subseteq {\mathcal D}^\epsilon_{\rm feas}
$
for any $\epsilon > 0$. To this end, define the function $c_d$ as follows 
\begin{align}\label{eq:cdexp}
c_d(s,r)=\mathbb{E}_{q, g} \big[ \min\big\{\abs{q} - \abs{ r g + s q },0\big\}^2 \big],
\end{align}
where $q,g \overset{\mathrm{i.i.d.}}{\sim} \mathcal{N}(0,1)$. 

\begin{theorem}[Sufficient feasibility set]\label{lem:fb}
Assume that the oversampling ratio $\alpha>1$. Define the deterministic set ${\mathcal D}_{\rm feas}^{\epsilon}$ as follows
\begin{align}
{\mathcal D}_{\rm feas}^{\epsilon}=\lbrace (s,r)\in\mathbb{R}^2:~ r\geq 0,~\alpha~c_d(s,r) \leq r^2+\epsilon \rbrace,
\end{align}
where $\epsilon >0$. Then, for all $\eps>0$ it holds that
$$
\lim_{n\rightarrow\infty} \Pro\Big( \mathcal{S}_{\text{feas}} \subseteq \mathcal{D}_{\text{feas}}^{\epsilon}  \Big) = 1.
$$
\end{theorem}

\begin{figure}[t!]
    \centering
    \subfigure[]{%\label{fig:rp1}
    \includegraphics[width=0.45\linewidth]{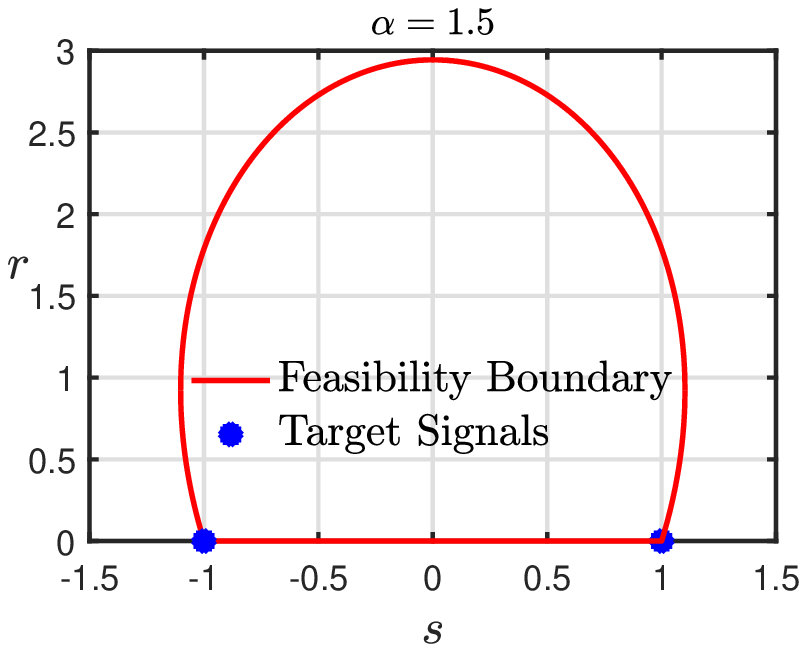}
    }
    \subfigure[]{%\label{fig:rp2}
        \includegraphics[width=0.45\linewidth]{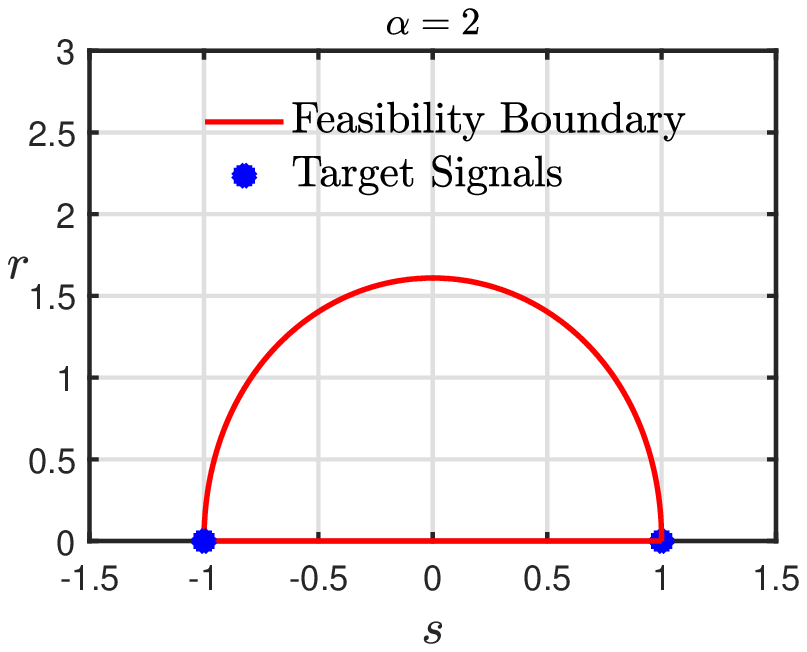}}\\
    \subfigure[]{%\label{fig:rp3}
    \includegraphics[width=0.45\linewidth]{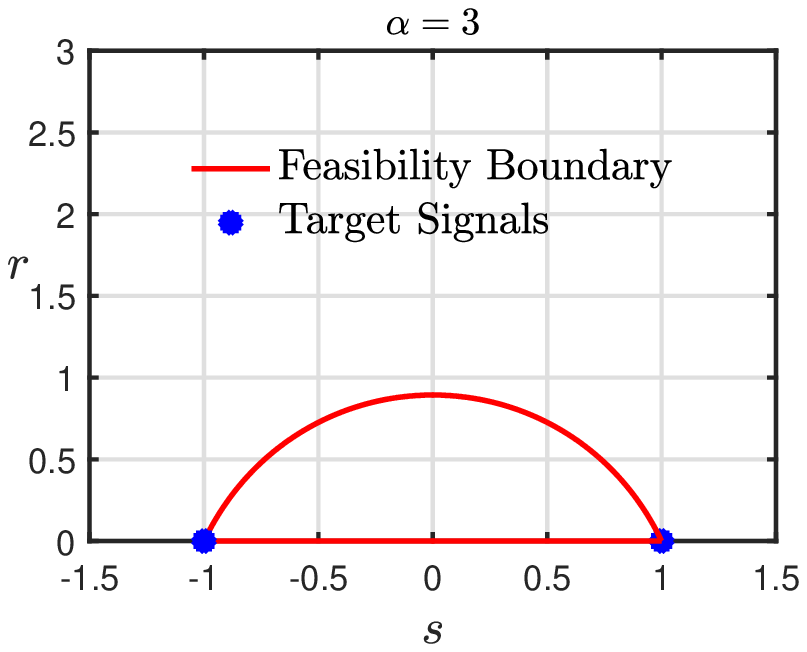}
    }
    \subfigure[]{%\label{fig:rp4}
        \includegraphics[width=0.45\linewidth]{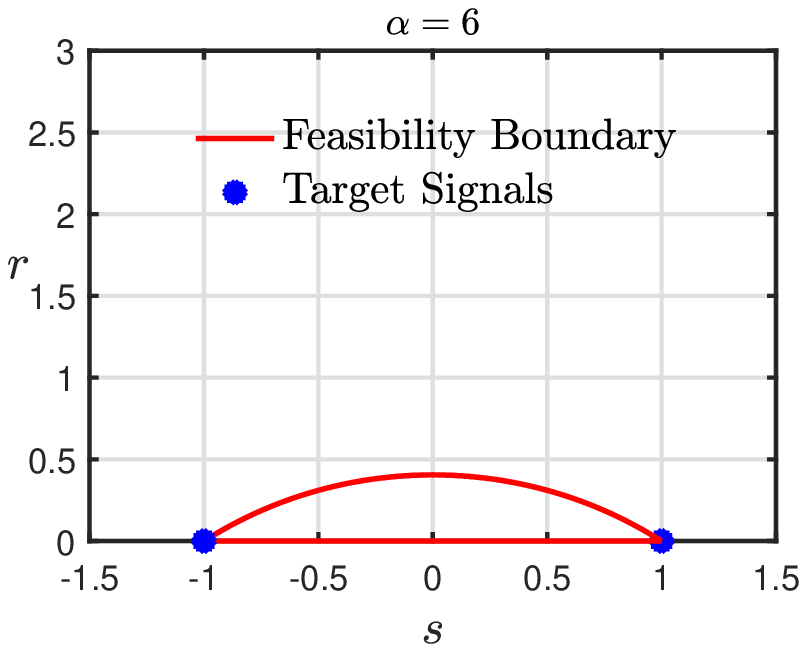}}
    \caption{Visualization of the deterministic set ${\mathcal D}_{\rm feas}$ for different values of the oversampling ratio $\alpha$. The blue dots denote the target signal vectors $\vxi$ and $-\vxi$ and the red curve denotes the boundary of the set ${\mathcal D}_{\rm feas}$. (a) $\alpha=1.5$, (b) $\alpha=2$, (c) $\alpha=3$, (d) $\alpha=6$.}
        \label{fig:fig_d_feas}
\end{figure}

%\begin{IEEEproof}
%The detailed proof is given in appendix \ref{plem:fb}.
%\end{IEEEproof}
The take away message of Theorem \ref{lem:fb}, whose proof is detailed in Appendix \ref{plem:fb}, is that the random feasibility set $\mathcal{S}_{\text{feas}}$ is essentially a subset (with high-probability in the large system limit) of any $\epsilon$-perturbation of the following \emph{deterministic} set
\begin{align}
{\mathcal D}_{\rm feas}=\lbrace (s,r)\in\mathbb{R}^2:~ r\geq 0,~\alpha~c_d(s,r) \leq r^2 \rbrace,
\end{align}
Hence, in order to understand the properties of $\mathcal{S}_{\text{feas}}$, it is 
essential to study the properties of ${\mathcal D}_{\rm feas}$.

We start with a visualization of ${\mathcal D}_{\rm feas}$ for different values of the oversampling ratio $\alpha$ in \fref{fig_d_feas}. Observe that, for sufficiently large oversampling ratio $\alpha$, ${\mathcal D}_{\rm feas}$ looks like a needle pointed towards the target signal vectors $\vxi$ and $-\vxi$. Recall, that this is consistent with the observations of Section \ref{hitworks}. Furthermore, note that  ${\mathcal D}_{\rm feas}$ is always convex and bounded.
%satisfies the second, third and fourth points in lemma \ref{prop} which validates the rigorous results stated in lemma \ref{prop}.

The following lemma, which is proved in Appendix \ref{proppf}, formalizes these observations.
%\subsection{Properties of the set ${\mathcal D}_{\rm feas}$ }
%The deterministic set ${\mathcal D}_{\rm feas}$ serves as a high probability super-set on the projected feasibility set ${\mathcal S}_{\rm feas}$ of the PhaseMax method. To understand the properties of the projected feasibility set of PhaseMax in the high--dimensional limit, it is essential to study the properties of the set ${\mathcal D}_{\rm feas}$.
%\christos{@Oussama: Please edit the lemma if needed.}
\begin{lemma}[Properties of ${\mathcal D}_{\rm feas}$]\label{prop}
The deterministic set ${\mathcal D}_{\rm feas}$ satisfies the following properties:
\begin{itemize}
\item[(P.1)] It is a convex set in $\mathbb{R}^2$.
%%%%%%%%%%%%%%%%%%%%%%%%%%%%%%%%%%%%%
%%%%%%%%%%%%%%%%%%%%%%%%%%%%%%%%%%%%%
\item[(P.2)] For $\alpha > 1$, the set ${\mathcal D}_{\rm feas}$ is compact. Additionally, $[-1,~1]\times \lbrace 0 \rbrace \subset {\mathcal D}_{\rm feas}$ and there exists $z > 0$ such that ${\mathcal D}_{\rm feas} \subseteq [-1,~1]\times [0,~z]$, for $\alpha \geq 2$. For $\alpha \geq 2$, the intersection between the set $\lbrace (s,r)\in\mathbb{R}^2:~s=1 \rbrace$ and $\mathcal{D}_{\text{feas}}$ is $\lbrace (1,0) \rbrace$.
%%%%%%%%%%%%%%%%%%%%%%%%%%%%%%%%%%%%%
%%%%%%%%%%%%%%%%%%%%%%%%%%%%%%%%%%%%%
%\item[(P.3)]  For $1 < \alpha < 2$, there exist $b > 1$ and $z > 0$ such that the set ${\mathcal D}_{\rm feas}$  is a subset of $[-b,~b]\times [0,~z]$. Additionally, there exists $a >0$ such that $(b,a) \in {\mathcal D}_{\rm feas}$.
%%%%%%%%%%%%%%%%%%%%%%%%%%%%%%%%%%%%%
%%%%%%%%%%%%%%%%%%%%%%%%%%%%%%%%%%%%%
\item[(P.3)]  For $\alpha \geq 2$ and $s \in [-1,~1]$, the maximum radius $r_{\text{max}}(s)$ of the set ${\mathcal D}_{\rm feas}$ satisfies $c_{d}(s,r_{\text{max}}(s))=r_{\text{max}}(s)/\alpha$. Moreover, for $s=0$, $r_{\text{max}}(0)>0$.
%%%%%%%%%%%%%%%%%%%%%%%%%%%%%%%%%%%%%
%%%%%%%%%%%%%%%%%%%%%%%%%%%%%%%%%%%%%
\item[(P.4)]  The slope of the boundary curve $\mathrm{bd}\left( {\mathcal D}_{\rm feas} \right)$ at $s=1$ is the unique solution of the following equation
\begin{equation}\label{eq:fxptpf1}
\begin{aligned}
\frac{\pi}{\alpha} c^2+c-(1+c^2) \atan(c)=0.
\end{aligned}
\end{equation}
\item[(P.5)] For any $\alpha>1$ and $\epsilon >0$, the set ${\mathcal D}^\epsilon_{\rm feas}$ is compact. Moreover, it satisfies $\mathcal{D}_{\text{feas}}^{\epsilon_1} \subseteq \mathcal{D}_{\text{feas}}^{\epsilon_2}$ for any $0<\epsilon_1 \leq \epsilon_2$ and we have
\begin{align}
\lim_{k \to \infty} {\mathcal D}^{\epsilon_k}_{\rm feas}=\bigcap_{k\geq 0} {\mathcal D}^{\epsilon_k}_{\rm feas}={\mathcal D}_{\rm feas},
\end{align}
for any decreasing sequence of positive numbers $\lbrace \epsilon_k \rbrace_{k\in\mathbb{N}}$ such that $\lim_{k\to \infty} \epsilon_k=0$.
\end{itemize}

\end{lemma}
%\begin{IEEEproof}
%The detailed proof is given in appendix \ref{proppf}.
%\end{IEEEproof}

%%%%%%%%%%%%%%%%%%%%%%%%%%%%%%%%%%%%%%%%%%

\subsection{Sufficient Condition for PhaseMax}\label{sec:suff_PM}
\label{sec:suf_PM}

With Theorem \ref{lem:fb} and Lemma \ref{prop} at hand, we have established an exact characterization of the high-dimensional geometry of the feasibility set of PhaseMax. Naturally, this leads to a sufficient condition under which its solution ${\widehat{\vx}}$ is the true unknown vector $\vxi$.
%
%
%In this section, we provide a sufficient condition for perfect recovery of the PhaseMax method. First, we introduce the following lemma which provides a supper set on the set of optimal solutions of PhaseMax.

All we need in addition to Theorem \ref{lem:fb} is the following simple observation regarding $\widehat{\vx}$, which follows directly by its optimality in the optimization problem in \eqref{eq:lp_form}.
\begin{lemma}\label{lem_ge_suf}
The optimal solution set of PhaseMax is a subset of the following deterministic set
\begin{align}\notag
{\mathcal{D}}_{\text{fp}}(\rho_{\text{init}})=\lbrace (s,r)\in\mathbb{R}^2:~r\geq 0,~\frac{\rho_{\text{init}}}{\sqrt{1-\rho_{\text{init}}^2}} (1-s) \leq r \rbrace.
\end{align}
%where $s$ and $r$ are as defined in section \ref{suffst}.
\end{lemma}
\begin{IEEEproof} %\christos{@Oussama: Please chage stars to hats (or vice versa), so that things are consistent.}
Without loss of generality, we can assume (due to symmetry) that $\rho_{\text{init}}\geq0$. Let $\widehat{\vx}$ be an optimal solution of \eqref{eq:lp_form} and partition it as follows
$
\widehat{\vx}^T={[s~~{{\widetilde{\vx}}}]}^T.
$
From optimality it holds:
\begin{align}
\eta_1 s+ \widetilde{\veta}^T\widetilde{\vx}=\vx_{\text{init}}^T\widehat{\vx} \geq \vx_{\text{init}}^T \vxi= \eta_1.
\end{align}
This implies that $\norm{\widetilde{\veta}}_2 r \geq  \eta_1- \eta_1 s$ with $r=\norm{{\widetilde{\vx}}}_2$. Recalling that
\begin{equation}
{\eta_1}\,\big/\,{\norm{\widetilde{\veta}}_2} = {\rho_{\text{init}}}\,\big/\,{\sqrt{1-\rho_{\text{init}}^2}},
\end{equation}
and rearranging terms, completes the proof of the lemma.
% This implies that PhaseMax solutions belong to the set
%\begin{align}\notag
%\widehat{\mathcal{D}}_{\text{fp}}(\rho_{\text{init}})=\lbrace (s,r)\in\mathbb{R}^2~\mathrm{s.t.}~r\geq 0,~\kappa (1-s) \leq r \rbrace,
%\end{align}
\end{IEEEproof}

\begin{figure}[t!]
    \centering
    \subfigure[]{\label{fig:phmax_suf1}
    \includegraphics[width=0.45\linewidth]{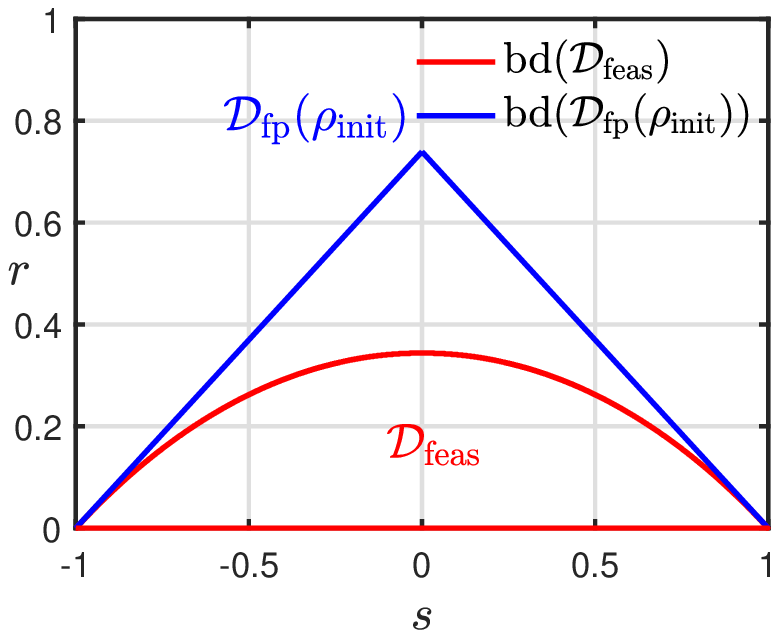}
    }
    \subfigure[]{\label{fig:phmax_suf2}
        \includegraphics[width=0.45\linewidth]{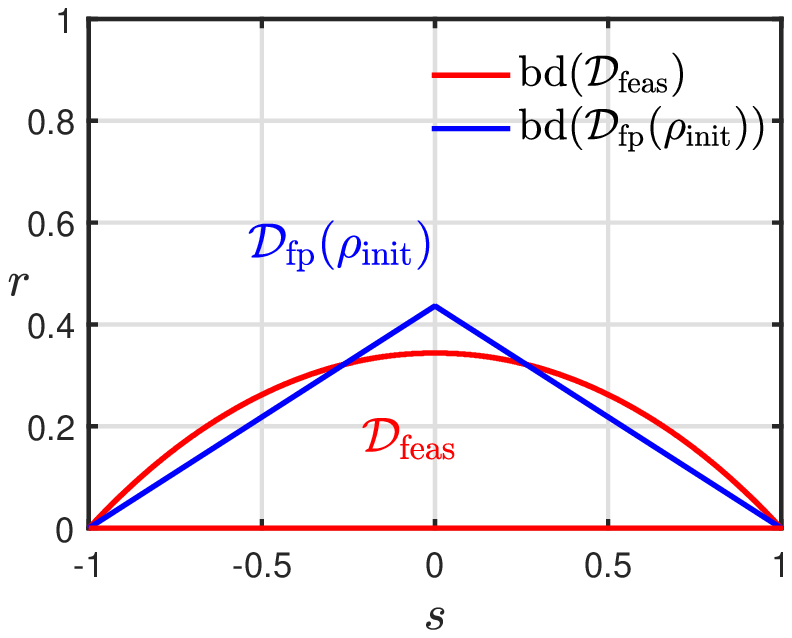}}

    \caption{Illustration of the set ${\mathcal{D}}_{\text{fp}}(\rho_{\text{init}})$ and the set ${\mathcal D}_{\rm feas}$ for $\alpha=7$ and (a) $\rho_{\text{init}}$ satisfies \eref{scpmax} with equality, (b) $\rho_{\text{init}}=0.4$. Note that we add the symmetric part.}
    \label{fig:figmax_suff}
\end{figure}

With these at hand, we have shown that in the high dimensional limit the solution of PhaseMax belongs to the intersection of the sets ${\mathcal{D}}_{\text{fp}}(\rho_{\text{init}})$ and ${\mathcal D}_{\rm feas}$. Therefore, a natural sufficient condition for perfect recovery is that this intersection only contains the desired points $\vxi$ and $-\vxi$. Proposition \ref{suffcondd} below formalizes this geometric condition and Figure \ref{fig:figmax_suff} serves as a numerical illustration of it.

%In order to see how Lemma \ref{lem_ge_suf} combined with Theorem \ref{lem:fb} naturally lead to a sufficient condition about recovery, we offer a visualization of the two sets $\widehat{\mathcal{D}}_{\text{fp}}$ and ${\mathcal D}_{\rm feas}$ in Figure \ref{fig:figmax_suff}.
%Now, we provide a simulation example to visualize the set $\widehat{\mathcal{D}}_{\text{fp}}$. \fref{figmax_suff} shows the supper set $\widehat{\mathcal{D}}_{\text{fp}}$ on the solutions of PhaseMax and the set ${\mathcal D}_{\rm feas}$. 
Note that in the high dimensional limit the solution of PhaseMax should belong to the intersection of the sets ${\mathcal{D}}_{\text{fp}}(\rho_{\text{init}})$ and ${\mathcal D}_{\rm feas}$. This fact leads us to a sufficient condition for perfect recovery of PhaseMax as stated in the following Proposition.
\begin{proposition}[Sufficient Condition]\label{suffcondd}
Assume that $\alpha > 2$ and let $c^\ast$ be the unique solution in \eref{fxptpf1}. PhaseMax perfectly recovers the target vector $\vxi$  (in the sense that $\mathrm{NMSE}_n \xrightarrow[]{n\to\infty} 0$, in probability) if the input cosine similarity $\rho_{\text{init}}$ satisfies 
\begin{equation}\label{eq:scpmax}
\rho_{\text{init}}  \geq \sqrt{ \frac{{c^\ast}^2}{{c^\ast}^2+1} }.
\end{equation}
\end{proposition}
%
%
%\fref{phmax_suf1} illustrates the sufficient condition in Proposition \ref{suffcondd}. Specifically, it can be noticed that if $\rho_{\text{init}}$ satisfies \eref{scpmax}, the intersection between the sets $\widehat{\mathcal{D}}_{\text{fp}}(\rho_{\text{init}})$ and ${\mathcal D}_{\rm feas}$ is only the target signal vector $\vxi$. In this case, PhaseMax perfectly recovers the target signal vector $\vxi$. One highlight of the sufficient condition in \eref{scpmax} is that it is valid for initial guess vectors $\vx_{\text{init}}$ constructed using the sensing vectors $\lbrace \va_i, 1 \leq i \leq m \rbrace$.
%
\begin{IEEEproof}%[of Proposition \ref{suffcondd}]
%\christos{@Oussama: the first part of this proof needs to be made rigorous. Note that we show perfect recovery only in probability...I.e., we need to show $\mathrm{bd}\left({\mathcal D}_{\rm feas}  \right) \cap \mathrm{bd}( \widehat{\mathcal D}_{\rm fp}(\rho_{\text{init}}) )=\lbrace (1,0) \rbrace$ whp!}
Let $\alpha>2$ and assume without loss  of generality that $\rho_{\text{init}} \geq 0$.
%Based on lemma \ref{lem_ge_suf}, any PhaseMax solution belongs to the  set
%\begin{align}\notag
%\widehat{\mathcal{D}}_{\text{fp}}(\rho_{\text{init}})=\lbrace (s,r)\in\mathbb{R}^2~\mathrm{s.t.}~r\geq 0,~\frac{\rho_{\text{init}}}{\sqrt{1-\rho_{\text{init}}^2}} (1-s) \leq r \rbrace.
%\end{align}
%Based on Theorem \ref{lem:fb}, the projected feasibility set ${\mathcal S}_{\rm feas}$ of the PhaseMax method is a subset of the deterministic set 
%\begin{align}
%{\mathcal D}_{\rm feas}=\lbrace (s,r)\in\mathbb{R}^2~\mathrm{s.t.}~r\geq 0,~ \alpha~c_d(s,r) \leq r^2 \rbrace,
%\end{align}
%in the high dimensional limit. Given that $\rho_{\text{init}} \geq 0$, 
First, we provide a sufficient condition such that the intersection between the sets ${\mathcal{D}}_{\text{fp}}(\rho_{\text{init}})$ and ${\mathcal D}_{\rm feas}$ is only the target signal vector $\lbrace (1,0) \rbrace$. 
Based on Lemma \ref{prop}, the slope of the boundary curve  $\mathrm{bd}\left( {\mathcal D}_{\rm feas} \right)$ at $s=1$ is the unique solution $c^\ast$ of the equation in \eqref{eq:fxptpf1}.
%\begin{equation}\label{eq:fxptpf}
%\begin{aligned}
%\frac{\pi}{\alpha} c^2+c-(1+c^2) \atan(c)=0.
%\end{aligned}
%\end{equation}
%This can be achieved by performing Taylor expansion at $s=1$ of the equation $\alpha c_d(s,r)=r^2$. 
Select $\rho_{\text{init}}$ such that $${\rho_{\text{init}}}/{\sqrt{1-\rho_{\text{init}}^2}}=c^\ast.$$ To show that $\mathrm{bd}\left({\mathcal D}_{\rm feas}  \right) \cap \mathrm{bd}( {\mathcal D}_{\rm fp}(\rho_{\text{init}}) )=\lbrace (1,0) \rbrace$, it suffices to prove that all the points satisfying $c^\ast (1-s) = r$ and $s \in [0,1)$ are not in the set ${\mathcal D}_{\rm feas}$, i.e., they satisfy $\alpha c_d(s,r)>r^2$, (recall the definition of $c_d$ in \eqref{eq:cdexp}). This is equivalent to showing that the following function 
\begin{align}
f(s)&:= \alpha c_d(s,c^\ast (1-s))-(c^\ast (1-s))^2,
\nonumber\\
&=(1+{c^\ast}^2) \atan(c^\ast) +\Big( \frac{(1+s)^2}{(1-s)^2}+{c^\ast}^2 \Big) \times\nonumber\\
&~~~~\atan\Big( \frac{c^\ast(1-s)}{1+s} \Big)-\frac{2c}{1-s} -\frac{\pi {c^\ast}^2}{\alpha},
\end{align}
is strictly positive in $[0,1)$. This can be checked to be true since the derivative of $f$ is strictly negative in $[0,1)$ and it follows from \eqref{eq:fxptpf1} that $f(0)=\frac{\pi}{\alpha} c^2$ and $\lim_{s\to 1} f(s)=0$. Thus, we have shown that selecting $\rho_{\text{init}}$ as in \eref{scpmax} 
%\begin{align}\label{eq:sufcondpf}
%\rho_{\text{init}} \geq \sqrt{\frac{{c^\ast}^2}{({c^\ast}^2+1)}},
%\end{align}
ensures that $\mathrm{bd}\left({\mathcal D}_{\rm feas}  \right) \cap \mathrm{bd}( {\mathcal D}_{\rm fp}(\rho_{\text{init}}) )=\lbrace (1,0) \rbrace$. 

Selecting the input cosine similarity in this way guarantees that for any $\epsilon>0$ there exists $\delta > 0$ such that
\begin{align}
\mathcal{D}_{\text{feas}}^{\epsilon} \cap {\mathcal D}_{\rm fp}(\rho_{\text{init}}) \subseteq \mathcal{B}^\delta,
\end{align}
where $\mathcal{B}^\delta$ is a ball of radius $\delta$ and center the target signal $\vxi$. Define the sequence $\lbrace \delta_{\epsilon_k} \rbrace_{k\in\mathbb{N}}$ as follows
\begin{align}\label{eq:depslon}
\delta_{\epsilon_k} :=\inf \lbrace \delta>0:~ \mathcal{D}_{\text{feas}}^{\epsilon_k} \cap {\mathcal D}_{\rm fp}(\rho_{\text{init}}) \subseteq \mathcal{B}^\delta \rbrace,
\end{align}
where $\lbrace \epsilon_k \rbrace_{k\in\mathbb{N}}$ is a decreasing  sequence of positive numbers with $\lim_{k\to\infty} \epsilon_k=0$. Based on P.5 Lemma \ref{prop}, the sequence $\lbrace \delta_{\epsilon_k} \rbrace_{k\in\mathbb{N}}$ is non-increasing and it is also lower bounded. This means that $\lim_{k\to\infty} \delta_k$ exists.
Next, we show that $\lim_{k\to \infty} \delta_{\epsilon_k}=0$. To this end, assume by contradiction that $\lim_{k\to \infty} \delta_{\epsilon_k}= \delta_0 > 0$. %Then, there exists a sequence $\lbrace \epsilon_k \rbrace_{k\in\mathbb{N}}$ such that 
%\begin{align}
%\begin{cases}
%\lim_{k\to \infty} \epsilon_k=0\\
%\lim_{k\to\infty} \delta_{\epsilon_k}=\delta_0 >0,
%\end{cases}
%\end{align}
%where the sequence $\lbrace \epsilon_k \rbrace_{k\in\mathbb{N}}$ can be assumed to be deceasing without loss of generality.
This means that there exists $k_0\in\mathbb{N}$ such that for all $k \geq k_0$, we have $\delta_{\epsilon_k} > \delta_0/2$. Hence, for any $k\geq k_0$, we have 
\begin{align}
\mathcal{D}_{\text{feas}}^{\epsilon_k} \cap {\mathcal D}_{\rm fp}(\rho_{\text{init}}) \subseteq \mathcal{B}^\delta \Rightarrow \delta > \delta_0/2,
\end{align}
which means that for any $k\geq k_0$ and $0 \leq \delta \leq \delta_0/2$, we have $\mathcal{D}_{\text{feas}}^{\epsilon_k} \cap {\mathcal D}_{\rm fp}(\rho_{\text{init}}) \not\subset \mathcal{B}^\delta$. Now, define the following set
\begin{align}
\mathcal{C}^k=\mathcal{D}_{\text{feas}}^{\epsilon_k} \cap {\mathcal D}_{\rm fp}(\rho_{\text{init}}) \cap \widehat{\mathcal{B}}^\delta,
\end{align}
where $\widehat{\mathcal{B}}^\delta=\lbrace \vx\in\mathbb{R}^2: \norm{\vx-\vxi} \geq \delta \rbrace$. Based on the previous analysis and P.5 in Lemma \ref{prop}, the set $\mathcal{C}^k$ is nonempty, compact and decreasing for any $k\geq k_0$ and $0 \leq \delta \leq \delta_0/2$. This implies that
\begin{align}
\bigcap_{k\geq k_0} C^k \neq \emptyset.
\end{align}
From P.5 in Lemma \ref{prop}, note that $\cap_{k\geq k_0} C^k=\mathcal{D}_{\text{feas}} \cap {\mathcal D}_{\rm fp}(\rho_{\text{init}}) \cap \widehat{\mathcal{B}}^\delta$ and $\mathcal{D}_{\text{feas}} \cap {\mathcal D}_{\rm fp}(\rho_{\text{init}})=\lbrace \vxi \rbrace$. This means that for any  $0 \leq \delta \leq \delta_0/2$, we have
$
\lbrace \vxi \rbrace \cap \widehat{\mathcal{B}}^\delta \neq \emptyset
$.
Hence, $\delta_0=0$ which leads to a contradiction. We then conclude that $\lim_{\epsilon_k \to 0} \delta_{\epsilon_k}=0$. Based on Theorem \ref{lem:fb}, i.e., 
$
\lim_{n\rightarrow\infty} \Pro( \mathcal{S}_{\text{feas}}  \subseteq \mathcal{D}_{\text{feas}}^{\epsilon}  ) = 1, \forall \epsilon>0
$, it holds
$$
\lim_{n\rightarrow\infty} \Pro\Big( \mathcal{S}_{\text{feas}} \cap   {\mathcal D}_{\rm fp}(\rho_{\text{init}})  \subseteq \mathcal{D}_{\text{feas}}^{\epsilon} \cap  {\mathcal D}_{\rm fp}(\rho_{\text{init}}) \Big) = 1, \forall \epsilon>0.
$$
Hence, for any decreasing sequence of positive numbers $\lbrace \epsilon_k \rbrace_{k\in\mathbb{N}}$ with $\lim_{k\to\infty} \epsilon_k=0$, there exists a  sequence of positive numbers $\lbrace \delta_k \rbrace_{k\in\mathbb{N}}$ such that
\begin{align}
\lim_{n\rightarrow\infty} \Pro\Big( \sup_{\widehat{\vx}\in \mathcal{S}_{\text{max}}} \norm{\widehat{\vx}-\vxi}_2 \leq \delta_k \Big) = 1,~\forall k \in \mathbb{N},
\end{align}
where $\lim_{k\to\infty} \delta_{\epsilon_k}=0$, the sequence $\lbrace \delta_{\epsilon_k} \rbrace_{k\in\mathbb{N}}$ is defined in \eref{depslon} and $\mathcal{S}_{\text{max}}$ denotes the set of optimal solutions of PhaseMax. 
%Based on property $5$ in Lemma \ref{prop}, we have $\lim_{\epsilon\to 0} \delta_\epsilon=0$. 
This implies that the set $\mathcal{S}_{\text{max}}$ converges to the set $\lbrace \vxi \rbrace$ in the sense that $\sup_{\widehat{\vx}\in \mathcal{S}_{\text{max}}} \norm{\widehat{\vx}-\vxi}_2$ converges to zero in probability, which completes the proof.
\end{IEEEproof}
 %On the other hand, the necessary and sufficient conditions that are derived in the next section will be under the assumption that $\vx_{\text{init}}$ is independent of the sensing vectors.

%!TEX root = phaselamp.tex

\section{Precise Analysis of PhaseMax}
\label{sec:main_results}

In Section \ref{sec:suff_PM} we derived a \emph{sufficient} condition for perfect recovery of PhaseMax. In this section, we establish a \emph{tight} such result by further assuming that the initial guess vector $\vx_\text{init}$ is independent of the sensing vectors $\set{\va_i}_{1 \le i \le m}$ and the target signal $\vxi$. In particular, we precisely characterize the minimum required number of measurements as a function of the input {cosine similarity} $\rho_\text{init}$ so that PhaseMax finds the true vector $\vxi$. Moreover, when this is not possible we precisely quantify the NMSE.

\subsection{Fundamental Limits}
%In this section, we characterize the asymptotic NMSE of the PhaseMax method under the stated assumptions. Additionally, we assume that the initial guess vector $\vx_\text{init}$ is independent from the sensing vectors $\set{\va_i}_{1 \le i \le m}$ and the target signal $\vxi$. Our results point out 

In order to state our results we need a few definitions. For any fixed {cosine similarity} $\rho_\text{init}$ and fixed oversampling ratio $\alpha > 2$, define $s^{\ast}$ as follows:
\begin{equation}\label{eq:fconprob}
\begin{aligned}
s^{\ast}&:=\underset{0\leq s \leq 1}{\arg\,\max}\ \ {\rho_{\text{init}}} s+\sqrt{(1-\rho_{\text{init}}^2)g_{\alpha}(s)},
\end{aligned}
\end{equation} 
where the function $g_\alpha$ (parametrized by $\alpha$) is given by
\begin{equation}\label{eq:fun_g}
g_{\alpha}(s):=-1-s^2+\frac{2\alpha \, r_{\alpha}(s)}{\pi}+\frac{2\alpha s}{\pi }\text{atan}\left( \frac{s}{r_{\alpha}(s)+c_{\alpha}} \right),
\end{equation}
%where the function $t$ in \eref{fun_g} can be expressed as follows
with $c_{\alpha}=1/\text{tan}\left( \pi/\alpha \right)$ and 
\begin{equation}\label{eq:optrfun}
r_{\alpha}(s):=\sqrt{c_{\alpha}^2+1-s^2}-c_{\alpha}.
\end{equation}
Moreover, define
\begin{equation}\label{eq:rstar}
r^\ast := r_{\alpha}(s^\ast).
\end{equation}

We are now ready to state the main result of this section. Its proof uses the recently developed CGMT framework \cite{chris:151,chris:152} and is deferred to Section \ref{phmax_cgmt}.
\vsp
%\christos{@oussama: please change stars to hats as needed.}
\begin{theorem}[Asymptotic properties of PhaseMax]\label{thm:the1}
Assume that $\vx_\text{init}$ is independent of the sensing vectors $\set{\va_i}_{1 \le i \le m}$ and of the target signal $\vxi$. For any fixed input {cosine similarity} $\rho_\text{init} > 0$ and any fixed oversampling ratio $\alpha > 2$, let $s^\ast, r^\ast$ be defined as in \eqref{eq:fconprob} and \eqref{eq:rstar}, respectively.  Then,  the NMSE of the PhaseMax method converges in probability as follows:
\begin{align}\label{eq:anmse}
\mathrm{NMSE}_n \xrightarrow[]{n\to\infty} 1+(s^\ast)^2+(r^\ast)^2-2\abs{s^\ast}.
\end{align}
Moreover, the optimal cost and the optimal solution $\widehat{\vx}$ of PhaseMax satisfy the following :
\begin{align}\label{eq:xstar}
&s(\widehat{\vx}) \xrightarrow[]{n\to\infty} s^\ast~~\text{and}~~ r(\widehat{\vx}) \xrightarrow[]{n\to\infty} r_\alpha(s^\ast), \\
& \vx_{\text{init}}^T \widehat{\vx} \xrightarrow[]{n\to\infty} {\rho_{\text{init}}} s^\ast+\sqrt{(1-\rho_{\text{init}}^2)g_{\alpha}(s^\ast)},
\end{align}
where $\widehat{\vx}=[s(\widehat{\vx})~~\widetilde{\vx}(\widehat{\vx})^T]^T$ and $r(\widehat{\vx})=\norm{\widetilde{\vx}(\widehat{\vx})}_2$.
\end{theorem}
%\begin{IEEEproof}
%The detailed proof is given in Section \ref{phmax_cgmt}.
%\end{IEEEproof}

\vspace{2pt}
Theorem \ref{thm:the1} accurately predicts the NMSE of PhaseMax in the large system limit. The formulae involve solving the one-dimensional deterministic maximization problem in \eqref{eq:fconprob}. In Section \ref{sspop} we show that this optimization is strictly concave, thus, $s^\ast$ is unique and can be efficiently determined by solving a fixed point equation.   
%
% The prediction is expressed in terms of $s^\ast$, the solution to the one-dimensional deterministic maximization problem in \eqref{eq:fconprob}. As shown in Section \ref{sspop}, this optimization problem is strictly concave, thus, $s^\ast$ can be uniquely determined by a fixed point equation. 

Clearly, we can use the formula on the NMSE given by Theorem \ref{thm:the1} to quantify necessary and sufficient conditions under which zero error is achieved. This is the content of the next theorem, which we prove in Section \ref{pr_th2}.

%%%%%%%%%%%%%%%%%%%%%%%%%%%%
%%%%%%%%%%%%%%%%%%%%%%%%%%%%
\begin{theorem}[Phase transition of PhaseMax]\label{thm:pt}
Let the same assumptions as in Theorem \ref{thm:the1} hold. Further assume  $\alpha > 2$. Then, PhaseMax perfectly recovers the target signal (in the sense that $\mathrm{NMSE}_n \xrightarrow[]{n\to\infty} 0$, in probability) \emph{if and only if}
\begin{align}\label{eq:phb}
\rho_\text{init} \geq \sqrt{1- \frac{{\pi/\alpha}}{\tan({\pi}/{\alpha} )}} =: \rho_c(\alpha).
%\frac{\frac{\pi}{\alpha}}{\tan(\frac{\pi}{\alpha} )} > 1 - \rho_\text{init}^2.
\end{align}
%Moreover, when $1<\alpha<2$, PhaseMax can not recover the target signal, i.e. $\mathrm{NMSE}_n \xrightarrow[]{n\to\infty} \mathrm{NMSE}>0$, for any $\rho_\text{init}$.
\end{theorem}
\noindent Theorem \ref{thm:pt} establishes a precise phase transition behavior on the performance of PhaseMax: for any fixed oversampling ratio $\alpha > 2$, there is a critical cosine similarity $\rho_c(\alpha)$ such that the algorithm perfectly recovers the target signal vector $\vxi$ if and only if $\rho_\text{init} \geq \rho_c(\alpha)$. 

\begin{figure}[t!]
    \centering
    \subfigure[]{\label{fig:nmse_1}
    \includegraphics[width=0.47\linewidth]{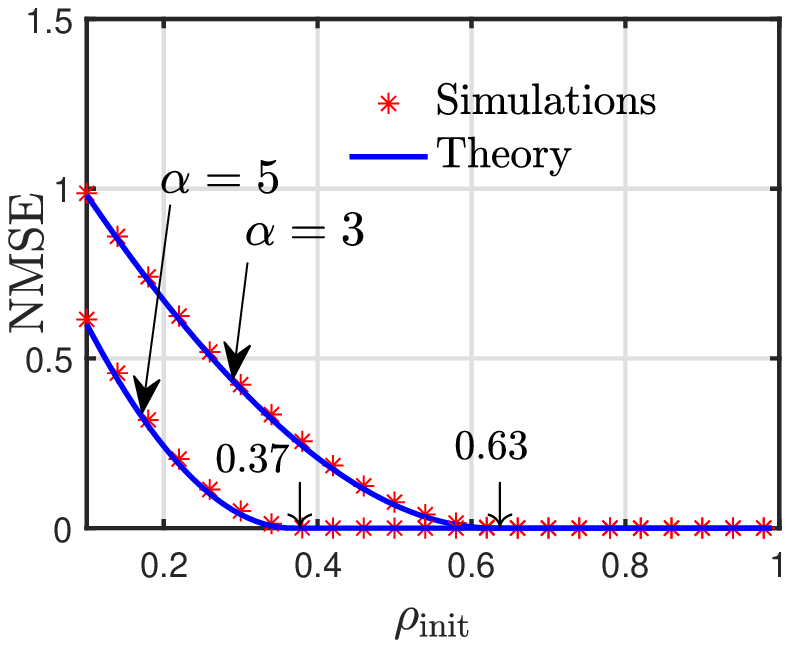}
    }
    \subfigure[]{\label{fig:nmse_2}
        \includegraphics[width=0.47\linewidth]{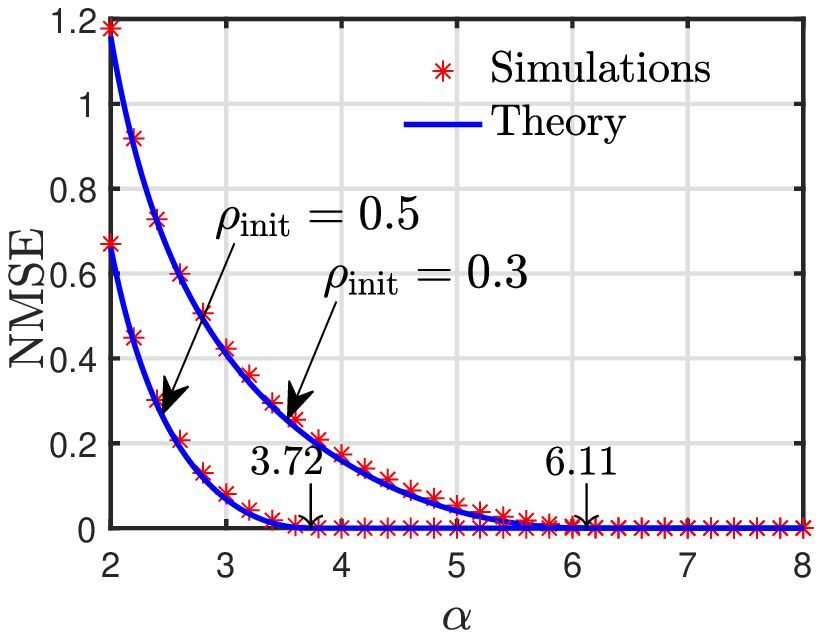}}

    \caption{Asymptotic predictions v.s. numerical simulations. (a) The NMSE of the PhaseMax method as a function of $\rho_\text{init}$, for two different values of $\alpha$; (b) The NMSE of the PhaseMax method as a function of $\alpha$, for two different values of $\rho_\text{init}$. The results are averaged over $50$ independent Monte Carlo trials. The asymptotic formulas are also in excellent agreement with the actual performance the PhaseMax method, for $n = 1000$.}
        \label{fig1}
\end{figure}

\subsection{Numerical Simulations}

The numerical results presented in this section aim to verify the validity of Theorems \ref{thm:the1} and \ref{thm:pt}. 
%
%We investigate the performance of our asymptotic predictions for PhaseMax given in \eref{anmse}. Specifically, we compare the asymptotic predictions against simulation results for different values of the input cosine similarity $\rho_{\text{init}}$ and the oversampling ratio $\alpha$. 
%
First, \fref{nmse_1} illustrates the NMSE of PhaseMax as a function of the input cosine similarity given in \eref{cosin}, for two different values of the oversampling ratio $\alpha$. For the simulations, we solve the convex optimization problem \eref{lp_form} using the techniques introduced in \cite{fasta} and we set the signal dimension as $n=1000$.  Note that the asymptotic prediction of Theorem \ref{thm:the1} is in excellent agreement with the actual performance of the PhaseMax method in finite dimensions. Of course, the same holds true for the recovery condition of Theorem \ref{thm:pt}: the theoretical values $\rho_{\text{init}}(\alpha=3) \approx 0.63$ and $\rho_{\text{init}}(\alpha=5)\approx 0.37$ perfectly match with the simulation results
Next, \fref{nmse_2} plots the NMSE of PhaseMax as a function of the oversampling ratio, for two different values of the input cosine similarity. Again, the figure highlights the sharpness of the results in Theorems \ref{thm:the1} and \ref{thm:pt}. %\christos{In figures, move NMSE label to y-axis (please do this in other figures as well as is standard. eg. Fig 6 and 7). Also, I suggest changing legend labels to: Simulations and Theory (instead of Monte Carlo and CGMT Prediction)}

%%%%%%%%%%%%%%%%%%%%%%%%%%%%%%%%%%%%%%%%%%%%%
%%%%%%%%%%%%%%%%%%%%%%%%%%%%%%%%%%%%%%%%%%%%%
%%%%%%%%%%%%%%%%%%%%%%%%%%%%%%%%%%%%%%%%%%%%%
%In the second example, we investigate the performance of the PhaseMax method when the oversampling ratio is smaller than $2$. Specifically, we compare our asymptotic predictions against simulation results for $\alpha < 2$. \fref{alps2} illustrates the NMSE of PhaseMax as a function of the input cosine similarity given in \eref{cosin}, for two different values of the oversampling ratio, $\alpha=1.8$ and $\alpha=1.6$.
%\begin{figure}[h!]
%    \centering
%    \includegraphics[width=0.55\linewidth]{figs/palps2.pdf}
%    \caption{The NMSE of the PhaseMax method as a function of $\rho_\text{init}$, for two different values of $\alpha$. The results are averaged over $50$ independent Monte Carlo trials.}
%        \label{fig:alps2}
%\end{figure}
%\fref{alps2} shows that our theoretical symptotic predictions are in excellent agreement with the actual performance of PhaseMax. %It can be noticed that the NMSE of the PhaseMax method is strictly positive  in the high dimensional limit for any input cosine similarity $\rho_\text{init}$ which validates the theoretical result stated in Proposition \ref{suffcondd}.
%%%%%%%%%%%%%%%%%%%%%%%%%%%%%%%%%%%%%%%%%%%%%
%%%%%%%%%%%%%%%%%%%%%%%%%%%%%%%%%%%%%%%%%%%%%
%%%%%%%%%%%%%%%%%%%%%%%%%%%%%%%%%%%%%%%%%%%%%
%%%%%%%%%%%%%%%%%%%%%%%%%%%%%%%%%%%%%%%%%%%%%%%%%%
%\section{Successive Linearization and Maximization}

%!TEX root = phaselamp.tex

\section{Nonconvex Formulation and New Algorithms}
\label{sec:slam_alg}

In this section, we propose and study an improved algorithm over PhaseMax, which we call PhaseMax. The natural idea behind PhaseLamp is to solve a \emph{sequence} of PhaseMax problems. Interestingly, we provide an interpretation of this algorithm as an iterative  method for solving the non-convex phase-retrieval problem formulation in \eref{qb_form}. This interpretation leads to strong performance guarantees for PhaseLamp in Section \ref{sec:th_PL}. Finally, in Section \ref{sec:WPL} we propose yet one more recovery algorithm, which is also based on optimization over polytopes and which appears to outperform both PhaseMax and PhaseLamp in numerical simulations.

\subsection{PhaseLamp}
%
%This section proposes an efficient iterative algorithm to solve the norm maximization problem formulated in \eref{qb_form}. 
We begin our exposition by arguing in Proposition \ref{gos} that, given enough measurements, the solution to the system of quadratic equations in \eqref{eq:abs} can be found by solving the optimization problem in \eref{qb_form}. The proof is  in Appendix \ref{gosp}.
%
%The following Proposition provides a sufficient condition under which the target signal vectors $\vxi$ and $-\vxi$ are the unique global optimal solutions of the PhaseLamp formulation \eref{qb_form}.
\begin{proposition}[New non-convex formulation]\label{gos}
Assume that $\alpha$ satisfies $\alpha>\frac{\pi}{\pi-2}$. Then, the set of optimal solutions of the PhaseLamp problem \eref{qb_form} converges to the set $\lbrace \vxi,-\vxi \rbrace$ in the sense that 
$
\sup_{\widehat{\vx}\in \mathcal{S}_{\text{lamp}}} (\min\lbrace \norm{\widehat{\vx}-\vxi}_2, \norm{\widehat{\vx}+\vxi}_2 \rbrace) 
$
converges to zero in probability, where $\mathcal{S}_{\text{lamp}}$ denotes the set of optimal solutions of the PhaseLamp problem given in \eref{qb_form}. 
%Then, the target signal vectors $$ and $-\vxi$ are the unique global optimal solutions of the optimization problem in \eref{qb_form} in the high-dimensional limit. \christos{the statement needs to be made rigorous. What does it mean optimal solution in the high-dimensional limit?}
\end{proposition}

According to the proposition, if the number of measurements satisfies $\alpha>\frac{\pi}{\pi-2} \approx 2.752$, then one can hope of solving the phase-retrieval problem by finding the optimal solution of the optimization problem in \eref{qb_form}. Unfortunately,  \eref{qb_form} is clearly non-convex since it involves maximizing a convex function over a convex set. 

In this section, we propose solving \eref{qb_form}  by using a standard minorization-maximization (MM) approach \cite{Lange:16} as follows. Start by observing that because of convexity the cost function $\norm{\vx}_2^2$ satisfies
\begin{equation}
\norm{\vx}_2^2 \geq \vx_k^T\vx_k+2\vx_k^T\left( \vx-\vx_k \right), \forall \vx_k,\vx \in\mathbb{R}^n. \label{eq:f_order_conv}
\end{equation}
Equivalently, the function $m_k:\vx \to \vx_k^T\vx_k+2\vx_k^T\left( \vx-\vx_k \right)$ is a minorizer of the function $m:\vx \to \norm{\vx}_2^2$. Moreover, the function $m_k$ satisfies $m_k(\vx_k)=m(\vx_k)$. Hence, it is natural to attempt solving \eref{qb_form}, via the following iterative scheme
\begin{equation*}
\begin{aligned}
{\vx}_{k+1}&=\underset{{\vx}}{\arg\,\max}~~~ m_k(\vx_k)\\	
&~~~~~~~~\text{s.t.}~~~~  \abs{\va_i^T \vx} \leq y_i, \text{ for }  1 \le i \le m,
\end{aligned}
\end{equation*}
which is of course equivalent to the following:
\begin{align}
\label{eq:itphmax_form}
{\vx}_{k+1}&=\underset{{\vx}}{\arg\,\max}~~~ \vx_k^T\vx\\	
&~~~~~~~~\text{s.t.}~~~~  \abs{\va_i^T \vx} \leq y_i, \text{ for }  1 \le i \le m.\nonumber
\end{align}
We call this iterative algorithm the PhaseLamp, which owes its name to the idea of successive \emph{linearization and maximization over a polytope}. PhaseLamp starts from an initial guess $\vx_0 =\vx_{\mathrm{init}}$ of the target vector $\vxi$ and terminates when the number of iterations exceeds a pre-specified number $I_{\text{max}}$ or when $\norm{\vx_{k+1}-\vx_{k}}_2 \leq \epsilon$ for some fixed threshold $\epsilon > 0$.  It can be shown that the sequence $\left\lbrace \vx_k \right\rbrace_{k=1}^{\infty}$ generated by iteratively solving \eqref{eq:itphmax_form} satisfies the following properties \cite[Theorem 4]{Gert:09}: (a) $\lbrace \norm{\vx_k}^2_2 \rbrace_{k=1}^{\infty}$ is a convergent nondecreasing sequence;
%\item [(3)] All the limit points of $\left\lbrace \vx_k \right\rbrace_{k=1}^{\infty}$ are stationary points of the norm maximization problem formulated in \eref{qb_form}.
(b) $\lim_{k\to \infty} \norm{\vx_k}^2_2=\norm{\vx_\ast}^2_2$, where $\vx_\ast$ is a stationary point of the norm maximization problem \eref{qb_form}.

\begin{figure}[t!]
    \centering
    \subfigure[]{\label{fig:slam_phmax21}
    \includegraphics[width=0.47\linewidth]{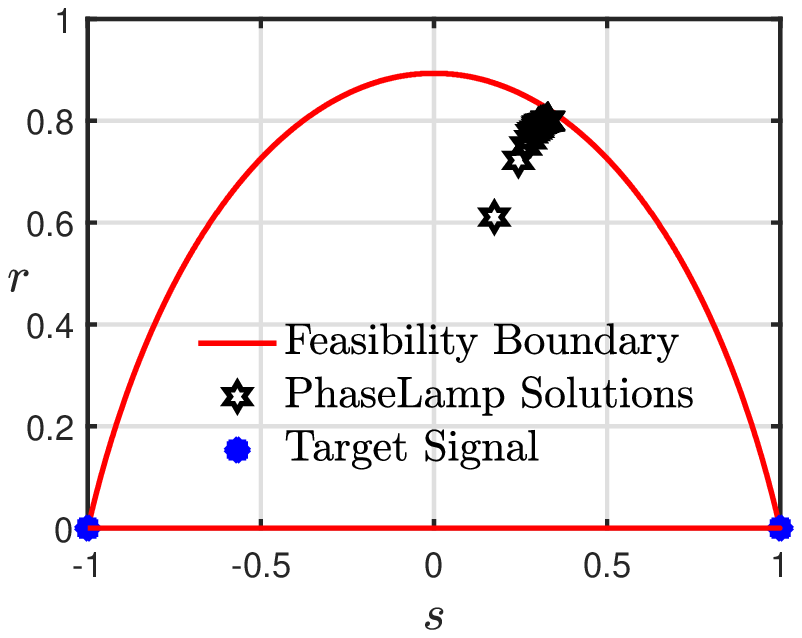}
    }
    \subfigure[]{\label{fig:slam_phmax22}
        \includegraphics[width=0.47\linewidth]{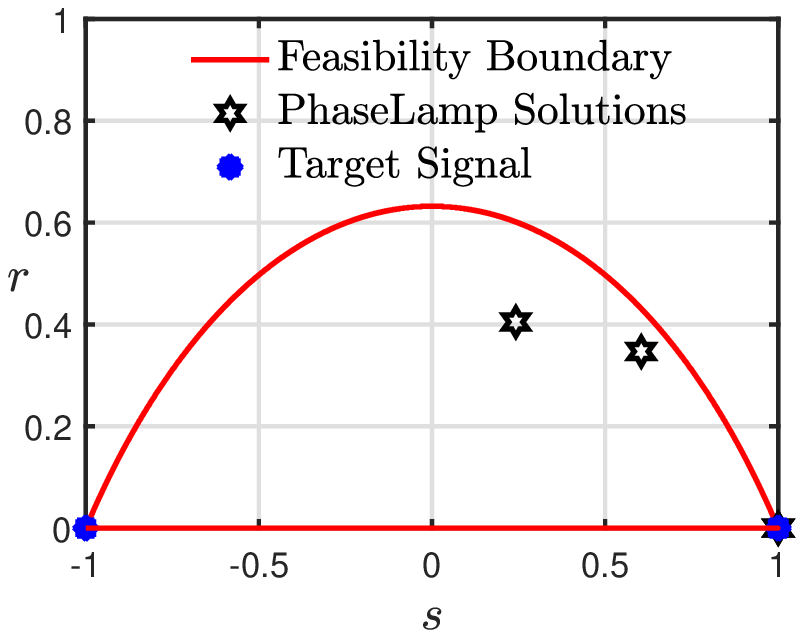}}

    \caption{The convergence behavior of PhaseLamp. $r$ as a function of $s$, for (a) $\alpha=3$ and $\rho_{\text{init}}=0.1$; (b) $\alpha=4$ and $\rho_{\text{init}}=0.1$. The stars denote the solutions of the sequence of PhaseMax problems given in \eref{itphmax_form}. The maximum number of iterations is set to $I_{\text{max}}=25$ and $\epsilon=10^{-4}$. For the simulations we have set $n=1000$.}
    \label{fig:fig21}
\end{figure}

However, due to the non-convexity of \eref{qb_form}, PhaseLamp is \emph{not} guaranteed in general to converge to the desired global optimal solution of \eref{qb_form}. The main theoretical result of this section involves identifying sufficient conditions under which this is indeed the case. Before formalizing those in Section \ref{sec:th_PL}, it is instructive to consider the performance of PhaseLamp on two different problem instances as shown in \fref{fig21}. Specifically, we present simulation results for the following two cases:  (a) $\alpha=3$ and $\rho_{\text{init}}=0.1$ (Figure \ref{fig:slam_phmax21}), and (b) $\alpha=4$ and $\rho_{\text{init}}=0.1$ (Figure \ref{fig:slam_phmax22}). First, in both instances $\alpha>2.752$; hence Proposition \ref{gos} guarantees that  the optimal solutions of \eref{qb_form} coincide with the target vectors $\vxi$ or $-\vxi$. However, as mentioned PhaseLamp is not always guaranteed to find the optimal solutions of \eref{qb_form}. For example, it fails to do so in \fref{slam_phmax21}, but it succeeds in \fref{slam_phmax22}. The sufficient conditions derived in the next section provide rigorous theoretical justifications to these observations. %\christos{@Oussama: as we said, might be nicer if there is an instance of the second plot where at least the first iteration of PhaseLamp gives a solution that is not that close to the target vector. Most important, explain on the caption of the figure what the stars represent.}

\subsection{Performance Guarantees for PhaseLamp}\label{sec:th_PL}
%%%%%%%%%%%%%%%%%%%%%%%%%%%%%%%%%%%%%%%%%%%%%%%%%%%%
%%%%%%%%%%%%%%%%%%%%%%%%%%%%%%%%%%%%%%%%%%%%%%%%%%%%
%%%%%%%%%%%%%%%%%%%%%%%%%%%%%%%%%%%%%%%%%%%%%%%%%%%%
%%%%%%%%%%%%%%%%%%%%%%%%%%%%%%%%%%%%%%%%%%%%%%%%%%%%
%
\begin{figure}[t!]
    \centering
    \includegraphics[width=0.32\textwidth]{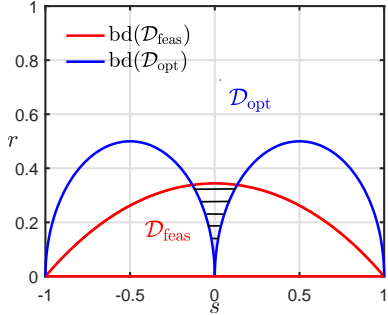}
    \caption{Illustration of the set of fixed points of PhaseLamp in the high-dimensional limit for $\alpha=7$. %{\color{blue}@Oussama: Please add grid and ylabel on the left of axis.}
    }
    \label{fig:stat}
\end{figure}

Clearly, PhaseLamp in the form of \eqref{eq:itphmax_form} can be naturally viewed as an iterative and \emph{bootstrapped} version of the PhaseMax method \eref{lp_form} where at each iteration $k \geq 1$ , the optimal solution at the previous iteration is used as an (improved) initial guess for a new iteration of PhaseMax. In other words, the cosine similarity $\rho^k_{\text{out}}$ between the PhaseMax solution at iteration $k$  and the target signal vector $\vxi$ serves as the input cosine similarity $\rho^{k+1}_{\text{init}}$ at iteration $k+1$. One may then imagine leveraging the analysis of PhaseMax in Section \ref{sec:main_results} to obtain similar sharp results for PhaseLamp. Unfortunately, more effort and several new arguments are required; the challenge becomes that, after the first iteration of PhaseLamp, the initial guess vector becomes dependent on the sensing vectors $\lbrace \va_i, 1 \leq i \leq m \rbrace$. 

In this section, we overcome these challenges, thus obtaining strong performance guarantees for PhaseLamp. Our arguments are geometric in nature, similar in nature (but somewhat more involved) to the proof of Proposition \ref{suffcondd} in Section \ref{sec:suf_PM}. 

On the one hand, the solution to each iteration of \eqref{eq:itphmax_form} is constrained to live in the feasibility set of PhaseMax. Thus, the same is true for the converging solution (cc. fixed point) of PhaseLamp. Combining this with Theorem \ref{lem:fb}, which obtains a sharp characterization of the high-dimensional geometry of this feasibility set (cc. its two dimensional projection), we conclude that the fixed points of PhaseLamp belong with probability approaching 1 as $n\rightarrow\infty$ to the set ${\mathcal D}_{\rm feas}^\epsilon$, for any $\epsilon>0$.

 On the other hand, any fixed point  $\widehat{\vx}$ of PhaseLamp satisfies
\begin{align}\label{eq:fp}
\widehat{\vx}&=\underset{{\vx}\in\mathbb{R}^n}{\arg\,\max}~~~ {\widehat{\vx}}^{T}\,{\vx}\\	
&~~~~~~~~\text{s.t.}~~~~  \abs{\va_i^T \vx} \leq y_i, \text{ for }  1 \le i \le m.\nonumber
\end{align}
From this and feasibility of the target vector $\vxi$, it holds
\begin{align}
s^2+r^2={\widehat{\vx}}^{T} {\widehat{\vx}} \geq \abs{ {\widehat{\vx}}^{T}\,\vxi }= \abs{ s }.
\end{align}
Concluding,  the fixed points of PhaseLamp belong to the following set
\begin{align}\notag
{\mathcal{D}}_{\text{opt}}=\left \lbrace (s,r)\in\mathbb{R}^2:~r\geq 0,~ s^2+r^2 \geq \abs{ s } \right\rbrace.
\end{align}
%For $s \geq 0$, the boundary of the set ${\mathcal{D}}_{\text{opt}}$ is a semi-circle where the origin is at $(0.5,0)$.
%\vsp

Overall, we have shown that in the limit of high-dimensions, the set of all possible fixed points of PhaseLamp belongs to the intersection of the two sets ${\mathcal D}^\epsilon_{\rm feas}$ and ${\mathcal{D}}_{\text{opt}}$. The sets ${\mathcal D}_{\rm feas}$ and ${\mathcal{D}}_{\text{opt}}$ are illustrated in \fref{stat} for $\alpha=7.$ Note that any $\epsilon$-perturbation of the shaded region union the points $(1,0)$ and $(-1,0)$ (ie., the set ${\mathcal D}^\epsilon_{\rm feas}\cap{\mathcal{D}}_{\text{opt}}$, for any $\epsilon>0$) represents the set of possible fixed points of PhaseLamp. Clearly PhaseLamp is successful when it escapes the shaded region of ``bad" stationary points. Hence, the question becomes: for given $\alpha$, what values of initial correlation $\rho_\text{init}$ guarantee escaping the bad region? We answer this in Section \ref{suffcond}; we defer the details to that latter section and only present the final result below.

\subsubsection{General initialization}
The theorem below provides an efficient sufficient condition for perfect recovery using PhaseLamp.
\begin{theorem}[Sufficient condition for perfect recovery]
\label{the2a}
For any $\alpha>2$, the PhaseLamp perfectly recovers the unknown signal, i.e., it holds in probability that
$\mathrm{NMSE}_n \xrightarrow[]{n\to\infty}0$, if
\begin{align}\label{eq:slampd}
\rho_\text{init} > \sin(\theta^\ast_\alpha)=:\widehat{\rho}_{s}(\alpha),
\end{align} %\left( 1+\cos^2(\theta^\ast_\alpha) \right) =: \widehat{\rho}_s(\alpha).
where $\theta^\ast_\alpha$ is the unique solution in the interval $(0, \pi/2)$ of the following equation:
\begin{equation}\label{eq:slam_fp}
\begin{aligned}
& \theta \cos^2\theta + (1 + 3 \sin^2\theta) \atan\left(\frac{\sin \theta \cos \theta}{1 + \sin^2 \theta}\right)= \nonumber\\
&\quad~\quad 2 \sin \theta \cos \theta + \Big(\frac{\pi}{\alpha}\Big) \sin^2 \theta \cos^2 \theta.
\end{aligned}
\end{equation}
\end{theorem}

\begin{figure}[t!]
    \centering
    \includegraphics[width=0.7\linewidth]{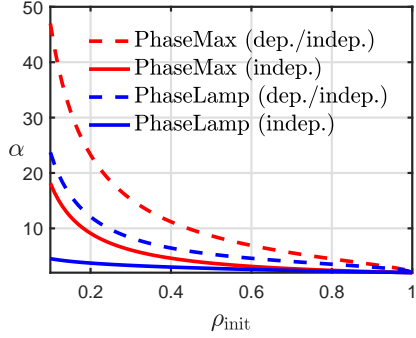}
    \caption{The oversampling ratio $\alpha$ as a function of the input cosine similarity $\rho_{\text{init}}$. The red dashed line shows the sufficient condition for PhaseMax given in \eref{scpmax}. The blue dashed line shows the sufficient condition for PhaseLamp given in \eref{slampd}. The red solid line shows the sharp phase transition boundary of PhaseMax given in \eref{phb}. The blue solid line shows the sufficient condition for PhaseLamp given in \eref{slampind}.%\color{blue}{Oussama: please edit legend}.
    }
    \label{fig:scond}
\end{figure}
\noindent Note that the sufficient condition for PhaseMax and PhaseLamp given in \eref{scpmax} and \eref{slampd}, respectively, are valid for \textit{any} initial guess vector $\vx_{\text{init}}$, which can depend on the sensing vectors $\lbrace \va_i, 1 \leq i \leq m \rbrace$ and the target signal $\vxi$. It can be noticed that the PhaseLamp largely improves the performance of the PhaseMax method for dependent and independent initial guess vector $\vx_{\text{init}}$.

For a better interpretation of the theorem,  we have depicted the sufficient recovery condition in Figure \ref{fig:scond}. In particular, the theorem guarantees that all pairs $(\alpha,\rho_{\text{init}})$ that are above the blue dashed curve lead to perfect recovery performance of PhaseLamp. In the same figure, we also depict in red dashed line the corresponding sufficient condition of PhaseMax from Proposition \ref{suffcondd}. Clearly, these results indicate that PhaseLamp outperforms PhaseMax in the sense that it achieves perfect recovery for a larger range of input parameters $(\alpha,\rho_{\text{init}})$.

\vspace{2pt}
\subsubsection{Independent initialization}
Similar to Section \ref{sec:main_results}, if the initial vector $\vx_0 = \vx_{\text{init}}$ is \emph{independent} of the sensing vectors and of the target vector, then we can obtain sharper recovery guarantees as shown in the proposition below. For the statement of the proposition it is convenient to first define the following:
\begin{align}
\widehat{s}_\alpha := \frac{\text{tan}(\theta^\ast_\alpha)}{\sqrt{1+c_{\alpha}^2+\text{tan}(\theta^\ast_\alpha)^{2}}+c_{\alpha}},
\end{align}
%\big[\sqrt{\text{tan}(\theta^\ast_\alpha)^{-2}+c_{\alpha}^2~\text{tan}(\theta^\ast_\alpha)^{-2}+1}+c_{\alpha}~\text{tan}(\theta^\ast_\alpha)^{-1}\big]^{-1},
where $c_{\alpha}=1/\text{tan}\left( \pi/\alpha \right)$, and 
\begin{align}
\ell_\alpha := \frac{\widehat{s}_\alpha-\frac{\alpha}{\pi}\text{atan}\left( \frac{\widehat{s}_\alpha}{\sqrt{ c_{\alpha}^2+1-\widehat{s}_\alpha^2} } \right)}{\sqrt{g_\alpha(\widehat{s}_\alpha)}},
\end{align}
where $g_\alpha(\cdot)$ is the function defined in \eref{fun_g}. 

\begin{theorem}[Sufficient condition: independent initialization]
\label{the2b}
Assume that $\vx_{\text{init}}$ is independent of the sensing vectors $\lbrace \va_i: 1\leq i \leq m\rbrace$ and of the target vector. Then, for any $\alpha>2$, it holds in probability that
$\mathrm{NMSE}_n \xrightarrow[]{n\to\infty}0$, if
\begin{align}\label{eq:slampind}
\rho_\text{init} > \frac{\ell_\alpha}{\sqrt{\ell^2_\alpha+1}} =: \rho_s(\alpha).
\end{align} 
\end{theorem}

The sufficient condition of the theorem is depicted in blue solid line in \fref{scond}. Observe that it is a "stronger" condition than that of Theorem \ref{the2a} when the initialization vector is independent of the sensing vectors and of the true signal. Also, observe by comparison with the red solid line, which represents the result of Theorem \ref{thm:pt}, that PhaseLamp outperforms Phasemax. In fact, this statement is provable since the condition of Theorem \ref{thm:pt} is not only sufficient, but also necessary.

Finally, despite the condition of Theorem~\ref{the2b} being only a sufficient one,  the simulation results in \fref{phase} suggest that it still provides a reasonably tight bound on the actual performance of PhaseLamp.

\subsection{Weighted PhaseLamp}
\label{sec:WPL}
%%%%%%%%%%%%%%%%%%%%%%%%%%%%%%%%%%%%%%
\begin{figure}[t!]
    \centering
    \includegraphics[width=0.35\textwidth]{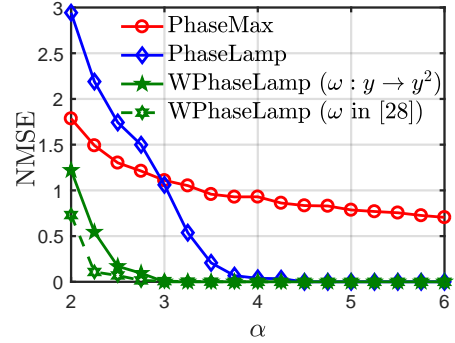}
    \caption{The NMSE of the PhaseMax, PhaseLamp and Weighted PhaseLamp methods as a function of the oversampling ratio $\alpha$ for real Gaussian sensing vectors. The algorithms are initialized randomly, i.e. $\vx_{\text{init}} \sim \mathcal{N}({\bf 0},\mI)$. The signal dimension is set to $n=200$, the maximum number of iterations is set to $I_{\text{max}}=25$ and the precision is set to $\epsilon=10^{-4}$. The results are averaged over $150$ independent Monte Carlo trials.}
    \label{fig:mpl}
\end{figure} 
%%%%%%%%%%%%%%%%%%%%%%%%%%%%%%%%%%%%%%%%
%%%%%%%%%%%%%%%%%%%%%%%%%%%%%%%%%%%%%%%%
%%%%%%%%%%%%%%%%%%%%%%%%%%%%%%%%%%%%%%%%
The Weighted PhaseLamp (WPhaseLamp) method is an alternative nonconvex formulation of the phase retrieval problem. Specifically, it consists of formulating the phase retrieval problem as a quadratic program
\begin{equation}\label{eq:wphlamp}
\begin{aligned}
\widehat{\vx}&=\underset{{\vx}\in\mathbb{R}^n}{\arg\,\max}~~~ {\vx}^{T} \mD_m \,{\vx}\\	
&~~~~~~~~\text{s.t.}~~~~  \abs{\va_i^T \vx} \leq y_i, \text{ for }  1 \le i \le m,
\end{aligned}
\end{equation} 
where
\begin{align}
\mD_m = \frac{1}{m} \sum_{i=1}^{m} \omega(y_i) \va_i \va_i^T,
\end{align}
and $\omega(\cdot)$ is a preprocessing function. Note that the cost function of the optimization problem \eref{wphlamp} is a weighted version of the PhaseLamp problem formulated in \eref{qb_form} where the weights depend on the sensing vectors $\lbrace \va_i, 0\leq i \leq m \rbrace$ and the target signal vector $\vxi$. It can also be noticed that the problem given in \eref{wphlamp} is the spectral initialization problem \cite{LuL:17,SIopt17} where only the unit norm constraint is replaced by the linear PhaseMax constraints.

%{\color{blue} @Christos: I generalized the Weighted PhaseLamp algorithm such that I can consider the optimal preproccessing function proposed by Montanari. I am running a simulation to observe the performance.}

In general, the preprocessing function $\omega(\cdot)$ can have negative output values \cite{SIopt17}. Hence, the matrix $\mD_m$ is indefinite in general. This means that the cost function of the problem \eref{wphlamp} is not convex or concave in general. %which means that the optimization problem \eref{wphlamp} is nonconcave. 
Write the matrix $\mD_m$ as follows
\begin{align}
\mD_m= \mD_m^{(1)}-\mD_m^{(2)},
\end{align}
where $\mD_m^{(1)}$ is constructed using the negative eigenvalues of $\mD_m$ and $\mD_m^{(2)}$ is constructed using the negatives of the positive eigenvalues of $\mD_m$. This means that the cost function of the Weighted PhaseLamp problem \eref{wphlamp} can be expressed as a difference of concave functions. Therefore, one can use the convex-concave procedure \cite{Gert:09} to efficiently solve the Weighted PhaseLamp problem \eref{wphlamp}.  Specifically, the proposed Weighted PhaseLamp algorithm consists of the following iterative scheme
\begin{align}\label{eq:wpro}
{\vx}_{k+1}&=\underset{{\vx}\in\mathbb{R}^n}{\arg\,\max}~~~ \vx^T \mD_m^{(1)} \vx -  2{\vx}_{k}^{T} \mD_m^{(2)} \,{\vx}\\	
&~~~~~~~~\text{s.t.}~~~~  \abs{\va_i^T \vx} \leq y_i, \text{ for }  1 \le i \le m,\nonumber
\end{align} 
for $k \geq 0$, where $\vx_0 =\vx_{\mathrm{init}}$ is an initial guess of the target vector $\vxi$. Note that when $\omega(\cdot)$ is a positive preprocessing function, $\mD_m^{(1)}$ is the zero matrix. In this case, the iterative procedure in \eref{wpro} solves a linear program in each iteration which is similar to the PhaseLamp algorithm \eref{itphmax_form}.

The analysis of the Weighted PhaseLamp method is left for future work. Next, we provide a simulation example to compare the recovery performance of the Weighted PhaseLamp, the PhaseLamp and the PhaseMax methods. To this end, we set the signal dimension to $n = 200$ and we initialize the algorithms randomly.
\fref{mpl} plots the NMSE as a function of the oversampling ratio $\alpha$. It can be noticed that the Weighted PhaseLamp method provides a better recovery performance as compared to the PhaseLamp method and the PhaseMax method for the considered preprocessing functions. Note that the critical oversampling ratio $\alpha_c$ needed by the Weighted PhaseLamp method for $\omega:y\to y^2$ is around $\alpha_c\approx 3$. Whereas, it is around $\alpha_c\approx 4$ for the PhaseLamp method. Additionally note that the optimal preprocessing function introduced in \cite{SIopt17} outperforms the preprocessing function $\omega:y\to y^2$.

\section{Additional Numerical Results}
\label{sec:addsim}

In this section, we present additional simulation results and we compare the performance of polytope-optimization based methods (i.e, PhaseMax, PhaseLamp, WPhaseLamp) to other existing recovery methods in the literature; in particular, Fienup \cite{Fienup:82}, Wirtinger Flow (WF) \cite{Candes:2015fv}, Truncated amplitude flow (TAF) \cite{WangGY:2016}, PhaseLift \cite{phlift13}. All algorithms are initialized using the optimal spectral initialization proposed in \cite{SIopt17} and the optimization problems are solved using the PhasePack \cite{phasepack17}. In our simulations we consider the following two cases on the measurement vectors: (1) random complex Gaussian measurements, and (2) coded diffraction patterns. 

\vspace{2pt}
\subsection{Complex Measurements}
%%%%%%%%%%%%%%%%%%%%%%%%%%%%%%%%%%%%%%%%
%%%%%%%%%%%%%%%%%%%%%%%%%%%%%%%%%%%%%%%%
%%%%%%%%%%%%%%%%%%%%%%%%%%%%%%%%%%%%%%%%

\begin{figure}[t!]
    \centering
    \includegraphics[width=0.32\textwidth]{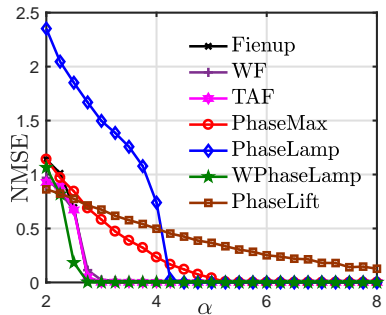}
    \caption{The NMSE as a function of $\alpha$ for complex Gaussian sensing vectors. For Weighted PhaseLamp, the preprocessing function is $\omega:y\to y^2$. The signal dimension is set to $n=400$, the maximum number of iterations is set to $I_{\text{max}}=40$ and the precision is set to $\epsilon=10^{-4}$. The results are averaged over $15$ independent Monte Carlo trials.}
    \label{fig:comp}
\end{figure}

First, we consider sensing vectors that follow a circularly symmetric normal distribution, i.e., $\va_i\sim\mathcal{CN}(0,{\mI_n}), 1 \leq i \leq m$. In \fref{comp}, we plot the NMSE values (average over independent problem realizations) as a function of the  oversampling ratio $\alpha$. Observe that WPhaseLamp appears to outperform the rest of the recovery methods. Also, note that PhaseLamp behaves worse than the PhaseMax for small values of $\alpha$ (cf. gets stuck in the bad regime of fixed points discussed in Section \ref{sec:th_PL}), but it achieves perfect recovery earlier than the latter.

%Our theoretical results are derived under the assumption that the sensing vectors and the target signal vector $\vx$ are real valued. Next, we provide simulation examples to study the recovery performance of PhaseLamp and Weighted PhaseLamp methods in the complex case. Assume that the target signal vector $\vx$ is complex valued and the sensing vectors $\lbrace \va_i, 1 \leq i \leq m \rbrace$ follow a complex Gaussian distribution. The signal dimension is set to $n = 400$. All algorithms are initialized using the optimal spectral initialization proposed in \cite{SIopt17} and the optimization problems are solved using the PhasePack \cite{phasepack17}.

%
% plots the NMSE as a function of the oversampling ratio $\alpha$. It can be noticed that the proposed PhaseLamp method outperforms the PhaseMax and the PhaseLift methods. \fref{comp} further shows that the Weighted PhaseLamp method provides the best recovery performance.
%%%%%%%%%%%%%%%%%%%%%%%%%%%%%%%%%%%%%%%%
%%%%%%%%%%%%%%%%%%%%%%%%%%%%%%%%%%%%%%%%
%%%%%%%%%%%%%%%%%%%%%%%%%%%%%%%%%%%%%%%%
\subsection{Fourier Measurements}
%%%%%%%%%%%%%%%%%%%%%%%%%%%%%%%%%%%%%%%%
%%%%%%%%%%%%%%%%%%%%%%%%%%%%%%%%%%%%%%%%
%%%%%%%%%%%%%%%%%%%%%%%%%%%%%%%%%%%%%%%%
Next, we consider a type of measurements that falls under the category of coded diffraction patterns, where the measurement vectors $\va_i$'s are the pointwise products between the $k^{th}$ Fourier vector $\vf_k$ and a random modulation pattern with i.i.d. symmetric Bernoulli entries $\vphi_l$, where $i=(k,l)$ and $1 \leq k \leq n$, $1 \leq l \leq \alpha$.
The simulation results are presented in \fref{fourier}. Note that the PhaseLamp and the WPhaseLamp methods provide similar recovery performance. Moreover, their performance is superior to the rest of the algorithms for $\alpha\geq3$.

%\begin{figure}[h!]
%    \centering
%    \subfigure[]{\label{fig:fourier1}
%    \includegraphics[width=0.47\linewidth]{figs/fourier_comp.eps}
%    }
%    \subfigure[]{\label{fig:fourier2}
%        \includegraphics[width=0.47\linewidth]{figs/fourier_comp.eps}}
%
%    \caption{The NMSE as a function of the oversampling ratio $\alpha$. Recovery performance of the Fienup \cite{Fienup:82}, Wirtinger Flow (WF) \cite{Candes:2015fv}, Truncated amplitude flow (TAF) \cite{WangGY:2016}, PhaseMax \cite{phmax,phmax2}, PhaseLin \cite{PhaseLin18}, PhaseLamp and weighted PhaseLamp (WPhaseLamp) and PhaseLift \cite{phlift13}. The signal dimension is set to $n=500$, the maximum number of iterations is set to $I_{\text{max}}=40$ and the precision is set to $\epsilon=10^{-4}$. For the PhaseLin method, we assume that the matrix $\mC_e$ is a scaled identity matrix, i.e. $\mC_e=0.01\times \mI$. The results are averaged over $10$ independent Monte Carlo trials.}
%    \label{fig:fourier}
%\end{figure}
\begin{figure}[t!]
   \centering
    \includegraphics[width=0.32\textwidth]{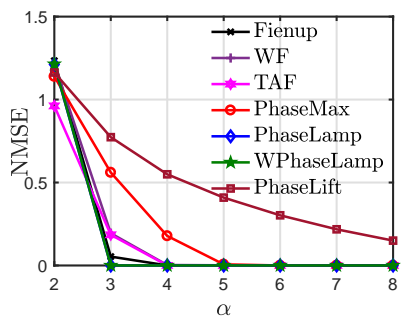}
    \caption{The NMSE as a function of $\alpha$. For Weighted PhaseLamp, the preprocessing function is $\omega:y\to y^2$. The signal dimension is set to $n=500$, the maximum number of iterations is set to $I_{\text{max}}=40$ and the precision is set to $\epsilon=10^{-4}$. The results are averaged over $50$ independent Monte Carlo trials.}
    \label{fig:fourier}
\end{figure}

\section{Technical Details: Gaussian Min-Max Inequalities}
\label{sec:tech}
%%%%%%%%%%%%%%%%%%%%%%%%%%%%%%%%%%%%%%
\subsection{Technical Tools}
\subsubsection{Gordon's Min-Max Theorem (GMT)}
The Gordon's Gaussian comparison inequality \cite{Gordon:85} compares the min-max value of two doubly indexed Gaussian processes based on how their autocorrelation functions compare. The inequality is quite general (see \cite{Gor88}), but for our purposes we only need its application to the following two Gaussian processes:
\begin{subequations}
\begin{align}
X_{\vw,\vu} &:= \vu^T \mC \vw + \psi(\vw,\vu),\\
Y_{\vw,\vu} &:= \norm{\vw}_2 \vg^T \vu + \norm{\vu}_2 \vh^T \vw + \psi(\vw,\vu),
\end{align}
\end{subequations}
where: $\mC\in\mathbb{R}^{m\times n}$, $\vg \in \mathbb{R}^m$, $\vh\in\mathbb{R}^n$, they all have entries iid Gaussian; the sets $\mathcal{S}_{\vw}\subset\R^n$ and $\mathcal{S}_{\vu}\subset\R^m$ are compact; and, $\psi: \mathbb{R}^n\times \mathbb{R}^m \to \mathbb{R}$. For these two processes, define the following (random) min-max optimization programs, which we refer to as the \emph{primary optimization} (PO) problem and the \emph{auxiliary optimization} (AO) -- for purposes that will soon become clear. 
\begin{subequations}
\begin{align}\label{eq:PO_loc}
\Phi(\mC)&=\min\limits_{\vw \in \mathcal{S}_{\vw}} \max\limits_{\vu\in\mathcal{S}_{\vu}} X_{\vw,\vu},\\
%\end{equation}
%and
%\begin{equation}
\label{eq:AO_loc}
\phi(\vg,\vh)&=\min\limits_{\vw \in \mathcal{S}_{\vw}} \max\limits_{\vu\in\mathcal{S}_{\vu}} Y_{\vw,\vu}.
\end{align}
\end{subequations}

According to Gordon's comparison inequality, for any $c\in\R$, it holds:
\begin{equation}\label{eq:gmt}
\mathbb{P}\left( \Phi(\mC) < c\right) \leq 2 \mathbb{P}\left(  \phi(\vg,\vh) < c \right).
\end{equation}
Put in words: a high-probability lower bound on the AO is a high-probability lower bound on the PO. The premise is that it is often much simpler to lower bound the AO rather than the PO. 
%\subsection{Convex Gaussian Min-Max Theorem (CGMT)}\label{cgmt_phmax}
%%%%%%%%%%%%%%%%%%%%%%%%%%%%%%%%%%%%%%
\subsubsection{Convex Gaussian Min-Max Theorem (CGMT)}
The proof of the technical results provided in Section  \ref{sec:main_results}  follows the CGMT framework\cite{chris:151,chris:152}. 
For ease of reference we summarize here the essential ideas of the framework; please see \cite[Section~6]{chris:151} for the formal statement of the theorem and further details.
The CGMT is an extension of the GMT and it asserts that the AO in \eref{AO_loc} can be used to tightly infer properties of the original (PO) in \eref{PO_loc}, including the optimal cost and the optimal solution.
%We think of the two problems in \eqref{eq:PO_loc} and \eqref{eq:AO_loc} as random optimization problems, in which $\mC$, $\vg$ and $\vh$ all have i.i.d standard normal entries. 
According to the CGMT \cite[Theorem 6.1]{chris:151}, if the sets $\mathcal{S}_{\vw}$ and $\mathcal{S}_{\vu}$ are convex and $\psi$ is continuous \emph{convex-concave} on $\mathcal{S}_{\vw}\times \mathcal{S}_{\vu}$, then, for any $\mu \in \mathbb{R}$ and $t>0$, it holds
\begin{equation}\label{eq:cgmt}
\mathbb{P}\left( \abs{\Phi(\mC)-\mu} > t\right) \leq 2 \mathbb{P}\left(  \abs{\phi(\vg,\vh)-\mu} > t \right).
\end{equation}
In words, concentration of the optimal cost of the AO problem around $\mu$ implies concentration of the optimal cost of the corresponding PO problem around the same value $\mu$. 

%Moreover, starting from \eqref{eq:cgmt} and under strict convexity conditions, the CGMT shows that concentration of the optimal solution of the AO problem implies concentration of the optimal solution of the PO to the same value. For example, if minimizers of \eqref{eq:AO_loc} satisfy $\norm{\vw^\ast(\vg,\vh)}_2 \to \zeta^\ast$ for some $\zeta^\ast>0$, then, the same holds true for the minimizers of \eqref{eq:PO_loc}: $\norm{\vw^\ast(\mC)}_2 \to \zeta^\ast$ \cite[Theorem 6.1(iii)]{chris:151}. Thus, one can analyze the AO to infer corresponding properties of the PO, the premise being of course that the former is simpler to handle than the latter.

\subsection{High-dimensional Analysis}
\label{pr_th1}
%%%%%%%%%%%%%%%%%%%%%%%%%%%%%%%%%%%%%%
\subsubsection{Approach}
We apply the CGMT and the GMT to characterize the asymptotic NMSE of the PhaseMax optimization in \eqref{eq:lp_form} as in \eqref{eq:anmse} and \eref{phb} and to prove Theorem \ref{lem:fb}, respectively. To show Theorem \ref{lem:fb}, we study the asymptotic behavior of the following optimization problem
\begin{align}\label{eq:rmax}
\mathrm{bd}^{\text{PO}} := \min_{s,r}~r^2-\alpha c_d(s,r)\quad\text{s.t.}\quad (s,r)\in{\mathcal S}_{\rm feas},
\end{align}
where the function $c_d$ is defined in \eref{cdexp} and its closed-form expression is given in \eref{cdref} and where ${\mathcal S}_{\rm feas}$ is a random set defined in Section \ref{suffst}.

To achieve the above goals, we start by writing the optimization problems \eref{lp_form}, \eref{qb_form} and \eref{rmax} in the form of a PO as in \eqref{eq:PO_loc}, which in turn leads to a corresponding AO optimization problem. Then, we analyze the AO problem. First, define the following general optimization problem
\begin{equation}\label{eq:general}
\begin{aligned}
\widehat{\vx}&=\underset{{\vx}\in\mathcal{G}_{\vx}}{\arg\,\min}~~~ p(\vx)\\	
&~~~~~~~~\text{s.t.}~~~~  \abs{\va_i^T \vx} \leq y_i, \text{ for }  1 \le i \le m,
\end{aligned}
\end{equation}
where $p$ and $\mathcal{G}_{\vx}$ are a general cost function and a general feasibility set, respectively. In this section, we are interested in the analysis of the following three cases:
\begin{itemize}
%%%%%%%%%%%%%%%%%%%%%%%%%%%%%%
%%%%%%%%%%%%%%%%%%%%%%%%%%%%%%
\item[(${\bf C_1}$)] $\mathcal{G}_{\vx}=\mathbb{R}^n$ and $p(\vx)=-{\vx}_\text{init}^{T}\,{\vx}$ which corresponds to the case when the problem \eref{general} is PhaseMax \eref{lp_form}. 
%%%%%%%%%%%%%%%%%%%%%%%%%%%%%%
%%%%%%%%%%%%%%%%%%%%%%%%%%%%%%
\item[(${\bf C_2}$)] $\mathcal{G}_{\vx}=\mathbb{R}^n$ and $p(\vx)=-\norm{{\vx}}^2_2$ which corresponds to the case when the problem \eref{general} is PhaseLamp \eref{qb_form}. 
%%%%%%%%%%%%%%%%%%%%%%%%%%%%%%
%%%%%%%%%%%%%%%%%%%%%%%%%%%%%%
\item[(${\bf C_3}$)] $\mathcal{G}_{\vx}=\mathbb{R}^n$ and $p(\vx)=\norm{\widetilde{\vx}}^2_2-\alpha c_d(x_1,\norm{\widetilde{\vx}}_2)$ which corresponds to the case when the problem \eref{general} is equivalent to the problem \eref{rmax}.
%%%%%%%%%%%%%%%%%%%%%%%%%%%%%%
%%%%%%%%%%%%%%%%%%%%%%%%%%%%%%
\end{itemize}
In this section, we assume that $\alpha > 2$. Next, the objective is to precisely analyze the problem \eqref{eq:general} in the large system limit when ${\bf C}_1$ holds using the CGMT framework. Moreover, the objective is to provide a high-probability lower bound on the problem \eqref{eq:general} when ${\bf C}_2$ or ${\bf C}_3$ holds using the GMT framework. Specifically, we show that the conditions of the CGMT (when ${\bf C}_1$ holds) and GMT (when ${\bf C}_2$/${\bf C}_3$ holds) are satisfied. Then, we formulate, simplify and analyze the corresponding AO.
\subsubsection{Formulating the PO}
Note that the GMT and CGMT assume that the feasibility sets are compact. We start our theoretical analysis by showing that the compactness assumption is guaranteed.
\begin{lemma}[Compactness]
\label{boundp}
Assume that $\alpha > 2$ and $\mathcal{K}_n$ is the feasibility set of the PhaseMax problem formulated in \eqref{eq:lp_form}. Then, there exists $\tau > 0$ such that $$\mathbb{P} \left\lbrace \norm{\mathcal{K}_n}_2 < \tau \right\rbrace \overset{n\to\infty}{\longrightarrow} 1,$$ where $\tau$ is a finite constant independent of $n$.
\end{lemma}
%\begin{IEEEproof}
%The detailed proof is given in appendix \ref{pboundp}.
%\end{IEEEproof}
The proof of Lemma \ref{boundp} is deferred to appendix \ref{pboundp}. Based on Lemma \ref{boundp}, the optimization problem given in \eqref{eq:general} is equivalent to the following problem with probability going to one as $n$ goes to $\infty$
\begin{equation}\label{eq:lp_form_2}
\begin{aligned}
\widehat{\vx}&=\underset{{\vx}\in\mathcal{S}_{\vx}}{\arg\,\min}~ p(\vx) ~\text{s.t.}~  \abs{\va_i^T \vx} \leq y_i,\text{ for }  1 \le i \le m,
\end{aligned}
\end{equation}
for $\mathcal{S}_{\vx}=\lbrace \vx\in\mathbb{R}^n:~\norm{\vx}^2_{2} \leq B \rbrace$ %if ${\bf C}_1$, ${\bf C}_3$ or ${\bf C}_4$ holds and $\mathcal{S}_{\vx}=\lbrace \vx\in\mathbb{R}^n:~x_1\mathrm{~is~fixed};~\norm{\vx}^2_{2} \leq B \rbrace$ if ${\bf C}_2$ holds
, where $B$ is a sufficiently large positive constant independent of $n$. The equivalence can be showed by taking $B>\tau$ and conditioning on the event $\lbrace \norm{\mathcal{K}_n}_2 < \tau \rbrace$. %Lemma \ref{boundp} is essential to guarantee the compactness assumption of the GMT and CGMT. 
The next step is to reformulated the optimization problem \eqref{eq:lp_form_2} as a min-max problem. To this end, assume that $V_n$ is the optimal objective value of the problem \eqref{eq:lp_form_2} and define the following optimization problem
\begin{align}\label{eq:pob_form}
V_n(\lambda_n)&=\min_{{\vx}\in\mathcal{S}_{\vx}} \max_{\abs{\vu} \leq  \lambda_n} p(\vx) + \vu^T \mA \vx - \abs{\vu}^T\vy,
\end{align}
where $\lambda_n$ is deterministic, finite and dependent on $n$. Moreover, assume that $\mathcal{V}_n$ denotes the set of optimal solutions of the problem \eref{lp_form_2} and $\mathcal{V}_n(\lambda_n)$ denotes the set of optimal solutions of the minimization problem in \eref{pob_form}. 
%The following proposition further simplifies the optimization problem \eref{lp_form_2}. 
\begin{proposition}[Min-max formulation]
\label{compactness}
If ${\bf C}_1$ holds, there exists a sequence of positive numbers $\lbrace \lambda_n \rbrace_{n\in\mathbb{N}}$ such that $\lambda_n < \infty$ and $\mathbb{P}\left( V_n(\lambda_n)=V_n \right)\overset{n\to\infty}{\longrightarrow} 1$. Moreover, we have 
$$
\mathbb{P}\left(  \mathcal{V}_n \subseteq \mathcal{V}_n(\lambda_n) \right) \overset{n\to\infty}{\longrightarrow}1.
$$
If ${\bf C}_2$ or ${\bf C}_3$ holds, we have $V_n(\lambda_n) \leq V_n$.
%deriving a high-probability lower bound on \eref{pob_form} leads to a high-probability lower bound on \eref{lp_form_2}.
\end{proposition}
%\begin{IEEEproof}
%The detailed proof is given in appendix \ref{prop3}.
%\end{IEEEproof}%or ${\bf C}_4$
The proof of the above proposition is deferred to Appendix \ref{prop3}. Proposition \ref{compactness} shows that under condition ${\bf C}_1$, the precise high-dimensional analysis of the optimization problem \eref{general} can be achieved by analyzing the problem \eref{pob_form}. Moreover, it shows that under conditions ${\bf C}_2$ or ${\bf C}_3$, deriving a high-probability lower bound on \eref{pob_form} leads to a high-probability lower bound on \eref{lp_form_2}. Note that the above proposition also guarantees the compactness assumption of the GMT and CGMT.

Based on Proposition \ref{compactness}, we proceed with analyzing the optimization problem \eref{pob_form}. Note that the set $\mathcal{S}_{\vx}$ can be rewritten as follows 
\begin{equation}
\mathcal{S}_{\vx}=
\lbrace (x_1,\widetilde{\vx}) \in \mathbb{R}^n:~x_1^2+\norm{\widetilde{\vx}}_2^2 \leq B \rbrace.
\end{equation}
%\begin{equation}
%\mathcal{S}_{\vx}=\begin{cases}
%\lbrace (x_1,\widetilde{\vx}):~x_1^2+\norm{\widetilde{\vx}}_2^2 \leq B %\rbrace & \mathrm{if}~{\bf C}_1, {\bf C}_3~\mathrm{or}~{\bf C}_4\\
%\lbrace \widetilde{\vx}:x_1~\mathrm{fixed};x_1^2+\norm{\widetilde{\vx}}_2^2 \leq B \rbrace &\mathrm{if}~ {\bf C}_2.
%\end{cases}
%\end{equation}
Define the set $\mathcal{S}_{\vu}(n)=\lbrace \vu\in\mathbb{R}^{m}:~\norm{\vu}_{\infty} \leq \lambda_n \rbrace$. Then, the optimization problem  \eref{pob_form} can be reformulated as follows
%%%-\eta_1 x_1 - \widetilde{\veta}^T \widetilde{\vx}
\begin{align}\label{eq:poff_form_2}
\hspace{-5mm}\underset{\substack{ ({x}_1,\widetilde{\vx})\in\mathcal{S}_{\vx} }}{\min} \max_{\substack{ \vu\in\mathcal{S}_{\vu}(n) }}&  \-p({x}_1,\widetilde{\vx}) + \vu^T \vq x_1 + \vu^T \mG \widetilde{\vx} - \abs{\vu}^T \abs{\vq},
\end{align}
where $p({x}_1,\widetilde{\vx})=p(\vx)$ and where $\vx=[x_1~\widetilde{\vx}^T]^T$.
At this point, observe that  \eqref{eq:poff_form_2} is in the desired form of a PO as in \eqref{eq:PO_loc} with $\mG\in\R^{m\times (n-1)}$ having i.i.d standard normal entries and the function $\psi$, defined as 
\begin{equation}
\psi(\vx,\vu)=p({x}_1,\widetilde{\vx}) +\vu^T \vq x_1 - \abs{\vu}^T \abs{\vq}.
\end{equation}
Further, note that the constraint sets are convex compact and $\psi$ is convex-concave on $\mathcal{S}_{\vx}\times \mathcal{S}_{\vu}(n)$ if ${\bf C}_1$ holds, i.e. $p({x}_1,\widetilde{\vx})=-\eta_1 x_1 - \widetilde{\veta}^T \widetilde{\vx}$, where $\vx=[x_1~\widetilde{\vx}^T]^T$.

\subsubsection{Formulating and simplifying the AO}
We are now ready to formulate the corresponding AO problem as follows
\begin{align}\label{eq:ao_form}
\hspace{-2mm}\underset{\substack{ ({x}_1,\widetilde{\vx})\in\mathcal{S}_{\vx} }}{\min} \max_{{ \vu\in\mathcal{S}_{\vu}(n) }}&\   \norm{\widetilde{\vx}}_2 \vg^T \vu + \norm{\vu}_2 \vh^T \widetilde{\vx} +p({x}_1,\widetilde{\vx})\nonumber\\
&  +\vu^T \vq x_1 - \abs{\vu}^T \abs{\vq}.
\end{align}
Following the GMT and the CGMT frameworks, we proceed onwards with analyzing \eqref{eq:ao_form}. Next, we focus on simplifying the optimization problem \eref{ao_form}. To this end, define the random  function $c_n:\R\times\R\rightarrow\R$ as follows
\begin{equation}\label{eq:cn}
c_n(s,r)=h\left( \abs{\vq} - \abs{ r \vg + s \vq}  \right),
\end{equation}
where the function $h:\R^n\rightarrow\R$ is defined as follows 
\begin{equation}\label{eq:funh}
h(\vc)=\begin{cases}
-\norm{\vc\wedge\mathbf{0}}_2 & ,\text{if}~\min(\vc) \leq 0, \\
\min(\vc) & ,\text{otherwise}. 
\end{cases}
\end{equation}
%function $f_n$ as follows
%\begin{align}\label{eq:funn}
%f_n:(s,r)\to \frac{{\norm{ (\abs{\vq} - \abs{ r \vg + s \vq}) \wedge {\bf 0} }_2}}{\sqrt{m(n)}},
%\end{align}
Moreover, define the following optimization problem
\begin{align}\label{eq:ao55_form}
\underset{ \substack{ (s,r)\in\mathcal{S}\\ \abs{z} \leq r} }{\min}&-\eta_1 s - \norm{\widetilde{\veta}}_2 z + \widetilde{\lambda}_n  \rho \Big\lbrace  \frac{\vh^T \widehat{\veta}}{\sqrt{m}} z \nonumber\\
&\hspace{-4mm}- \sqrt{\frac{\norm{\vh}^2_2}{m}-\left(\frac{\widehat{\veta}^T\vh}{\sqrt{m}}\right)^2} \sqrt{r^2-z^2}-\frac{c_n(s,r)}{\sqrt{m(n)}}   \Big\rbrace,
\end{align}
where the function $\rho:x\to \max(x,0)$ and where $\widetilde{\lambda}_n=\lambda_n \sqrt{m}$ and $\mathcal{S}=\lbrace (s,r)\in\mathbb{R}^2: r\geq 0,~s^2+r^2 \leq B \rbrace$. Also, consider the following problem 
\begin{align}\label{eq:ao_noncv}
\hspace{-2mm} \min_{(s,r)\in\mathcal{S}} p(s,r) + \widetilde{\lambda}_n \rho \left( -\frac{\norm{\vh}_2}{\sqrt{m(n)}} r - \frac{c_n(s,r)}{\sqrt{m(n)}}  \right),
\end{align}
where $p(s,r)=-s^2-r^2$ if ${\bf C}_2$ holds and $p(s,r)=r^2-\alpha c_d(s,r)$ if ${\bf C}_3$ holds. In addition, assume that ${Z}_n$ is the optimal objective and $\mathcal{Z}_n$ is the projected set of optimal solutions of the minimization problem in \eref{ao_form}, i.e.
$$
\mathcal{Z}_n=\lbrace (s,r)\in\mathbb{R}^2: [x_1~\vxt^T]\in {\mathcal S}_{n},x_1=s,\|\vxt\|_2=r \rbrace,
$$
where ${\mathcal S}_{n}$ is the set of optimal solutions of the minimization problem in \eref{ao_form}.
Also, assume that $\widetilde{Z}_{1,n}$ $\widetilde{Z}_{2,n}$ are the optimal objectives and $\mathcal{\widetilde{Z}}_{1,n}$ and $\mathcal{\widetilde{Z}}_{2,n}$ are the sets of optimal $(s,r)$ of the problems \eref{ao55_form} and \eref{ao_noncv}, respectively.

%The following proposition simplifies the AO problem \eref{ao_form}.
\begin{proposition}[Simplifying the AO]\label{prop_spao}
If ${\bf C}_1$ holds, we have $\mathbb{P}\left( Z_n=\widetilde{Z}_{1,n} \right)\overset{n\to\infty}{\longrightarrow} 1$. Moreover, we have 
$$
\mathbb{P}\left(  \mathcal{Z}_n \subseteq \mathcal{\widetilde{Z}}_{1,n} \right) \overset{n\to\infty}{\longrightarrow}1,
$$
for any sequence of positive numbers $\lbrace \lambda_n\rbrace_{n\in\mathbb{N}}$ that converges to infinity as $n \to \infty$.
If ${\bf C}_2$ or ${\bf C}_3$ holds, we have $\mathbb{P}\left( Z_n=\widetilde{Z}_{2,n} \right)\overset{n\to\infty}{\longrightarrow} 1$. Moreover, we have 
$$
\mathbb{P}\left(  \mathcal{Z}_n \subseteq \mathcal{\widetilde{Z}}_{2,n} \right) \overset{n\to\infty}{\longrightarrow}1,
$$
for any sequence of positive numbers $\lbrace \lambda_n \rbrace_{n\in\mathbb{N}}$ that converges to infinity as $n \to \infty$.
\end{proposition}
%\begin{IEEEproof}
%\end{IEEEproof}
The proof of the above proposition is deferred to Appendix \ref{prop4}. It essentially shows that under condition ${\bf C}_1$, it suffices to precisely analyze the optimization problem \eref{ao55_form} in the large system limit to determine the properties of the problem \eref{ao_form}. Moreover, it shows that under conditions ${\bf C}_2$ or ${\bf C}_3$, deriving a high-probability lower bound on \eref{ao_noncv} leads to a high-probability lower bound on \eref{ao_form}. 

Now that we have simplified the AO to a minimization problem as in \eqref{eq:ao55_form} and \eref{ao_noncv}, we are ready to study its asymptotic behavior in the regime $m,n\rightarrow\infty, m/n\rightarrow\alpha$. 
%%%%%%%%%%%%%%%%%%%%%%%%%%%%%%%%
\subsection{CGMT for the PhaseMax Method}
\label{phmax_cgmt}

In this part, we focus on the PhaseMax problem which means that we assume that the cost function $p$ is given by $p({\vx})=-\eta_1 x_1 - \widetilde{\veta}^T \widetilde{\vx}$, where $\vx=[x_1~\widetilde{\vx}^T]^T$. 
\subsubsection{Convergence analysis}
Note that Section \ref{pr_th1} shows that the precise high-dimensional analysis of the problem \eref{ao_form} can be achieved by precisely analyzing the problem \eref{ao55_form}. Hence, the main objective of this part is to study the asymptotic properties of the optimization problem \eref{ao55_form}. To this end, define the following \emph{deterministic} optimization problem
\begin{align}\label{eq:aod_form2}
&\underset{ \substack{ (s,r) \in \mathcal{S}\\ \abs{z} \leq r } }{\max}~\eta_1 s + \norm{\widetilde{\veta}}_2 z \\
&~~~\text{s.t.}~~~ -\sqrt{r^2- z^2 }+\sqrt{\alpha~c_d(s,r)} \leq 0, \nonumber 
\end{align}
where the function $c_d$ is defined in \eref{cdexp} and its closed-form expression is given by
\begin{align}\label{eq:cdref}
c_d(r,s) =\begin{cases}
\Upsilon(s,r) & \mathrm{if}~  r\neq 0
\\
(\abs{s}-1)^2 & \mathrm{if}~ \abs{s} \geq 1~\text{and}~ r=0
\\
0 & \mathrm{if}~ \abs{s} < 1~\text{and}~ r=0,
\end{cases}
\end{align}
and where the function $\Upsilon$ can be expressed as follows 
\begin{align}\label{eq:cdrefup}
&\Upsilon(s,r)=\frac{1}{\pi} \Bigg[  ( (1-s)^2+r^2 )\left({\pi}/{2}-{\atan}\left( {(1-s)}/{r} \right) \right)\nonumber\\
&+( (1+s)^2+r^2 )\left({\pi}/{2}-{\atan}\left( {(1+s)}/{r} \right) \right)-2r \Bigg].
\end{align}

\noindent The following proposition studies the asymptotic properties of the optimization problem \eref{ao55_form} in detail. The proof of the proposition is provided in Appendix \ref{pr_lem1}.
%%%%%%%%%%%%%%%%%%%%%%%%%%%%%%%%%

\begin{proposition}[Convergence analysis]
\label{lem1}
Assume that the oversampling ratio satisfies $\alpha > 2$. Let $\mathcal{V}^{\ast}_n$ and $V^{\ast}_n$ be the set of optimal solutions and the optimal objective value of the problem \eref{ao55_form} and let $\mathcal{V}^{\ast}$ and $V^{\ast}$ be the set of optimal solutions and the optimal objective value of the {deterministic} problem formulated in \eref{aod_form2}. Then, we have 
\begin{equation}
V^{\ast}_n  \overset{n\to \infty}{\longrightarrow} V^{\ast}~\mathrm{and}~\mathbb{D}( \mathcal{V}^{\ast}_n,\mathcal{V}^{\ast} )  \overset{n\to \infty}{\longrightarrow} 0.
\end{equation}
%Fix any $s\in\R$. Then, for all $\eps>0$ it holds that
%\begin{align}
%\lim_{n\rightarrow\infty} \Pro\Big( r_{\max}(s) < r^*(s) + \eps \Big) = 1.
%\end{align}
%where $r_{\max}(s)$ is defined in \eref{ao_noncv} and $r^*(s)$ is defined in \eref{rs}.
\end{proposition}
%\begin{IEEEproof}
%The detailed proof is given in appendix \ref{pr_lem1}.
%\end{IEEEproof}
%%%%%%%%%%%%%%%%%%%%%%%%%%%%%%%%
%%%%%%%%%%%%%%%%%%%%%%%%%%%%%%%%

The above proposition 
%first proves theorem \ref{lem:fb} when $p({x}_1,\widetilde{\vx})=-\norm{ \widetilde{\vx}}_2$ and it also 
shows that the set of optimal solutions and the optimal objective value of the problem \eref{ao55_form} concentrate around the set of optimal solutions and the optimal objective value of the deterministic problem \eref{aod_form2}. %This means that the precise high-dimensional analysis of the PhaseMax performance can be achieved by solving the deterministic problem \eref{aod_form2}.
\subsubsection{Solving the scalar performance optimization}
\label{sspop}

%Note that the constraint sets are convex compact and $\psi$ is convex-concave on $\mathcal{S}_{\vx}\times \mathcal{S}_{\vu}(n)$. 
%Based on Proposition \ref{lem1}, the set of optimal solutions and the optimal objective value of the problem \eref{ao55_form} concentrate around the set of optimal solution  and the optimal objective value of the problem \eref{aod_form2}. 
In what follows, we focus on simplifying the deterministic problem \eref{aod_form2}. The following lemma, which is proved in Appendix \ref{puniqz}, simplifies the deterministic optimization problem \eref{aod_form2}.
\begin{lemma}[Simplifying the deterministic problem]
\label{uniqz}
The optimization problem \eref{aod_form2} admits a unique solution in the variable $z$ which is given by 
$$
z^\ast=\sqrt{r^2-\alpha~c_d(s,r)}.
$$
Additionally, it is equivalent to the two-dimensional problem
\begin{align}\label{eq:aod_form}
&\underset{ \substack{ (s,r) \in \mathcal{S} } }{\max}~\eta_1 s + \norm{\widetilde{\veta}}_2 \sqrt{r^2-\alpha~c_d(r,s)  } \\
&~~~\text{s.t.}~~~ c_d(s,r) \leq r^2/\alpha. \nonumber 
\end{align}
\end{lemma}
%\begin{IEEEproof}
%The detailed proof is given in appendix \ref{puniqz}.
%\end{IEEEproof}

We call the deterministic two-dimensional optimization problem in \eqref{eq:aod_form} as the scalar performance optimization (SPO).%; according to the CGMT solving the SPO allows us to conclude on the asymptotic performance of PhaseMax (cc. the PO). 
%Next, we assume that $\alpha>2$. But, it can be checked that the optimization problem has the same solutions in the case when $1 < \alpha \leq 2$. Recall that the SPO in \eqref{eq:aod_form} is the converging limit of the problem in \eqref{eq:ao55_form}.
%Next, we consider the case when the oversampling ratio $\alpha$ satisfies $\alpha>1$. 
 Recall that the SPO in \eqref{eq:aod_form} is the converging limit of the problem in \eqref{eq:ao55_form}.
%Specifically, the optimization variables $s$ and $r$ in \eqref{eq:aod_form}  correspond exactly to  $x_1$ and $\|\widetilde\vx\|_2$ in \eqref{eq:ao_form}. From this and uniform convergence discussed previously, the optimal values of $s$ and $r$ are the converging limits of $x_1$ and of $\|\widetilde\vx\|_2$, respectively. 
In what follows, we solve the SPO problem for the optimal $s$ and $r$.
%%%%%%%%%%%%%%%%%%%%%%%%%%%%%%%%
%%%%%%%%%%%%%%%%%%%%%%%%%%%%%%%%
%%%%%%%%%%%%%%%%%%%%%%%%%%%%%%%%
%\begin{lemma}
%\label{lem2}
%For any fixed $\alpha > 2$, the feasible set of the optimization problem  \eqref{eq:aod_form} is nonempty if and only if $\abs{s} \leq 1$.
%\end{lemma}
%\begin{IEEEproof}
%The detailed proof is given in Appendix \ref{pr_lem2}.
%\end{IEEEproof}
%%%%%%%%%%%%%%%%%%%%%%%%%%%%%%%%
%%%%%%%%%%%%%%%%%%%%%%%%%%%%%%%%
%%%%%%%%%%%%%%%%%%%%%%%%%%%%%%%%
%\noindent Based on Lemma \ref{prop}, the optimization problem \eref{aod_form} can be reformulated as follows
%\begin{align}\label{eq:detphmax}
%&\underset{ \substack{ |s|\leq 1 \\ (s,r) \in \mathcal{S} } }{\max}~\eta_1 s + \norm{\widetilde{\veta}}_2 \sqrt{r^2-\alpha~c_d(r,s)  } \\
%&~~~\text{s.t.}~~ c_d(s,r) \leq r^2/\alpha. \nonumber 
%\end{align}
%Based on (P.2) in Lemma \ref{prop}, the optimization problem \eref{aod_form} can be equivalently formulated as follows
%\begin{align}\label{eq:aod_forms}
%&\underset{ \substack{ (s,r) \in \mathcal{S} } }{\max}~\eta_1 s + \norm{\widetilde{\veta}}_2 \sqrt{r^2-\alpha~c_d(r,s)  } \\
%&~~~\text{s.t.}~~~ c_d(s,r) \leq r^2/\alpha. \nonumber 
%\end{align}
The following lemma, which is proved in Appendix \ref{pr_lem3}, further simplifies the optimization problem \eref{aod_form} by showing that it has a unique optimal $r$ for any feasible variable $s$.
%%%%%%%%%%%%%%%%%%%%%%%%%%%%%%%%
%%%%%%%%%%%%%%%%%%%%%%%%%%%%%%%%
%%%%%%%%%%%%%%%%%%%%%%%%%%%%%%%%
\begin{lemma}[Simplifying the deterministic problem]
\label{lem3}
Fix $s$ such that $\abs{s} \leq 1$ and $\alpha > 2$. Then, the following optimization problem
\begin{align}\label{eq:aod2_form}
&\underset{  r\geq 0}{\max}~ r^2-\alpha\,c_d(s,r), 
\end{align} 
admits a unique global optimal solution given by 
\begin{equation}\label{eq:vstar}
r_{\alpha}(s)=\sqrt{{1}/{{\tan}\left( {\pi}/{\alpha} \right)^2}+(1-s^2)}-{1}/{{\tan}\left( {\pi}/{\alpha} \right)}.
\end{equation}
\end{lemma}
%\begin{IEEEproof}
%The detailed proof is given in appendix \ref{pr_lem3}.
%\end{IEEEproof}

Based on P.2 in Lemma \ref{prop}, the set $\mathcal{D}_{\text{feas}}$ is compact. Hence, we can always find a large enough constant $\widetilde B>0$ such that $s^2+r_{\alpha}(s)^2<\widetilde B$, for all $s$ such that $\abs{s} \leq 1$. Therefore, choosing $B$ in \eqref{eq:aod_form} such that $B=\widetilde B$ guarantees that the optimal value of $r$ in \eqref{eq:aod_form} is given by \eqref{eq:vstar}. Substituting this value back in \eqref{eq:aod_form} and using P.2 in Lemma \ref{prop}, we can now optimize over $s$ by solving the following:
\begin{align}\label{eq:aod4_form0}
&\max_{ \abs{s} \leq 1 }~\eta_1 s + \norm{\widetilde{\veta}}_2 \sqrt{ g_\alpha(s) },
%&~~~~~~\text{s.t.}~~~~~~ g_\alpha(s)\geq 0. \nonumber 
\end{align}
where $g_\alpha(s)=(r_{\alpha}(s))^2-\alpha~c_d(r_{\alpha}(s),s)$. A few algebraic manipulations show that the function $g_\alpha$ is as given in \eref{fun_g} and show that \eqref{eq:aod4_form0} is equivalent to \eref{fconprob} in the statement of Theorem \ref{thm:the1}. To show the equivalence, further note that $\eta_1$ and $\widetilde{\veta}$ in \eqref{eq:aod4_form0} are related to the input cosine similarity $\rho_{\text{init}}$, defined in \eref{cosin}, as follows (recall: $\boldsymbol{\xi}=\ve_1$.),
\begin{equation}
\frac{\eta_1}{\norm{\widetilde{\veta}}_2} = \frac{\rho_{\text{init}}}{\sqrt{1-\rho_{\text{init}}^2}}.
\end{equation}
%Next, we assume that $\alpha \geq 2$. Based on (P.2) in Lemma \ref{prop}, the optimization problem \eqref{eq:aod4_form0} can be reformulated as follows
%%%%%
%\begin{align}\label{eq:aod4_form}
%&{\max_{|s|\leq 1}}~\rho_{\text{init}} s + \sqrt{1-\rho_{\text{init}}^2} \sqrt{g_\alpha(s)  }.
%&~~~\text{s.t.}~~ \alpha~c_d(s,r^\ast(s)) \leq (r^\ast(s))^2 \nonumber 
%\end{align}
%where the function $g_\alpha$ is given in \eref{fun_g}.
Finally, note that the optimization in \eqref{eq:aod4_form0} is a strictly concave program as shown in the following lemma.
\begin{lemma}[Properties of the deterministic problem]
\label{lem4}
For any fixed $\alpha > 2$, the optimization problem formulated in \eqref{eq:aod4_form0} is strictly concave.
\end{lemma}
%\begin{IEEEproof}
%The detailed proof is given in appendix \ref{pr_lem4}.
%\end{IEEEproof}

The proof of the above lemma is detailed in Appendix \ref{pr_lem4}. Based on Lemmas \ref{uniqz}, \ref{lem3} and \ref{lem4}, the deterministic optimization problem \eref{aod_form2} has a unique global optimal solution. Based Propositions \ref{prop_spao} and \ref{lem1}, the optimal objective value and the projected set of optimal solutions of the AO problem \eref{ao_form} concentrate around the optimal objective value and the set of optimal $(s,r)$ of the deterministic problem \eref{aod_form2}. Again, given the uniqueness of the solution of the problem \eref{aod_form2}, based on the proof of Proposition \ref{lem1} and using the CGMT, the optimal objective value and the projected set of optimal solutions of the PO problem \eref{pob_form} concentrate around the optimal objective value and the set of optimal $(s,r)$ of the deterministic problem \eref{aod_form2}. Now, using the result stated in Proposition \ref{compactness}, the optimal objective value and the projected set of optimal solutions of PhaseMax \eref{lp_form} concentrate around the optimal objective value and the set of optimal $(s,r)$ of the deterministic problem \eref{aod_form2}.

Therefore, the optimal objective value of the PhaseMax problem \eref{lp_form} converges in probability to the optimal objective value of the problem \eref{aod4_form0}, i.e,
\begin{align}
\vx_{\text{init}}^T \widehat{\vx} \xrightarrow[]{n\to\infty} &{\rho_{\text{init}}} s^\ast+\sqrt{(1-\rho_{\text{init}}^2)}\sqrt{g_\alpha(s^\ast)  }.
\end{align}
Moreover, any optimal solution $\widehat{\vx}$ of the PhaseMax problem \eref{lp_form} satisfies the following 
\begin{align}
&s(\widehat{\vx}) \xrightarrow[]{n\to\infty} s^\ast~~\text{and}~~ r(\widehat{\vx}) \xrightarrow[]{n\to\infty} r_{\alpha}(s^\ast),
\end{align}
where $s^\ast$ is the solution of the problem \eref{aod4_form0}, $r_{\alpha}(s)$ is given in \eref{vstar}, $r(\widehat{\vx})=\norm{\widetilde{\vx}(\widehat{\vx})}_2$, and $\widehat{\vx}=[s(\widehat{\vx})~~\widetilde{\vx}(\widehat{\vx})^T]^T$. Note that the above convergence results are valid for $\alpha>2$. This then gives us the statement of Theorem \ref{thm:the1}.
\vsp
\subsubsection{Phase transition calculations}
\label{pr_th2}
In this section, we compute the phase transition boundary of the PhaseMax method. Our goal is to find necessary and sufficient conditions under which the solution $\widehat\vx$ of PhaseMax is, with high probability, equal to $\boldsymbol{\xi}=\ve_1$. Mapping this to the SPO in \eqref{eq:aod_form}, we seek conditions under which $s^\ast=1$ and $r^\ast=0$. 

%First, assume that $1<\alpha<2$. Note that the optimal $r$ of the optimization problem \eref{aod_form} for any fixed $s$ is as given in \eref{vstar}. Since $1<\alpha<2$, it can be noticed that $r_\alpha(s)$ is always strictly positive. This means that there is no perfect recovery in the case when $1<\alpha<2$.

Assume that $\alpha > 2$. From the strict concavity result in Lemma \ref{lem4}, perfect recovery happens if and only if the 
derivative of the cost function of the optimization problem \eref{fconprob} at $s=1$ is nonnegative.  By performing a Taylor expansion of the function $s\to\sqrt{(r_{\alpha}(s))^2-\alpha~c_d(r_{\alpha}(s),s)  }$ at $s=1$, the derivative of the cost function of the optimization problem \eref{aod4_form0} at $s=1$ can be expressed as follows
$$\eta_1 -  \norm{\widetilde{\veta}}_2 \sqrt{\frac{\alpha}{\pi}\text{tan}\left(  {\pi}/{\alpha}\right)-1}.$$
Hence, the necessary and sufficient condition for perfect recovery of the PhaseMax method is given by 
\begin{equation}\label{eq:nsuffphmax}
\frac{\rho_{\text{init}}}{\sqrt{1-\rho_{\text{init}}^2}} \geq \sqrt{\frac{\alpha}{\pi}\text{tan}\left(  {\pi}/{\alpha}\right)-1},
\end{equation}
for $\alpha > 2$. 
%Note that $\eta_1$ and $\widetilde{\veta}$ are related to the input cosine similarity $\rho_{\text{init}}$, defined in \eref{cosin}, as follows
%\begin{equation}
%\frac{\eta_1}{\norm{\widetilde{\veta}}_2} = \frac{\rho_{\text{init}}}{\sqrt{1-\rho_{\text{init}}^2}}.
%\end{equation}
Equivalently, the oversampling ratio $\alpha$ and the input cosine similarity given in \eref{cosin} must satisfy the condition given in \eref{phb}.
%\begin{equation}\label{eq:ptb}
%\frac{\pi}{\alpha~\text{tan}\left(  \pi / \alpha \right)} \geq 1-\rho_{\text{init}}^2.
%\end{equation}
This then gives us the statement of Theorem \ref{thm:pt}.

\subsection{Sufficient Condition for PhaseLamp}
\label{suffcond}
In this subsection, we focus on the PhaseLamp problem. We prove the sufficient conditions for perfect recovery of PhaseLamp stated in Theorems \ref{the2a} and \ref{the2b}. To this end, fix the oversampling ratio $\alpha$ such that $\alpha>2$. 
%Based on Theorem \ref{lem:fb}, the projected feasibility set $\mathcal{S}_{\text{feas}}$ of the PhaseMax method is a subset of the following deterministic set 
%\begin{align}
%{\mathcal D}_{\rm feas}=\lbrace (s,r)\in\mathbb{R}^2:~r\geq 0,~ \alpha~c_d(s,r) \leq r^2 \rbrace,
%\end{align}
%in the high dimensional limit. 

%One particular property of the fixed points of the proposed PhaseLamp algorithm is that they satisfy the following
%\begin{align}\label{eq:fp}
%\widehat{\vx}&=\underset{{\vx}\in\mathbb{R}^n}{\arg\,\max}~~~ {\widehat{\vx}}^{T}\,{\vx}\\	
%&~~~~~~~~\text{s.t.}~~~~  \abs{\va_i^T \vx} \leq y_i, \text{ for }  %1 \le i \le m.\nonumber
%\end{align}
%where $\widehat{\vx}$ is any fixed point of PhaseLamp.
%Note that a fixed point $\widehat{\vx}$ of PhaseLamp is a solution of a PhaseMax problem where $\widehat{\vx}$ is used as an initial guess of the target signal vector $\vxi$.
%Given that the target signal vector $\vxi$ is feasible for the problem \eref{fp}, any fixed point of the proposed PhaseLamp algorithm satisfies the following
%\begin{align}
%s^2+r^2={\widehat{\vx}}^{T} {\widehat{\vx}} \geq \abs{ {\widehat{\vx}}^{T}\,\vxi }= \abs{ s }.
%\end{align}
%
%\vsp
%\textbf{\emph{Property 1}}: 

\subsubsection{Fixed points of PhaseLamp}
Note that the fixed points of the PhaseLamp algorithm are elements of the following deterministic set
\begin{align}\notag
{\mathcal{D}}_{\text{opt}}=\left \lbrace (s,r)\in\mathbb{R}^2:~r\geq 0,~ s^2+r^2 \geq \abs{ s } \right\rbrace.
\end{align}

Based on Lemma \ref{prop}, the set ${\mathcal D}_{\rm feas}$ is a subset of the set $\lbrace (s,r)\in\mathbb{R}^2:~r\geq 0,-1\leq s \leq 1 \rbrace$. Given the symmetry, we only consider the case when $s \geq 0$. Based on P.3 in Lemma \ref{prop}, the intersection between the boundary of the set ${\mathcal D}_{\rm feas}$ and the boundary of the optimality set ${\mathcal{D}}_{\text{opt}}$ for $s\in(0,1)$ satisfies 
\begin{align}\label{eq:dfdo}
%\begin{cases}
c_d(s,r)=\frac{r^2}{\alpha};~r^2+s^2=s;~s\in(0,~1),~r\geq 0.
%\end{cases}
\end{align}
The following lemma, which is proved in Appendix \ref{pLamp_01}, analyzes the system given in \eref{dfdo}.

\begin{lemma}
\label{Lamp_01}
The system given in \eref{dfdo} has a unique solution. Moreover, $\text{bd}(\mathcal{D}_{\text{opt}}) \not\subset \mathcal{D}_{\text{feas}}$ .
\end{lemma}
Note that the point $(0,0)$ is in the set $\text{bd}(\mathcal{D}_{\text{feas}})$ and it is also in the set $\text{bd}(\mathcal{D}_{\text{opt}})$. Moreover, observe that the target signal vector $(1,0)$ is in $\text{bd}(\mathcal{D}_{\text{feas}})$ and $\text{bd}(\mathcal{D}_{\text{opt}})$. Based on Lemma \ref{Lamp_01}, the intersection between $\text{bd}({\mathcal D}_{\text{feas}} )$ and $\text{bd}({\mathcal{D}}_{\text{opt}})$ is $\lbrace (0,0),(\widehat{s},\widehat{r}),(1,0) \rbrace$ where $(\widehat{s},\widehat{r})$ is the unique solution of the system in \eref{dfdo}. Given that the solutions of \eref{dfdo} satisfies $r^2+s^2=s$ and $s\in(0,~1)$, we have $\widehat{s}\in(0,~1)$ and $\widehat{r}>0$.   Also, Lemma \ref{prop} shows that the maximum radius of $\mathcal{D}_{\text{feas}}$ is strictly positive for $s=0$. Hence, Lemma \ref{Lamp_01} essentially shows that all the points $(s,r)$ satisfying $\widehat{s} < s < 1 $ are not elements of the following set $\mathcal{D}_{\text{feas}} \cap \mathcal{D}_{\text{opt}}$.
%can not be fixed points of PhaseLamp.

Assuming that $s=\sin(\theta)^2$ where $\theta\in(0,\pi/2)$, the system in \eref{dfdo} can be rewritten as follows
\begin{align}\label{eq:slam_fppf}
& \theta \cos^2\theta + (1 + 3 \sin^2\theta) \atan\left(\frac{\sin \theta \cos \theta}{1 + \sin^2 \theta}\right)= \nonumber\\
&\quad~\quad 2 \sin \theta \cos \theta + \Big(\frac{\pi}{\alpha}\Big) \sin^2 \theta \cos^2 \theta.
\end{align}
Based on Lemma \ref{Lamp_01}, equation \eref{slam_fppf} has a unique solution in $(0,~\pi/2)$. Denote this solution by $\theta_\alpha^\ast$. 

\subsubsection{PhaseMax properties}
The PhaseLamp method solves a PhaseMax problem as given in \eref{itphmax_form} at iteration $k+1$.
Given that the target signal vectors $\vxi$ and $-\vxi$ are feasible for the problem \eref{itphmax_form}, the optimal solution ${\vx}_{k+1}$ at iteration $k+1$ satisfies the following inequality
\begin{align}\label{eq:ineq1lamp}
\eta_{1k} s+ \widetilde{\veta}_k^T \widetilde{\vx}_{k+1}={\vx}_{k}^T {\vx}_{k+1} \geq \abs{ {\vx}_{k}^T \vxi }=  \abs{\eta_{1k}},
\end{align}
where we express ${\vx}_{k}^T=[\eta_{1k}~~{\widetilde{\veta}_k}^T]$ and ${\vx}_{k+1}^T={[s~~{{\widetilde{\vx}}_{k+1}}^T]}$. Given that the vectors $\vxi$ and $-\vxi$ are both valid targets, one can assume without loss of generality that $\eta_{1k} \geq 0$, for all $k \geq 0$. Based on the Cauchy Schwarz inequality,  \eref{ineq1lamp} can be rewritten as follows
$$
\norm{ \widetilde{\veta}_k }_2 r \geq {\eta_{1k}}- \eta_{1k} s,
$$
where $r=\norm{{{\widetilde{\vx}}_{k+1}}}_2$. Now, define the input cosine similarity $\rho_{\text{init}}^k$ at iteration $k+1$ as follows 
$$
\rho_{\text{init}}^k=\frac{{{\vx}_{k}^T\vxi}}{\norm{{\vx}_{k}}_2\norm{\vxi}_2},
$$
where $\vx_0=\vx_{\text{init}}$ denotes the initial guess of PhaseMax and $\rho_{\text{init}}^0$ is the input cosine similarity of PhaseMax, i.e. $\rho_{\text{init}}^0=\rho_{\text{init}}$. Note that the following equality holds for any $k \geq 0$ (recall: $\boldsymbol{\xi}=\ve_1$.)
\begin{equation}\label{eq:lam_def}
\frac{{\eta_{1k}}}{\norm{\widetilde{\veta}_k}_2} = \frac{\rho_{\text{init}}^k}{\sqrt{1-{\rho_{\text{init}}^k}^2}}=:		\chi_k.
\end{equation}
Hence, any optimal solution of PhaseLamp at iteration $k+1$ satisfies the following inequality $r \geq \chi_k (1-s)$.
This implies that any optimal solution of PhaseLamp at iteration $k+1$ belongs to the following set
\begin{align}\notag
{\mathcal{D}}_{\text{fp}}(\rho_{\text{init}}^k)=\left\lbrace (s,r)\in\mathbb{R}^2:~r\geq 0,~\chi_k (1-s) \leq r \right\rbrace.
\end{align}

Now, we provide another property which guarantees that PhaseLamp escapes the bad set of stationary points and converge to the target signal vector. To this end, fix the iteration index $k \geq 0$. Based on P.3 in Lemma \ref{prop}, the intersection between the boundary of the set ${\mathcal D}_{\rm feas}$ and the boundary of the set ${\mathcal{D}}_{\text{fp}}(\rho_{\text{init}}^k)$ for $s\in(0,1)$ and $r>0$ satisfies
\begin{align}\label{eq:dfdr}
%\begin{cases}
c_d(s,r)=\frac{r^2}{\alpha};~\chi_k (1-s) = r;~s\in(0,~1),~r> 0,
%\end{cases}
\end{align}
where $\chi_k$ is defined in \eref{lam_def} and it satisfies $\chi_k\geq 0$. Note that if $\chi_k = 0$, the boundary of the set ${\mathcal{D}}_{\text{fp}}(\rho_{\text{init}}^k)$ is the set of $(s,r)$ such that $r=0$. 
%Based on Lemma \ref{Lamp_01}, the points $(s,r)$ such that $s\in[0,1)$ and $r=0$ are in the interior of the set ${\mathcal D}_{\rm feas}$. 
Therefore the system given in \eref{dfdr} has no solutions.
The following lemma, which is proved in Appendix \ref{pLamp_02}, analyzes the system given in \eref{dfdr} in further details.
\begin{lemma}
\label{Lamp_02}
The system given in \eref{dfdr} has at most one solution. 
When \eref{dfdr} has a solution, the intersection between the set $\text{bd}({\mathcal{D}}_{\text{fp}}(\rho_{\text{init}}^k))$ and the line $s=0$ is not in the set ${\mathcal D}_{\rm feas}$.
\end{lemma}
Lemma \ref{Lamp_02} is essential to prove our sufficient conditions for perfect recovery of PhaseLamp stated in Theorems \ref{the2a} and \ref{the2b}. Note that the intersection between $\mathrm{bd}\left({\mathcal D}_{\rm feas} \right)$ and $\mathrm{bd}\left({\mathcal{D}}_{\text{opt}}\right)$ is $\lbrace (0,0),(\widehat{s},\widehat{r}),(1,0) \rbrace$. Now, select $\chi_k$ such that $\chi_k (1-\widehat{s}) = \widehat{r}$. This means that $\mathrm{bd}\left({\mathcal D}_{\rm feas} \right)$ and $\mathrm{bd}\left({\mathcal{D}}_{\text{fp}}(\rho_{\text{init}}^k)\right)$ intersect at $(\widehat{s},\widehat{r})$ where $\widehat{s}\in(0,~1)$ and $\widehat{r}>0$. Based on Lemma \ref{Lamp_02}, all the points $(s,r)$ satisfying $0 \leq s < \widehat{s}$ are not elements of the set $\mathcal{D}_{\text{feas}} \cap \mathcal{D}_{\text{opt}} \cap {\mathcal{D}}_{\text{fp}}(\rho_{\text{init}}^k)$.
%can not be solutions of PhaseLamp at iteration $k+1$. 

\subsubsection{Sufficient condition for general initialization}
Now, define $\widehat{\rho}_{s}(\alpha)$ such that 
\begin{align}\label{eq:rhos}
\frac{\widehat{\rho}_{s}(\alpha)}{\sqrt{1-\widehat{\rho}_{s}(\alpha)^2}} (1-\widehat{s})=\widehat{r}.
\end{align}
Note that $\widehat{\rho}_{s}(\alpha)$ represents the input cosine similarity that guarantees that $\mathrm{bd}\left({\mathcal D}_{\rm feas} \right)$ and $\mathrm{bd}\left({\mathcal{D}}_{\text{fp}}(  \widehat{\rho}_{s}(\alpha)  )\right)$ intersect at $(\widehat{s},\widehat{r})$. We know that the unique solution of \eref{dfdo} satisfies $\widehat{s}=\sin^2(\theta_\alpha^\ast)$ and $\widehat{r}=\sqrt{\sin^2(\theta_\alpha^\ast)-\sin^4(\theta_\alpha^\ast)}$.
Based on \eref{rhos}, note that $\widehat{\rho}_{s}(\alpha)$ can be expressed as follows
\begin{align}
\widehat{\rho}_{s}(\alpha)&=\frac{\widehat{r}}{\sqrt{\widehat{r}^2+(1-\widehat{s})^2}}\nonumber\\
&=\frac{\sqrt{\sin^2(\theta_\alpha^\ast)-\sin^4(\theta_\alpha^\ast)}}{\sqrt{  \sin^2(\theta_\alpha^\ast)-\sin^4(\theta_\alpha^\ast)+(1-\sin^2(\theta_\alpha^\ast))^2   }}\nonumber\\
&=\sin(\theta_\alpha^\ast).
\end{align}
Given that $\theta_\alpha^\ast \in (0~\pi/2)$, we have $0 < \widehat{\rho}_{s}(\alpha) <1$. The following lemma shows that selecting the input cosine similarity of PhaseMax such that it is higher than $\widehat{\rho}_{s}(\alpha)$ guarantees that all the input cosine similarities of the PhaseLamp procedure are higher than $\widehat{\rho}_{s}(\alpha)$.
\begin{lemma}
\label{lemm9}
Select the input cosine similarity of PhaseMax $\rho_{\text{init}}$ such that $\rho_{\text{init}} > \widehat{\rho}_{s}(\alpha)$. Then,
the input cosine similarity $\rho_{\text{init}}^k$ at iteration $k+1$ of PhaseLamp satisfy the following
\begin{align}
\rho_{\text{init}}^{k} > \widehat{\rho}_{s}(\alpha), \forall k \geq 0.
\end{align}
\end{lemma}

The proof of the above lemma is deferred to Appendix \ref{plemm9}. 
Based on Lemma \ref{lemm9}, we obtain $\rho_{\text{init}}^k > \widehat{\rho}_{s}(\alpha)$ for any $k \geq 0$. Therefore, we conclude that $0 \leq s \leq \widehat{s}$ are not elements of the set $\mathcal{D}_{\text{feas}} \cap \mathcal{D}_{\text{opt}} \cap {\mathcal{D}}_{\text{fp}}(\rho_{\text{init}}^k)$.
%can not be fixed points of PhaseLamp. 
Now, based on Lemma \ref{Lamp_01}, all the points $(s,r)$ satisfying $\widehat{s}<s < 1 $ are not elements of the set $\mathcal{D}_{\text{feas}} \cap \mathcal{D}_{\text{opt}}$.
%can not be fixed points of PhaseLamp. 
Based on P.2 in Lemma \ref{prop}, we conclude that selecting the input cosine similarity of PhaseMax $\rho_{\text{init}}$ in this way guarantees that 
\begin{align}\label{eq:inters}
\mathcal{D}_{\text{feas}} \cap \mathcal{D}_{\text{opt}} \cap {\mathcal{D}}_{\text{fp}}(\rho_{\text{init}}^k)=\lbrace \vxi \rbrace,~\forall k \geq 0.
\end{align}

\subsubsection{Sufficient condition for independent initialization}
Note that the sufficient condition $\rho_{\text{init}} > \widehat{\rho}_{s}(\alpha)$ is valid for any initial guess vectors $\vx_{\text{init}}$, which can dependent on the sensing vectors $\lbrace \va_i: 1 \leq i \leq m \rbrace$ and the target signal vector $\vxi$. Next, we focus on the case when the initial guess vector $\vx_{\text{init}}$ is independent of the sensing vectors $\lbrace \va_i: 1 \leq i \leq m \rbrace$ and the target signal vector $\vxi$. To improve the above condition, we further exploit the properties of the problem \eref{itphmax_form} and PhaseMax \eref{lp_form} as given in the following property.

\vsp
\textbf{\emph{Property}}: The optimization problem \eref{itphmax_form} is scale invariant for any $k \geq 0$. Based on Theorem \ref{thm:the1}, the optimal solution $\widehat{\vx}$ of PhaseMax satisfies the following 
\begin{align}
r(\widehat{\vx}) \xrightarrow[]{n\to\infty} r_{\alpha}(s)\bydef\sqrt{c_{\alpha}^2+1-s^2}-c_{\alpha},
\end{align}
with $s \in [0,~1]$, $c_{\alpha}=1/\text{tan}\left( \pi/\alpha \right)$, $\widehat{\vx}=[s(\widehat{\vx})~~\widetilde{\vx}(\widehat{\vx})^T]^T$ and $r(\widehat{\vx})=\norm{\widetilde{\vx}(\widehat{\vx})}_2$. Based on Section \ref{phmax_cgmt}, we know that $r_{\alpha}(s)$ is the unique solution to the optimization problem \eref{aod2_form}, for any $s \in [0,~1]$. This means that $\forall~s \in [0,~1]$, $(s,r_{\alpha}(s)) \in \mathcal{D}_{\text{feas}}$.
\vsp

The above property shows that it suffices to select $\rho_{\text{init}}> \rho_{s}(\alpha)$ to guarantee \eref{inters}, where $\rho_{s}(\alpha)$ is determined such that the optimal solution of the following optimization problem 
\begin{equation}\label{eq:fconprobp}
\begin{aligned}
\underset{0 \leq {s} \leq 1}{\max}\ \ \frac{\rho_{s}(\alpha)}{\sqrt{1-\rho_{s}(\alpha)^2}} s+\sqrt{g_{\alpha}(s)},
\end{aligned}
\end{equation} 
is $\widehat{s}_\alpha$ the unique solution of the following system of equations
\begin{equation}\label{eq:sys_opt_s}
\begin{cases}
r=\sqrt{c_\alpha^2+1-s^2}-c_\alpha\\
r=\frac{\sqrt{1-\widehat{\rho}_{s}(\alpha)^2}}{\widehat{\rho}_{s}(\alpha)} ~s\\
s\in(0,~1),~r\geq 0,
\end{cases}
\end{equation}
where the function $g_\alpha$ is defined in \eref{fun_g}. 
\begin{figure}[t!]
    \centering
    \includegraphics[width=0.35\textwidth]{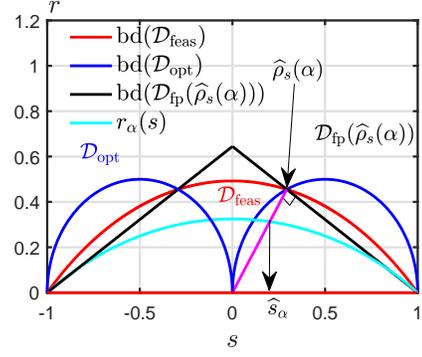}
    \caption{Illustration of the sufficient condition given in \eref{suffp} for $\alpha=5$. The cyan curve represents the function $r_\alpha$ given in \eref{optrfun} which denotes the optimal $r$ of the PhaseMax problem for any fixed $s\in[0,~1]$, in the high dimensional limit.}
    \label{fig:fc2}
\end{figure}
\fref{fc2} illustrates the above sufficient condition for $\alpha=5$. Note that due to the scale invariance of the optimization problem \eref{itphmax_form}, it is sufficient to select $\rho_{s}(\alpha)$ such that the solution of the PhaseMax problem in the large system limit is determined by the intersection between the equations $r_\alpha(s)=\sqrt{c_\alpha^2+1-s^2}-c_\alpha$ (cyan curve) and $r={\sqrt{1-\widehat{\rho}_{s}(\alpha)^2}/\widehat{\rho}_{s}(\alpha)} ~s$ (magenta curve).

Note that the unique solution $\widehat{s}_\alpha$ of \eref{sys_opt_s} can be expressed as follows
\begin{align}
\widehat{s}_\alpha=\frac{\sqrt{a_\alpha^2 c_\alpha^2+a_\alpha^2+1}-a_\alpha c_\alpha}{a_\alpha^2+1},
\end{align}
where $a_\alpha=\frac{\sqrt{1-\widehat{\rho}_{s}(\alpha)^2}}{\widehat{\rho}_{s}(\alpha)}$. Given that $\widehat{\rho}_{s}(\alpha)$ is selected such that \eref{rhos} is satisfied, we have $\widehat{\rho}_{s}(\alpha)^2=\sin(\theta_\alpha^\ast)^2$ where $\theta_\alpha^\ast$ is the unique solution of \eref{slam_fppf}. This means that $a_\alpha=\tan(\theta_\alpha^\ast)^{-1}$ and $\widehat{s}_\alpha$ can be rewritten as follows
\begin{align}
\widehat{s}_\alpha=\frac{ \tan(\theta_\alpha^\ast) }{ \sqrt{ c_\alpha^2+\tan(\theta_\alpha^\ast)^2+1}+ c_\alpha }.
\end{align}
To ensure that $\widehat{s}_\alpha$ is the optimal solution of the optimization problem \eref{fconprobp}, the first derivative of the cost function of the problem \eref{fconprobp} should be zero at $\widehat{s}_\alpha$. Note that the first derivative of the cost function of problem \eref{fconprobp} can be expressed as 
\begin{align}
\frac{\rho_{s}(\alpha)}{\sqrt{1-\rho_{s}(\alpha)^2}}-\frac{2{s}-2\frac{\alpha}{\pi}\text{atan} \frac{{s}}{\sqrt{ c_{\alpha}^2+1-{s}^2} } }{2\sqrt{g_\alpha({s})}}.
\end{align}
This means that the sufficient input cosine similarity $\rho_{s}(\alpha)$ satisfies the following
\begin{equation}\label{eq:suffp}
\rho_{s}(\alpha)=\frac{\ell_\alpha}{\sqrt{\ell^2_\alpha+1}},
\end{equation}
where $\ell_\alpha$ is given by
\begin{align}
\ell_\alpha \bydef\frac{\widehat{s}_\alpha-\frac{\alpha}{\pi}\text{atan} \frac{\widehat{s}_\alpha}{\sqrt{ c_{\alpha}^2+1-\widehat{s}_\alpha^2} } }{\sqrt{g_\alpha(\widehat{s}_\alpha)}}.
\end{align}
\subsubsection{Convergence analysis}
Now, assume that the input cosine similarity satisfies $\rho_{\text{init}} > \widehat{\rho}_{s}(\alpha)$ for general initial guess and it satisfies $\rho_{\text{init}} > {\rho}_{s}(\alpha)$ for independent initial guess. This means that \eref{inters} is satisfied. Based on Lemma \ref{prop}, an input cosine similarity selected in this way ensures that for any $\epsilon>0$ there exists $\delta > 0$ such that
\begin{align}
\mathcal{D}_{\text{feas}}^{\epsilon} \cap \mathcal{D}_{\text{opt}} \cap {\mathcal D}_{\rm fp}(\rho^k_{\text{init}}) \subseteq \mathcal{B}^\delta, \forall k \geq 0,
\end{align}
where $\mathcal{B}^\delta$ is a ball of radius $\delta$ and center the target signal $\vxi$. Define the sequence $\lbrace \delta_{\epsilon_j} \rbrace_{j\in\mathbb{N}}$ as follows
\begin{align}\label{eq:depslon2}
\hspace{-5mm}\delta_{\epsilon_j} :=\inf \lbrace \delta>0:\mathcal{D}_{\text{feas}}^{\epsilon_j} \cap \mathcal{D}_{\text{opt}} \cap {\mathcal D}_{\rm fp}(\rho^k_{\text{init}}) \subseteq \mathcal{B}^\delta, \forall k \geq 0 \rbrace,
\end{align}
where $\lbrace \epsilon_j \rbrace_{j\in\mathbb{N}}$ is a decreasing sequence of positive numbers such that $\lim_{j\to\infty} \epsilon_j=0$.
Based on the proof of Proposition \ref{suffcondd}, it can be checked that $\lim_{j \to \infty} \delta_{\epsilon_j}=0$. Now, we have
\begin{align}
\bigcup_{k\geq 0}	\lbrace \mathcal{D}_{\text{feas}}^{\epsilon} \cap \mathcal{D}_{\text{opt}} \cap {\mathcal D}_{\rm fp}(\rho^k_{\text{init}}) \rbrace \subseteq \mathcal{B}^{\delta},~\forall \epsilon>0.
\end{align}
Based on Theorem \ref{lem:fb}, %, i.e.
%$
%\lim_{n\rightarrow\infty} \Pro\Big( \mathcal{S}_{\text{feas}}  \subseteq \mathcal{D}_{\text{feas}}^{\epsilon}  \Big) = 1, \forall \epsilon>0
%$, 
for any decreasing sequence of positive numbers $\lbrace  {\epsilon_j} \rbrace_{j\in\mathbb{N}}$ with $\lim_{j\to\infty} \epsilon_j=0$, there exists a sequence of positive numbers $\lbrace \delta_j \rbrace_{j\in\mathbb{N}}$ such that
$$
\lim_{n\rightarrow\infty} \Pro\Big( \bigcup_{k\geq 0}	\lbrace  \mathcal{S}_{\text{feas}} \cap \mathcal{D}_{\text{opt}} \cap {\mathcal D}_{\rm fp}(\rho^k_{\text{init}}) \rbrace  \subseteq \mathcal{B}^{\delta_j} \Big) = 1,~\forall j \in\mathbb{N}.
$$
%Therefore, we have
%$$
%\lim_{n\rightarrow\infty} \Pro\Big(  \mathcal{S}_{\text{feas}} \cap \mathcal{D}_{\text{opt}} \cap {\mathcal D}_{\rm fp}(\rho^k_{\text{init}})  \subseteq \mathcal{B}^{\delta_j}, 
%\forall k \geq 0 \Big) = 1,~\forall j \in\mathbb{N}.
%$$
Therefore, we have
\begin{align}
\lim_{n\rightarrow\infty} \Pro\Big( \sup_{\widehat{\vx}\in\mathcal{F}_{\text{lamp}}}\norm{\widehat{\vx}-\vxi}_2 \leq \delta_j \Big) = 1,
\end{align}
for any $j\geq0$, where $\lim_{j\to 0} \delta_{\epsilon_j}=0$ and $\mathcal{F}_{\text{lamp}}$ denotes the set of fixed points of the PhaseLamp algorithm.
%Given the following properties $\mathcal{D}_{\text{feas}}^{\epsilon_1} \subseteq \mathcal{D}_{\text{feas}}^{\epsilon_2}$ for $\epsilon_1 \leq \epsilon_2$ and $\mathcal{D}_{\text{feas}}^{0}=\mathcal{D}_{\text{feas}}$, 
This implies that the set $\mathcal{F}_{\text{lamp}}$ converges to the set $\lbrace \vxi \rbrace$ in the sense that $\sup_{\widehat{\vx}\in\mathcal{F}_{\text{lamp}}}\norm{\widehat{\vx}-\vxi}_2$ converges to zero in probability. 
This then gives us the statement of Theorems \ref{the2a} and \ref{the2b}.

\section{Conclusion}
\label{sec:conclusion}

We presented in this paper an asymptotically exact characterization of the performance of the PhaseMax method for phase retrieval. Specifically, our analysis reveals a sharp phase transition behavior in the performance of the method as one varies the oversampling ratio and the input cosine similarity. Our analysis is based on the CGMT, and the results match previous predictions derived from the non-rigorous replica method. Moreover, we also presented a new nonconvex formulation of the phase retrieval problem and PhaseLamp, an iterative algorithm based on  linearization and maximization over a polytope. We provided a sufficient condition for PhaseLamp to perfectly retrieve the target vector. Simulation results confirm the validity of our theoretical predictions. They also show that the proposed iterative algorithm significantly improves the recovery performance of the PhaseMax method.

%!TEX root = PhaseLamp.tex

\appendices%\label{app_a}
%\renewcommand{\thesection}{\arabic{section}}
%\renewcommand{\thesubsection}{\arabic{subsection}}
%%%%%%%%%%%%%%%%%%%%%%%%%%%%%%%%%%%%%%
%\input{PhaseMax_Proof.tex}

%\input{PhaseLamp_Proof.tex}

\section{Probabilistic Analysis}
%%%%%%%%%%%%%%%%%%%%%%%%%%%%%%%%%%%%%%%%%%%%%%%%%%
%%%%%%%%%%%%%%%%%%%%%%%%%%%%%%%%%%%%%%%%%%%%%%%%%%
%%%%%%%%%%%%%%%%%%%%%%%%%%%%%%%%%%%%%%%%%%%%%%%%%%
%%%%%%%%%%%%%%%%%%%%%%%%%%%%%%%%%%%%%%%%%%%%%%%%%%
\subsection{Proof of Theorem \ref{lem:fb}}
\label{plem:fb}
Based on Section \ref{pr_th1}, deriving a high-probability lower bound on the optimal objective of $\mathrm{bd}^{\text{AO}}$ leads to a high-probability lower bound on \eref{rmax} in the high dimensional limit, where $\mathrm{bd}^{\text{AO}}$ can be expressed as follows
\begin{align}\label{eq:ao_noncvc2}
 \mathrm{bd}^{\text{AO}}&= \min_{(s,r)\in\mathcal{S}}  r^2- \alpha c_d(s,r)\nonumber\\
 &+ \widetilde{\lambda}_n \rho \Big( -\frac{\norm{\vh}_2}{\sqrt{m(n)}} r - \frac{c_n(s,r)}{\sqrt{m(n)}}  \Big),
\end{align}
and where $\rho:x\to \max(x,0)$.
Define the following problem
\begin{align}\label{eq:rs}
 \mathrm{bd}^{\ast}&=\min_{(s,r)\in\mathcal{S}} r^2- \alpha c_d(s,r) ~\text{s.t.}~ \sqrt{c_d(s,r)} \leq r/\sqrt{\alpha}.  
\end{align}
%%%%%%%%%%%%%%%%%%%%%%%%%%%%%%%%
%%%%%%%%%%%%%%%%%%%%%%%%%%%%%%%%
%%%%%%%%%%%%%%%%%%%%%%%%%%%%%%%%
 Next, we study the asymptotic properties of the problem \eref{ao_noncvc2}. Specifically, we study the convergence properties (with growing $n$) of the formulation given in \eref{ao_noncvc2}. Based on the proof of Proposition \ref{lem1} provided in Appendix \ref{pr_lem1}, we have the following convergence 
\begin{align}\notag
\underbrace{-\frac{\norm{\vh}_2}{\sqrt{m(n)}} r - \frac{c_n(s,r)}{\sqrt{m(n)}}}_{\widehat{d}_n(s,r)} \xrightarrow[]{u.p} \underbrace{-r/\sqrt{\alpha} + \sqrt{c_d(s,r)}}_{\widehat{d}(s,r)},
\end{align}
where $\xrightarrow[]{u.p}$ denotes the convergence uniformly in probability. Since the function $(s,r) \to r^2- \alpha c_d(s,r)$ is continuous and the set $\mathcal{S}=\lbrace (s,r)\in\mathbb{R}^2: r\geq 0,~ s^2+r^2 \leq B \rbrace$ is compact, the function $(s,r) \to r^2- \alpha c_d(s,r)$ is uniformly continuous on the set $\mathcal{S}$. Also, the functions $\widehat{d}_n$ and $\widehat{d}$ are continuous. 
%Based on lemma \ref{prop}, the feasibility set is a subset of the set $[-1,~1]\times \mathbb{R}$, for $\alpha >2$. Assuming that $\alpha>2$,
Note that the set $\lbrace (s,r)\in\mathbb{R}^2:~ r\geq0,~s^2+r^2\leq B,~\widehat{d}(s,r)\leq 0\rbrace$ has a nonempty interior.
%\vsp 

%\noindent \textit{First case}: Assume that $\abs{s} < 1$, then, the set $\lbrace r\in\mathbb{R}~\mathrm{s.t.}~ r\geq0,~s^2+r^2\leq B,~\widehat{d}(r)\leq 0\rbrace$ has a nonempty interior.\\
%\textit{Second case}: Assume that $s=\pm 1$, then, based on \eref{eq3lem9} and property 2 in appendix \ref{proppf}, the set $\lbrace r\in\mathbb{R}~\mathrm{s.t.}~ r\geq0,~s^2+r^2\leq B,~\widehat{d}(r)\leq 0\rbrace$ has a unique element which is $\lbrace 0 \rbrace$. Also, note that in this case, $\widehat{d}(0)=0$ and $\widehat{d}_n(0)=0$.
%\vsp

Then, based on Lemma \ref{gresult}, the optimal objective value of the problem \eref{ao_noncvc2} converges in probability to the optimal objective value $ \mathrm{bd}^{\ast}$ given in \eref{rs}. This means that for any $\epsilon > 0$, we have
\begin{align}%\label{eq:gmt_bd}
\lim_{n\rightarrow\infty} \Pro\Big( \mathrm{bd}^{\text{AO}} < \mathrm{bd}^{\ast} - \eps \Big) = 0.
\end{align}
Using GMT and based on Section \ref{pr_th1}, %we have the following inequality
%\begin{align}%\label{eq:gmt_bd}
%\Pro\Big( \mathrm{bd}^{\text{PO}} < \mathrm{bd}^{\ast} - \eps \Big) \leq  2\Pro\Big( \mathrm{bd}^{\text{AO}} < \mathrm{bd}^{\ast} - \eps \Big), \forall \epsilon >0,
%\end{align}
%where $\mathrm{bd}^{\text{PO}}$ is given in \eref{rmax}. 
%Hence, for all $\eps>0$, 
it holds that
\begin{align}\label{eq:gmt_bd}
\lim_{n\rightarrow\infty} \Pro\Big( \mathrm{bd}^{\text{PO}} \geq \mathrm{bd}^{\ast} - \eps \Big) = 1, \forall \epsilon >0,
\end{align}
where $\mathrm{bd}^{\text{PO}}$ is given in \eref{rmax}.
Based on P.3 in Lemma \ref{prop}, we have $\mathrm{bd}^\ast=0$. Then, the convergence result in \eref{gmt_bd} can be rewritten as follows: for any $\epsilon >0$,
\begin{align}\label{eq:gmt_bd2}
\lim_{n\rightarrow\infty} \Pro\Big( r^2-\alpha c_d(s,r) \geq - \eps,~\forall (s,r)\in \mathcal{S}_{\text{feas}} \Big) = 1. 
\end{align}
Now, define the deterministic set ${\mathcal D}_{\rm feas}^{\epsilon}$ as follows
\begin{align}
{\mathcal D}_{\rm feas}^{\epsilon}=\lbrace (s,r)\in\mathbb{R}^2:~ r\geq 0,~\alpha~c_d(s,r) \leq r^2+\epsilon \rbrace.
\end{align}
Therefore, the convergence result in \eref{gmt_bd2} is equivalent to the following
\begin{align}\label{eq:gmt_bd3}
\lim_{n\rightarrow\infty} \Pro\Big(  \mathcal{S}_{\text{feas}} \subseteq {\mathcal D}_{\rm feas}^{\epsilon} \Big) = 1, \forall \epsilon >0.
\end{align}
This completes the proof of Theorem \ref{lem:fb}.
%%%%%%%%%%%%%%%%%%%%%%%%%%%%%%%%%%%%%%%%%%%%%%%%%%
%%%%%%%%%%%%%%%%%%%%%%%%%%%%%%%%%%%%%%%%%%%%%%%%%%
%%%%%%%%%%%%%%%%%%%%%%%%%%%%%%%%%%%%%%%%%%%%%%%%%%
%%%%%%%%%%%%%%%%%%%%%%%%%%%%%%%%%%%%%%%%%%%%%%%%%%
\subsection{Proof of Proposition \ref{compactness}}
\label{prop3}
First, we appropriately write the optimization problem in \eqref{eq:lp_form_2} as a min-max program. Start with the following equivalent formulation: 
\begin{align}\label{eq:po_form}
\min_{{\vx}\in\mathcal{S}_{\vx}} p(\vx) + \sum_{i=1}^{m} \mathds{1} \Big\lbrace \abs{\va_i^T \vx} - y_i  \Big\rbrace,
\end{align}
where the function $x\to\mathds{1}\lbrace x \rbrace$ is defined as follows: $\mathds{1}\lbrace x \rbrace=0$ if $x \leq 0$ and $\mathds{1}\lbrace x \rbrace=+\infty$ if $x > 0$. Therefore, \eref{po_form} is equivalent to the following optimization problem
\begin{align}\label{eq:pof_form}
V_n&=\min_{{\vx}\in\mathcal{S}_{\vx}} p(\vx) + \sum_{i=1}^{m} \max_{u_i \geq 0} \Big\lbrace \Big( \abs{\va_i^T \vx} - y_i \Big) u_i \Big\rbrace\nonumber\\
&=\min_{{\vx}\in\mathcal{S}_{\vx}} p(\vx) + \sum_{i=1}^{m} \max_{u_i \in \mathbb{R}} \Big\lbrace \left( \va_i^T \vx \right) u_i - y_i \abs{u_i} \Big\rbrace\nonumber\\
&=\min_{{\vx}\in\mathcal{S}_{\vx}} \max_{\vu \in \mathbb{R}^m} p(\vx) + \vu^T \mA \vx - \abs{\vu}^T\vy.
\end{align}
The GMT and the CGMT assumes that the feasibility sets of the optimization variables $\vx$ and $\vu$ are compact. Clearly, this assumption is not satisfied by the min-max problem \eref{pof_form} since the feasibility set of the variable $\vu$ is not compact.\\ 
{\bf Case 1}: Assume that ${\bf C}_2$ or ${\bf C}_3$ holds. It can be noticed that the optimal objective of \eref{pob_form} is smaller than the optimal objective of \eref{pof_form} with probability one, i.e. $V_n(\lambda_n) \leq V_n$.\\ %This implies that deriving a high-probability lower bound on $V_n(\lambda_n)$ leads to a high-probability lower bound on $V_n$.\\
{\bf Case 2}: Assume that ${\bf C}_1$ holds. %Next, we show that the high dimensional analysis of the problem \eref{pof_form} is equivalent to the study of the asymptotic behavior of the optimization problem \eref{pob_form}. To this end, 
Define the following optimization problem
\begin{equation}\label{eq:bdmax}
\begin{aligned}
C_n&=\underset{\norm{\vx}_{\infty}\leq w_n}{\min} -\vx_{\text{init}}^T\vx ~~\text{s.t.}~~  \abs{\va_i^T \vx} \leq y_i, \text{ for }  1 \le i \le m,
\end{aligned}
\end{equation}
where $\lbrace w_n \rbrace_{n\in\mathbb{N}}$ is a sequence of positive numbers. Based on Lemma \ref{boundp}, $\norm{\mathcal{K}_n}_2$ is bounded by a constant $\tau$ independent of $n$ with probability going to one as $n$ goes to infinity. This means that there exists at least one sequence of positive numbers $\lbrace w_n \rbrace_{n\in\mathbb{N}}$ satisfying $w_n > \tau,~\forall n$ such that $\mathbb{P}\left( V_n=C_n \right) \overset{n\to\infty}{\longrightarrow}1$. Now, define the following linear program 
\begin{equation}\label{eq:bdmax2}
\begin{aligned}
C_n(\lambda_n)&=\underset{ \substack{ \norm{\vx}_{\infty}\leq w_n\\ z_i \geq 0}}{\min} -\vx_{\text{init}}^T\vx+\lambda_n\sum_{i=1}^{m}z_i\\	
&~~~~~~~~\text{s.t.}~~~~  \abs{\va_i^T \vx} \leq y_i+z_i, \text{ for }  1 \le i \le m.
\end{aligned}
\end{equation}
Let $\widetilde{\mathcal{K}}_n$ be the feasibility set of the optimization problem \eref{bdmax2}. Clearly, the feasibility set $\widetilde{\mathcal{K}}_n$ is a polytope with nonempty extreme point set. Moreover, the cost function of the problem in \eref{bdmax2} is lower bounded in the feasibility set $\widetilde{\mathcal{K}}_n$. Then, using the result in \cite[Corollary 32.3.4]{rockafellar70}, the optimal objective value $C_n(\lambda_n)$ is achieved at one of the vertices of the polytope $\widetilde{\mathcal{K}}_n$. Define the set $\widetilde{\mathcal{E}}$ as follows $$\widetilde{\mathcal{E}}=\Big\lbrace \vz\in\mathbb{R}^m:\sum_{i=1}^{m}z_i\neq 0,~\exists \vx\in\mathbb{R}^n: (\vx^T,\vz^T)\in\mathcal{E}(\widetilde{\mathcal{K}}_n) \Big\rbrace,$$ where the set $\mathcal{E}(\widetilde{\mathcal{K}}_n)$ denotes the set of all extreme points of the polytope $\widetilde{\mathcal{K}}_n$. Since the polytope $\widetilde{\mathcal{K}}_n$ has a finite number of extreme points, the set $\widetilde{\mathcal{E}}$ has a finite cardinality. 

Assume that $\lambda_n \geq \norm{\vx_{\text{init}}}_2 w_n\sqrt{n}/\zeta_n$ where $\zeta_n$ is defined as follows
\begin{equation}
\zeta_n=\begin{cases}
\min\limits_{\vz\in\widetilde{\mathcal{E}}} \sum_{i=1}^{m} z_i & \mathrm{if}~\widetilde{\mathcal{E}}\neq \varnothing \\
1 & \mathrm{otherwise}.
\end{cases}
\end{equation}
Note that $C_n(\lambda_n) \leq C_n$, for any sequence of positive numbers $\lbrace \lambda_n \rbrace_{n\in\mathbb{N}}$. Next, the objective is to show that $C_n(\lambda_n) \geq C_n$. To this end, we consider two different cases:

{\bf Case 2.a}: Assume that the set $\widetilde{\mathcal{E}}$ is empty. This implies that all the extreme points of the polytope $\widetilde{\mathcal{K}}_n$ are of the form $(\vx^T, \vz^T={\bf 0})$ where $\vx\in \mathcal{K}_n$, and where $\mathcal{K}_n$ is the PhaseMax feasibility set. Therefore, $C_n(\lambda_n) \geq C_n$ which means that $C_n(\lambda_n) = C_n$ for any sequence of positive numbers $\lbrace w_n \rbrace_{n\in\mathbb{N}}$ and $\lbrace \lambda_n \rbrace_{n\in\mathbb{N}}$.

{\bf Case 2.b}: Assume that the set $\widetilde{\mathcal{E}}$ is nonempty. Then, for any extreme point $(\vx^T,\vz^T)$ of the polytope $\widetilde{\mathcal{K}}_n$ which belongs to the set $\widetilde{\mathcal{E}}$, we have 
$$
-\vx_{\text{init}}^T\vx+\lambda_n\sum_{i=1}^{m}z_i \geq -\vx_{\text{init}}^T\vx+\norm{\vx_{\text{init}}}_2 w_n \sqrt{n} \geq 0,
$$
where the last inequality follows since $\norm{\vx}_2 \leq w_n \sqrt{n}$ for any $\vx$ satisfying $\norm{\vx}_\infty\leq w_n$. Since the all zero vector is in the feasibility set of the PhaseMax problem, then, $C_n \leq 0$. This implies that $C_n \leq C_n(\lambda_n)$ which leads to the following equality $C_n = C_n(\lambda_n)$ for any sequence of positive numbers $\lbrace w_n \rbrace_{n\in\mathbb{N}}$ and $\lbrace \lambda_n \rbrace_{n\in\mathbb{N}}$ satisfying $\lambda_n\geq \norm{\vx_{\text{init}}}_2 w_n\sqrt{n}/\zeta_n$. 

Now, we discuss the existence of the sequence $\lbrace \lambda_n \rbrace_{n\in\mathbb{N}}$ satisfying $\lambda_n\geq \norm{\vx_{\text{init}}}_2 w_n\sqrt{n}/\zeta_n$. First, note that $\zeta_n$ is a well-definite random variable with $\zeta_n >0,~\forall n$. This implies that we can construct a sequence of positive numbers $\lbrace \delta_n \rbrace_{n\in\mathbb{N}}$ such that  $\mathbb{P}\left( 0 \leq \zeta_n \leq \delta_n \right) \leq 2^{-n}$. We can then choose $\lambda_n$ as $\lambda_n=\norm{\vx_{\text{init}}}_2 w_n\sqrt{n}/ \delta_n$. This leads to the following $$\mathbb{P}\left(  C_n=C_n(\lambda_n)\right) \geq 1-2^{-n},$$ where this is true for any sequence of positive numbers $\lbrace w_n \rbrace_{n\in\mathbb{N}}$.

Finally, consider the event $E_n=\lbrace \mA:~\norm{\mathcal{K}_n}_2 \leq \tau \rbrace$, then, we have the following
\begin{align}
\mathbb{P}\left( \lbrace C_n=C_n(\lambda_n) \rbrace \cap E_n \right) \leq \mathbb{P} \left( C_n=V_n(\lambda_n)  \right),
\end{align}
for any sequence of positive numbers $\lbrace \lambda_n \rbrace_{n\in\mathbb{N}}$ and $\lbrace w_n \rbrace_{n\in\mathbb{N}}$ satisfying $w_n > \tau,~\forall n$. Moreover, we have
\begin{align}\notag
\mathbb{P}\left( \lbrace C_n=C_n(\lambda_n) \rbrace \cap E_n \right) \geq \mathbb{P}\left( C_n=C_n(\lambda_n) \right)-\mathbb{P}\left( E^c_n \right),
\end{align}
which implies that
\begin{align}\notag
\mathbb{P}\left( C_n=V_n(\lambda_n) \right) \overset{n\to\infty}{\longrightarrow} 1.
\end{align}
We know that there exists at least one sequence of positive numbers $\lbrace w_n \rbrace_{n\in\mathbb{N}}$ satisfying $w_n > \tau,~\forall n$ such that $\mathbb{P}\left( V_n=C_n \right) \overset{n\to\infty}{\longrightarrow}1$. Then, there exists a sequence of positive numbers $\lbrace \lambda_n \rbrace_{n\in\mathbb{N}}$ such that $\mathbb{P}\left( V_n=V_n(\lambda_n) \right) \overset{n\to\infty}{\longrightarrow}1$.

Note that $V_n$ is given by
\begin{align}\label{eq:vn}
V_n=\min_{{\vx}\in\mathcal{S}_{\vx}} -{\vx}_\text{init}^{T}\,{\vx} + \sum_{i=1}^{m} \mathds{1} \Big\lbrace \abs{\va_i^T \vx} - y_i  \Big\rbrace,
\end{align}
and $V_n(\lambda_n)$ is given by
\begin{align}\label{eq:vnlm}
V_n(\lambda_n)&=\min_{{\vx}\in\mathcal{S}_{\vx}} -{\vx}_\text{init}^{T}\,{\vx} + \lambda_n \sum_{i=1}^{m} \rho\Big( \abs{\va_i^T \vx} - y_i \Big),
\end{align}
where the function $\rho:x\to\max(x,0)$. Denote by $\widehat{V}_n$ the cost function of the problem \eref{vn} and $\widetilde{V}_n$ the cost function of the problem \eref{vnlm}. Further, assume that $\widehat{\vx}_n$ is an optimal solution of the problem \eref{vn} and $\widetilde{\vx}_n$ is an optimal solution of the problem \eref{vnlm}. It is clear that $\widehat{V}_n(\widehat{\vx}_n)=\widetilde{V}_n(\widehat{\vx}_n)$ and also 
\begin{equation}
\widehat{V}_n(\widetilde{\vx}_n)=\begin{cases}
\widetilde{V}_n(\widetilde{\vx}_n) & \mathrm{if}~\abs{\va_i^T \widetilde{\vx}_n} \leq y_i\\
+\infty & \mathrm{otherwise}.
\end{cases}
\end{equation}
Since the all zero vector is in the polytope $\mathcal{K}_n$, $V_n$ is finite with probability one. Moreover, since 
$$\mathbb{P}\left( V_n=V_n(\lambda_n) \right) \overset{n\to\infty}{\longrightarrow}1,$$ 
we obtain the following convergence result 
$$\mathbb{P}\left( \mathcal{V}_n \subseteq \mathcal{V}_n(\lambda_n) \right) \overset{n\to\infty}{\longrightarrow}1,$$ 
where $\mathcal{V}_n$ denotes the set of optimal solutions of the problem \eref{vn} and $\mathcal{V}_n(\lambda_n)$ denotes the set of optimal solutions of the problem \eref{vnlm}.

\noindent The above two cases give us the statement in Proposition \ref{compactness}.

%forcing the compactness of the feasibility set of the variable $\vu$ does not change the optimal solution $V_n$ with high probability in the high dimensional limit. 
%Define $V_n(\lambda_n)$ as follows 
%\begin{align}%\label{eq:pob_form}
%V_n(\lambda_n)&=\min_{{\vx}\in\mathcal{S}_{\vx}} \max_{\abs{\vu} \leq  \lambda_n} ~p(\vx) + \vu^T \mA \vx - \abs{\vu}^T\vy.
%\end{align}
%where $\lambda_n$ is a deterministic sequence, finite and dependent on $n$. 
%\begin{lemma}
%\label{boundd}
%Assume that ${\bf C}_1$ holds. There exists a sequence of positive numbers $\lambda_n < \infty$ such that $\mathbb{P}\left( V_n(\lambda_n)=V_n \right)\overset{n\to\infty}{\longrightarrow} 1$. Moreover, we have 
%$$\mathbb{P}\left(  \mathcal{V}_n \subseteq \mathcal{V}_n(\lambda_n) \right) \overset{n\to\infty}{\longrightarrow}1,$$ 
%where $\mathcal{V}_n$ denotes the set of optimal solutions of the problem \eref{pof_form} and $\mathcal{V}_n(\lambda_n)$ denotes the set of optimal solutions of the problem \eref{pob_form}. 
%\end{lemma}
%\begin{IEEEproof}
%The detailed proof is given in appendix \ref{pboundd}.
%\end{IEEEproof}
%Lemma \ref{boundd} shows that when ${\bf C}_1$ holds, the high dimensional analysis of the problem \eref{pof_form} is equivalent to the study of the asymptotic behavior of the optimization problem \eref{pob_form}. This completes the proof of Proposition \ref{compactness}.

%%%%%%%%%%%%%%%%%%%%%%%%%%%%%%%%%%%%%%%%%%%%%%%%%%
%%%%%%%%%%%%%%%%%%%%%%%%%%%%%%%%%%%%%%%%%%%%%%%%%%
%%%%%%%%%%%%%%%%%%%%%%%%%%%%%%%%%%%%%%%%%%%%%%%%%%
%%%%%%%%%%%%%%%%%%%%%%%%%%%%%%%%%%%%%%%%%%%%%%%%%%
\subsection{Proof of Proposition \ref{prop_spao}}
\label{prop4}
It can be noticed that the optimization problem \eref{ao_form} can be rewritten as follows:
\begin{align}\label{eq:ao1_form}
\hspace{-2mm}\underset{\substack{ ({x}_1,\widetilde{\vx})\in\mathcal{S}_{\vx} }}{\min} \max_{ \vu\in\mathcal{S}_{\vu}(n) }&  \left( \norm{\widetilde{\vx}}_2 \vg + x_1\vq \right)^T \vu + \norm{\vu}_2 \vh^T \widetilde{\vx} \nonumber\\
&+p({x}_1,\widetilde{\vx})  -\abs{\vu}^T \abs{\vq}.
\end{align}
Next, observe that if we fix $\abs{\vu}$, then the optimal $\vu$ satisfies $\text{sign}(\vu)=\text{sign}\left(\norm{\widetilde{\vx}}_2 \vg + \vq x_1\right)$ which simplifies the optimization to the following
\begin{align}\label{eq:ao1_form_2}
\hspace{-2mm}\underset{\substack{ ({x}_1,\widetilde{\vx})\in\mathcal{S}_{\vx} }}{\min} \max_{ 0\leq\vu\leq \lambda_n }& -[|\vq| -|\norm{\widetilde{\vx}}_2 \vg + \vq x_1|]^T\vu  +  \norm{\vu}_2 \vh^T \widetilde{\vx} \nonumber\\
&+p({x}_1,\widetilde{\vx}). 
\end{align}

Define the following optimization problem 
\begin{align}\label{eq:ao2_form_2}
\hspace{-2mm}\underset{\substack{ ({x}_1,\widetilde{\vx})\in\mathcal{S}_{\vx} }}{\min} \max_{ \vu\in\widetilde{\mathcal{S}}_{\vu}(n) }&  -[|\vq| -|\norm{\widetilde{\vx}}_2 \vg + \vq x_1|]^T\vu  +  \norm{\vu}_2 \vh^T \widetilde{\vx} \nonumber\\
&+p({x}_1,\widetilde{\vx}), 
\end{align}
where $\widetilde{\mathcal{S}}_{\vu}(n)=\lbrace \vu\in\mathbb{R}^{m}:~\vu \geq 0,~\norm{\vu}_{2} \leq \lambda_n \rbrace$. Moreover, define the vector $\vx=[x_1~\widetilde{\vx}^T]^T$ and the functions $\vx \to \Delta_{\infty}(\lambda_n,\vx)$ and $\vx \to \Delta_{2}(\lambda_n,\vx)$ as follows
\begin{align}
\Delta_{\infty}(\lambda_n,\vx)&=\max_{0\leq \vu\leq \lambda_n} -[|\vq| -|\norm{\widetilde{\vx}}_2 \vg + \vq x_1|]^T\vu  \nonumber\\
&+  \norm{\vu}_2 \vh^T \widetilde{\vx}
+p({x}_1,\widetilde{\vx}),
\end{align}
and 
\begin{align}
\Delta_{2}(\lambda_n,\vx)&=\max_{\substack{ \vu\geq0\\ \norm{\vu}_2\leq \lambda_n}} -[|\vq| -|\norm{\widetilde{\vx}}_2 \vg + \vq x_1|]^T\vu  \nonumber\\
&+  \norm{\vu}_2 \vh^T \widetilde{\vx}
+p({x}_1,\widetilde{\vx}). 
\end{align}
Next, we show that the analysis of the optimization problem 
\begin{align}\label{eq:norinp}
\Delta_{\infty}^\ast(\lambda_n)=\min_{(x_1,\widetilde{\vx})\in\mathcal{S}_{\vx}} \Delta_{\infty}(\lambda_n,\vx),
\end{align}
can be achieved by analyzing the following problem 
\begin{align}\label{eq:nor2p}
\Delta_{2}^\ast(\lambda_n)=\min_{(x_1,\widetilde{\vx})\in\mathcal{S}_{\vx}} \Delta_{2}(\lambda_n,\vx),
\end{align}
in the high dimensional limit, if the sequence of positive numbers $\lambda_n \overset{n\to\infty}{\longrightarrow}\infty$. Next, we assume that the sequence of positive numbers $ \lbrace \lambda_n \rbrace_{n\in\mathbb{N}}$ satisfies $\lambda_n \overset{n\to\infty}{\longrightarrow}\infty$. First, it is clear that 
\begin{align}
\Delta_{2}^\ast(\lambda_n) \leq \Delta_{\infty}^\ast(\lambda_n) \leq \Delta_{2}^\ast(\sqrt{n}\lambda_n),
\end{align}
which means that 
\begin{align}\label{eq:f1}
\mathbb{P}\left( \Delta_{2}^\ast(\lambda_n)=\Delta_{\infty}^\ast(\lambda_n) \right)\overset{n\to\infty}{\longrightarrow}1.
 \end{align}
%%%%%%%%%%%%%%%%%%%%%%%%%%%%%%%%%%
%%%%%%%%%%%%%%%%%%%%%%%%%%%%%%%%%%
%%%%%%%%%%%%%%%%%%%%%%%%%%%%%%%%%%
Moreover, assume that $\mathcal{X}_{2}^\ast(\lambda_n)$ is the set of optimal solutions of the problem \eref{nor2p} with sequence $\lbrace \lambda_n \rbrace_{n\in\mathbb{N}}$ and $\mathcal{X}_{\infty}^\ast(\lambda_n)$ is the set of optimal solutions of the problem \eref{norinp} with sequence $\lbrace \lambda_n \rbrace_{n\in\mathbb{N}}$. Assume that there exists $\vx_n \in \mathcal{X}_{\infty}^\ast(\lambda_n)$ such that $\vx_n \notin \mathcal{X}_{2}^\ast(\lambda_n)$ which implies that 
\begin{equation}\label{eq:f31}
\Delta_{2}(\lambda_n,\vx_n) < \Delta_{\infty}(\lambda_n,\vx_n) \leq \Delta_{2}(\sqrt{n}\lambda_n,\vy),~\forall \vy.
\end{equation}
Therefore, we have
\begin{equation}\label{eq:f32}
\Delta_{2}(\lambda_n,\vy_n) < \Delta_{\infty}(\lambda_n,\vx_n) \leq \Delta_{2}(\sqrt{n}\lambda_n,\widetilde{\vy}_n),
\end{equation}
where $\vy_n \in \mathcal{X}_{2}^\ast(\lambda_n)$ and $\widetilde{\vy}_n \in \mathcal{X}_{2}^\ast(\sqrt{n}\lambda_n)$. This implies that 
\begin{equation}\label{eq:f33}
\Delta_{2}^\ast(\lambda_n) < \Delta_{2}^\ast(\sqrt{n}\lambda_n).
\end{equation}
Now, assume that the probability of the event $\lbrace \mathcal{X}_{\infty}^\ast(\lambda_n) \subseteq \mathcal{X}_{2}^\ast(\lambda_n) \rbrace$ does not converge to one as $n$ goes to infinity. This means that there exists $\delta > 0$, a sequence $\lambda_{n_j}, j\geq 0$, and $j_0 \in \mathbb{N}$ such that for all $j \geq j_0$, 
\begin{equation}\label{eq:f34}
\mathbb{P} \left( \Delta_{2}^\ast(\lambda_{n_j}) < \Delta_{2}^\ast(\sqrt{n_j}\lambda_{n_j}) \right) > \delta.
\end{equation}
Furthermore, we know that
\begin{equation}\label{eq:f35}
\mathbb{P} \left( \Delta_{2}^\ast(\lambda_{n}) = \Delta_{2}^\ast(\sqrt{n}\lambda_{n}) \right) \overset{n\to\infty}{\longrightarrow}1,
\end{equation}
which means that for any $\epsilon > 0$, there exists $n_0\in\mathbb{N}$ such that for all $n \geq n_0$, 
\begin{equation}\label{eq:f36}
\mathbb{P} \left( \Delta_{2}^\ast(\lambda_{n}) = \Delta_{2}^\ast(\sqrt{n}\lambda_{n}) \right) > 1-\epsilon.
\end{equation}
Now, for $n_j \geq \max(n_{j_0},n_0)$, we have
\begin{align}\label{eq:f37}
1&=\mathbb{P} \left( \Delta_{2}^\ast(\lambda_{n_j}) = \Delta_{2}^\ast(\sqrt{n_j}\lambda_{n_j}) \right)\nonumber\\&+\mathbb{P} \left( \Delta_{2}^\ast(\lambda_{n_j}) < \Delta_{2}^\ast(\sqrt{n_j}\lambda_{n_j}) \right) \nonumber\\
&> 1-\epsilon+\delta,
\end{align}
where this is true for any $\epsilon>0$. Taking $\epsilon \leq \delta$ gives a contradiction. This implies that
\begin{align}\label{eq:f38}
\mathbb{P} \left( \mathcal{X}_{\infty}^\ast(\lambda_n) \subseteq \mathcal{X}_{2}^\ast(\lambda_n) \right)\overset{n\to\infty}{\longrightarrow}1.
\end{align}
In what follows, we analyze the problem \eref{ao2_form_2} where the sequence $\lambda_n < \infty$ satisfies $\lambda_n\overset{n\to\infty}{\longrightarrow} \infty$.
%\begin{lemma}
%\label{refana}
%The analysis of the problem \eref{ao1_form_2} is equivalent to the analysis of the problem \eref{ao2_form_2} in the high dimensional limit,
%when the sequence of positive numbers $\lambda_n$ converges to infinity as $n \to \infty$.
%\end{lemma}
%\begin{IEEEproof}
%The detailed proof is given in appendix \ref{prefana}.
%\end{IEEEproof}
%Lemma \ref{refana} essentially shows that the optimization problems \eref{ao1_form_2} and \eref{ao2_form_2} have the same optimal objective value and the same optimal solution set in the high dimensional limit when the sequence $\lambda_n\overset{n\to\infty}{\longrightarrow} \infty$. In what follows, we analyze the problem \eref{ao2_form_2} where the sequence $\lambda_n < \infty$ satisfies $\lambda_n\overset{n\to\infty}{\longrightarrow} \infty$.
In the optimization problem \eref{ao2_form_2}, one can fix the norm of $\vu$  and optimize over its direction. This leads to the following optimization problem
\begin{align}\label{eq:ao2_form}
\underset{ ({x}_1,\widetilde{\vx})\in\mathcal{S}_{\vx} }{\min}\max_{ 0 \leq \lambda \leq \lambda_n }&p({x}_1,\widetilde{\vx})+ \lambda \vh^T \widetilde{\vx}   \nonumber\\
&-\lambda h\left( \abs{\vq} - \abs{ \norm{\widetilde{\vx}}_2 \vg + x_1 \vq}  \right),
\end{align}
where the function $h$ is defined in \eref{funh}. 
%\begin{equation}\label{eq:funh}
%h(\vc)=\begin{cases}
%-\norm{\vc\wedge\mathbf{0}}_2 & ,\text{if}~\min(\vc) \leq 0, \\
%\min(\vc) & ,\text{otherwise}. 
%\end{cases}
%\end{equation}
Therefore, \eref{ao2_form} is equivalent to the following problem
\begin{align}\label{eq:ao3_form}
\underset{ ({x}_1,\widetilde{\vx})\in\mathcal{S}_{\vx} }{\min}&p({x}_1,\widetilde{\vx}) \nonumber\\
&+ \lambda_n \rho \Big( \vh^T \widetilde{\vx} - h\left( \abs{\vq} - \abs{ \norm{\widetilde{\vx}}_2 \vg + x_1 \vq}  \right)  \Big), 
\end{align}
where the function $\rho:x\to \max(x,0)$. Now, we distinguish between two cases:\\
\textbf{Case 1}: Assume that ${\bf C}_1$ holds, i.e. $p({x}_1,\widetilde{\vx})=-\eta_1 x_1 - \widetilde{\veta}^T \widetilde{\vx}$. The final step in simplifying the AO problem is as follows. For fixed value of $x_1$ (say $x_1=s>0$), and for fixed norm of $\widetilde{\vx}$ (say, $\norm{\widetilde{\vx} }_2=r$), we optimize over the direction of $\widetilde{\vx}$. First, fix $\norm{\widetilde{\vx} }_2=r$ such that $s^2+r^2 \leq B,~r\geq 0$ and fix $z=\widetilde{\veta}^T\widetilde{\vx}/\norm{\widetilde{\veta}}_2$ and solve the following optimization problem
\begin{align}\label{eq:ao3_form2}
&\underset{ \widetilde{\vx} }{\min}~ \vh^T \widetilde{\vx}  \\
&~~\text{s.t.}~~ \norm{\widetilde{\vx}}_2=r,~z=\widetilde{\veta}^T\widetilde{\vx}/\norm{\widetilde{\veta}}_2,~\abs{z} \leq r. \nonumber
\end{align}
To solve the optimization problem \eref{ao3_form2}, we write $\vh$ as follows $\vh =  \left( \frac{\vh^T \widetilde{\veta}}{\norm{\widetilde{\veta}}_2} \right) \frac{\widetilde{\veta}}{\norm{\widetilde{\veta}}_2}+\left( \vw^T \vh \right) \vw$, where $\frac{\widetilde{\veta}}{\norm{\widetilde{\veta}}_2}$ and $\vw$ form an orthonormal basis for the two dimensional subspace spanned by $\widetilde{\veta}$ and $\vh$. Thus, \eref{ao3_form2} becomes 
\begin{align}\label{eq:ao3_form3}
&\underset{ \widetilde{\vx} }{\min}~ \Big[ \Big( \frac{\vh^T \widetilde{\veta}}{\norm{\widetilde{\veta}}_2} \Big) \frac{\widetilde{\veta}}{\norm{\widetilde{\veta}}_2}+\left( \vw^T \vh \right) \vw \Big]^T \widetilde{\vx}  \\
&~~\text{s.t.}~~ \norm{\widetilde{\vx}}_2=r,~ z=\widetilde{\veta}^T\widetilde{\vx}/\norm{\widetilde{\veta}}_2,~\abs{z} \leq r. \nonumber
\end{align}
It is clear that the optimal $\widetilde{\vx}$ should be in the span of $\widetilde{\veta}$ and $\vh$ which means that 
$$
\abs{\vw^T\widetilde{\vx}}=\sqrt{\norm{\widetilde{\vx}}^2_2-\left( {\widetilde{\veta}^T\widetilde{\vx}}/{\norm{\widetilde{\veta}}_2} \right)^2}.
$$
 Therefore, the optimal objective value of the problem \eref{ao3_form3} can be expressed as follows 
$$
\left( {\vh^T \widetilde{\veta}}/{\norm{\widetilde{\veta}}_2} \right) z - \sqrt{\norm{\vh}^2_2-\left( {\widetilde{\veta}^T\vh}/{\norm{\widetilde{\veta}}_2} \right)^2} \sqrt{r^2-z^2}
,
$$ 
where $\abs{z} \leq r$. Assume that $\widehat{\veta}=\frac{\widetilde{\veta}}{\norm{\widetilde{\veta}}_2}$, then, the optimization problem reduces to the following problem
\begin{align}\label{eq:ao4_form}
\underset{ \substack{ (s,r)\in\mathcal{S}\\ \abs{z} \leq r} }{\min}&-\eta_1 s - \norm{\widetilde{\veta}}_2 z + \lambda_n  \rho \Big\lbrace \vh^T \widehat{\veta} ~z \nonumber\\
& - \sqrt{\norm{\vh}^2_2-(\widehat{\veta}^T\vh)^2} \sqrt{r^2-z^2} - c_n(s,r)  \Big\rbrace,
\end{align}
where the function $c_n$ is defined in \eref{cn} 
%\begin{equation}\label{eq:cn}
%c_n(s,r)=h\left( \abs{\vq} - \abs{ r \vg + s \vq}  \right),
%\end{equation}
and where the set $\mathcal{S}=\lbrace (s,r)\in\mathbb{R}^2:~s^2+r^2\leq B,~r\geq 0 \rbrace$. Note that the optimization problem \eref{ao4_form} is equivalent to the problem \eref{ao55_form}.\\
%%%
\textbf{Case 2}: Assume that ${\bf C}_2$ or ${\bf C}_3$. The last step is to optimize over the direction of $\vxt$: it will align itself with $\vh$. Doing this, and calling $r=\|\vxt\|_2$ we have arrived at the following simple formulation of the AO
\begin{align}%\label{eq:ao_noncv}
\min_{(s,r)\in\mathcal{S}} p(s,r) + \widetilde{\lambda}_n \rho \Big( -\frac{\norm{\vh}_2}{\sqrt{m(n)}} r - \frac{c_n(s,r)}{\sqrt{m(n)}}  \Big),
\end{align}
where the function $c_n$ is defined in \eref{cn}, $\widetilde{\lambda}_n=\sqrt{m(n)} \lambda_n$, and where $p(s,r)=-s^2-r^2$ if ${\bf C}_2$ holds and $p(s,r)=r^2-\alpha c_d(s,r)$ if ${\bf C}_3$ holds.

\noindent 
Note that \eref{f1} and \eref{f38} hold for all cases ${\bf C}_1$, ${\bf C}_2$ and ${\bf C}_3$. This, then, leads us to the statement in Proposition \ref{prop_spao}.

\subsection{Proof of Proposition \ref{lem1}}
\label{pr_lem1}

Assume that the oversampling ratio satisfies $\alpha > 2$. We show Proposition \ref{lem1} in three steps. The first two steps  study the asymptotic properties of the random function $c_n$. Then,  these properties are used to prove Proposition \ref{lem1} in the final step. To this end, consider the random function $f_n$ defined as follows
\begin{align}\label{eq:funn}
f_n:(s,r)\to \frac{{\norm{ (\abs{\vq} - \abs{ r \vg + s \vq}) \wedge {\bf 0} }_2}}{\sqrt{m(n)}},
\end{align}
and defined on the set $\mathcal{S}=\lbrace (s,r)\in\mathbb{R}^2: r\geq 0, s^2+r^2 \leq B \rbrace$.

{\bf Step 1}:  We start by showing that the functions $f_n$ and $\widetilde{c}_n:(s,r)\to-c_n(s,r)/\sqrt{m(n)}$ have the same pointwise limit. Moreover, the function $f_n$ converges pointwise to the function $(s,r)\to \sqrt{c_d(s,r)}$, where the function $c_d$ is defined in \eref{cdexp} and its closed-form expression is given in \eref{cdref}.
To prove the above property, fix $s$ and $r$ such that $(s,r)\in\mathcal{S}$. Using the weak law of large number (WLLN), we have
\begin{equation}
\frac{{\norm{ (\abs{\vq} - \abs{ r \vg + s \vq}) \wedge {\bf 0} }^2_2}}{m(n)} \xrightarrow[]{n\to\infty} \mathbb{E}(\min(\abs{q} - \abs{ r g + s q},0)^2),\nonumber
\end{equation}
where $q$ and $g$ are i.i.d. standard Gaussian random variables. This means that $f_n(s,r)$ converges in probability to $\sqrt{c_d(s,r)}$ where $c_d(s,r)=\mathbb{E}(\min(\abs{q} - \abs{ r g + s q},0)^2)$. 

Now, fix $\epsilon>0$ and consider the probability event $E_n= \lbrace (\vq,\vg):~\abs{ \widetilde{c}_n(s,r)-\sqrt{c_d(s,r)} } > \epsilon \rbrace$. Then, we have
\begin{align}
\mathbb{P}\left( E_n \right) &= \mathbb{P}\left( E_n | A \right) \mathbb{P}\left( A \right) + \mathbb{P}\left( E_n | A^c \right) \mathbb{P}\left( A^c \right),
\end{align}
where $A$ is the event $\lbrace (\vq,\vg):~ \min(\abs{\vq} - \abs{ r \vg + s \vq}) \leq 0 \rbrace$. The probability of the event $A^c$ is given by
\begin{align}
\mathbb{P}\left( A^c \right)&=\mathbb{P}\left( \min(\abs{\vq} - \abs{ r \vg + s \vq}) > 0 \right)\nonumber\\
&=\left( \mathbb{P}\left(  \abs{q} - \abs{ r g + s q} > 0 \right) \right)^{m(n)}.
\end{align}
Next, we distinguish between two different cases:\\
\textbf{Case 1}: if $r=0$ and $\abs{s} < 1$. Then, the condition $\abs{q} - \abs{ r g + s q} > 0$ is satisfied with probability one which means that $\mathbb{P}\left( A^c \right)=1$. Note that $c_d(s,r)$ is zero in this case. This means that
\begin{align}\label{eq:act}
\mathbb{P}\left( E_n \right) &=  \mathbb{P}\Big( \abs{ {\min(\abs{\vq} - \abs{ r \vg + s \vq})}/{\sqrt{m(n)}} } > \epsilon  \Big).
\end{align}
Equation \eref{act} can be rewritten as follows
\begin{align}\nonumber
\mathbb{P}\Big(  \frac{(1-\abs{s})}{\sqrt{m(n)}} \min(\abs{\vq})   > \epsilon \Big) &= \mathbb{P}\Big(  \abs{q_i}  > \frac{\sqrt{ m(n)}{\epsilon}}{(1-\abs{s})},\forall i \Big).
\end{align}
Given that $\lbrace q_i \rbrace_{i=1}^{m(n)}$ are i.i.d. standard Gaussian random variable, the above equation can be rewritten as follows 
\begin{align}\nonumber
\mathbb{P}\Big(  \frac{(1-\abs{s}) }{\sqrt{m(n)}}\min(\abs{\vq})   > \epsilon \Big) = \Big( 2 \Phi\Big(-\frac{\sqrt{m(n)}{\epsilon}}{(1-\abs{s})} \Big) \Big)^{m(n)},
\end{align}
where $\Phi$ denotes the cumulative distribution function of the standard normal random variable. We know that $\epsilon > 0$ and $\abs{s} < 1$, hence, we obtain the following inequality $2\Phi\left(-{\sqrt{m(n)}{\epsilon}}/{(1-\abs{s})} \right) < 1$, for all $m(n) > 0$. This leads to the following
\begin{equation}
\mathbb{P}\left( E_n \right)  \overset{n\to\infty}{\longrightarrow} 0.
\end{equation}
Hence, we can conclude that $\widetilde{c}_n(s,r)$ converges in probability to $\sqrt{c_d(s,r)}$ for any $s$ and $r$ such that $r=0$ and $\abs{s} < 1$.\\
\textbf{Case 2}: if $r=0$ and $\abs{s} \geq 1$ or $r\neq 0$. In this case, note that
\begin{align}
\mathbb{P}\left( A^c \right)&=\mathbb{P}\left( \min(\abs{\vq} - \abs{ r \vg + s \vq}) > 0 \right)\nonumber\\
&=\left( \mathbb{P}\left(  \abs{q} - \abs{ r g + s q} > 0 \right) \right)^{m(n)} \overset{n\to\infty}{\longrightarrow} 0,
\end{align}
where the convergence follows since $\mathbb{P}\left(  \abs{q} - \abs{ r g + s q} > 0 \right) < 1$ and $m(n)=\alpha n$, where $\alpha > 0$ is independent of $n$. Therefore, we have
\begin{align}
\mathbb{P}\left( E_n \right) &= \mathbb{P}\left( \abs{ f_n(s,r)-\sqrt{ c_d(s,r)} } > \epsilon  \right)  \overset{n\to\infty}{\longrightarrow} 0.
\end{align}
We can conclude that $-c_n(s,r)/ \sqrt{m(n)}$ converges in probability to $\sqrt{c_d(s,r)}$ in this case. 

Based on \textbf{Case 1} and \textbf{Case 2}, the functions $f_n$ and $\widetilde{c}_n$ have the same pointwise limit which is the function $(s,r)\to \sqrt{ c_d(s,r)}$. Moreover, the function $c_d$ is given by
\begin{equation}
c_d(s,r)=\mathbb{E}( \min(\abs{q} - \abs{ r g + s q},0)^2 ).
\end{equation}
It can be checked that the function $c_d$ is as given in \eref{cdref}.

{\bf Step 2}: The first step mainly shows that the functions $f_n$ and $\widetilde{c}_n:(s,r)\to-c_n(s,r)/\sqrt{m(n)}$ have the same pointwise limit. In the second step, we show that they also converge uniformly in probability to the same function. First, assume that the function $f_n$ converges uniformly to some function $f^\ast$ and fix $\epsilon >0$.  This means that 
\begin{equation}\label{eq:assp}
\mathbb{P}\Big( \sup_{(s,r)\in\mathcal{S}} \abs{ f_n(s,r)-f^\ast(s,r) } > \epsilon \Big) \overset{n\to\infty}{\longrightarrow} 0.
\end{equation}
Consider the following three functions $f_{1,n}:(s,r)\to \widetilde{c}_n(s,r)-f^\ast(s,r)$, $f_{2,n}:(s,r)\to \widetilde{c}_n(s,r)-f_n(s,r)$ and $f_{3,n}:(s,r)\to f_n(s,r)-f^\ast(s,r)$. It is clear that
\begin{align}\label{eq:inesup}
&\sup_{(s,r)\in\mathcal{S}} \abs{ f_{1,n}(s,r) } = \sup_{(s,r)\in\mathcal{S}} \abs{ f_{2,n}(s,r)+f_{3,n}(s,r) }\nonumber\\
&\leq \sup_{(s,r)\in\mathcal{S}} \abs{ f_{2,n}(s,r) }+\sup_{(s,r)\in\mathcal{S}} \abs{ f_{3,n}(s,r) }.
\end{align}
Consider the following probability events 
\begin{equation}
\begin{cases}
E_{1,n}(\epsilon)= \lbrace (\vq,\vg):~\sup\limits_{(s,r)\in\mathcal{S}} \abs{ f_{1,n}(s,r) } > \epsilon \rbrace\\
E_{2,n}(\epsilon)= \lbrace (\vq,\vg):~\sup\limits_{(s,r)\in\mathcal{S}} \abs{ f_{2,n}(s,r) } > \epsilon \rbrace\\
E_{3,n}(\epsilon)= \lbrace (\vq,\vg):~ \sup\limits_{(s,r)\in\mathcal{S}} \abs{ f_{3,n}(s,r) } > \epsilon \rbrace.
\end{cases}
\end{equation}
Based on \eref{inesup}, we have
\begin{align}
\mathbb{P}\left( E_{1,n}(\epsilon) \right) &\leq \mathbb{P}\Big( \sup_{(s,r)\in\mathcal{S}} \abs{ f_{2,n}(s,r) }+\hspace{-2mm}\sup_{(s,r)\in\mathcal{S}} \abs{ f_{3,n}(s,r) } > \epsilon \Big)\nonumber\\
&\leq  \mathbb{P}\left( E_{2,n}(\epsilon/2) \right) + \mathbb{P}\left( E_{3,n}(\epsilon/2) \right).\nonumber
\end{align}
By assumption \eref{assp}, we have  
\begin{align}
 \mathbb{P}\Big( \sup_{(s,r)\in\mathcal{S} } \abs{ f_{3,n}(s,r) } > \frac{\epsilon}{2} \Big) \overset{n\to\infty}{\longrightarrow} 0.
\end{align}
Consider the following two functions $g_{1,n}:(s,r)\to \min(\abs{\vq} - \abs{ r \vg + s \vq})$ and $g_{2,n}:(s,r)\to \norm{(\abs{\vq} - \abs{ r \vg + s \vq})\wedge\mathbf{0}}_2$. The function $f_{2,n}$ can be expressed as follows
\begin{equation}
f_{2,n}(s,r)=\begin{cases}
0 & ,\text{if}~ g_{1,n}(s,r) \leq 0,\notag \\
\frac{-g_{1,n}(s,r)-g_{2,n}(s,r)}{\sqrt{m(n)}} & ,\text{otherwise}. \notag
\end{cases}
\end{equation}
Define the set $\mathcal{C}$ as $\mathcal{C}=\lbrace (s,r)\in\mathcal{S}:~ g_{1,n}(s,r)>0\rbrace$ and $P$ as the probability of the  event $\lbrace (\vq,\vg):\sup\limits_{\substack{ (s,r)\in\mathcal{S} }} \abs{ f_{2,n}(s,r) } > \frac{\epsilon}{2} \rbrace$. Then, we have
\begin{align}
P&=\mathbb{P}\Big(   \sup_{(s,r)\in\mathcal{C}} \abs{ \frac{-g_{1,n}(s,r)-g_{2,n}(s,r)}{\sqrt{m(n)}} } > \frac{\epsilon}{2} \Big)\nonumber\\
&=\mathbb{P}\Big( \frac{1}{\sqrt{m(n)}}  \sup_{(s,r)\in\mathcal{C}}  \abs{ g_{1,n}(s,r) } > \frac{\epsilon}{2} \Big)\nonumber\\
&\leq \mathbb{P}\Big(  \frac{\min(\abs{\vq})}{\sqrt{m(n)}}   > \frac{\epsilon}{2} \Big).
\end{align}
Moreover, we have
\begin{align}\notag
\mathbb{P}\Big(  \frac{1}{\sqrt{m(n)}}\min(\abs{\vq})   > \frac{\epsilon}{2} \Big) &= \mathbb{P}\Big(  \abs{q_i}  > \sqrt{{m(n)}}\frac{\epsilon}{2},\forall i \Big).
\end{align}
Given that $\lbrace q_i \rbrace_{i=1}^{m(n)}$ are i.i.d. standard Gaussian random variable, we get
\begin{align}\notag
\mathbb{P}\Big(  \frac{1}{\sqrt{m(n)}}\min(\abs{\vq})   > \frac{\epsilon}{2} \Big) = \Big( 2 \Phi\Big(-\sqrt{{m(n)}}\frac{\epsilon}{2} \Big) \Big)^{m(n)},  
\end{align}
where $\Phi$ denotes the cumulative distribution function of the standard normal random variable. Since $\epsilon > 0$, we obtain the following inequality $2\Phi\left(-\sqrt{{m(n)}}\frac{\epsilon}{2} \right) < 1,~\forall m(n)>0$, which means that
\begin{equation}
\mathbb{P}\Big(  \frac{1}{\sqrt{m(n)}}\min(\abs{\vq})   > \frac{\epsilon}{2} \Big) \overset{n\to\infty}{\longrightarrow} 0.
\end{equation}
Therefore, we obtain
\begin{equation}
\mathbb{P}\Big( \sup_{(s,r)\in\mathcal{S}} \abs{ \widetilde{c}_n(s,r)-f_n(s,r) } > \frac{\epsilon}{2} \Big) \overset{n\to\infty}{\longrightarrow} 0.
\end{equation}
This implies that
\begin{align}
\mathbb{P}\Big( \sup_{(s,r)\in\mathcal{S}} \abs{ \widetilde{c}_n(s,r) -f^\ast(s,r) } > \epsilon \Big) \overset{n\to\infty}{\longrightarrow} 0,
\end{align}
which means that the function $\widetilde{c}_n$ converges uniformly to the function $f^\ast$. Now, if we repeat the above steps with $f_n$ replaced by $\widetilde{c}_n$, we obtain the second direction.

{\bf Step 3}: The final step is the prove Proposition \ref{lem1} by exploiting the properties introduced in the first two steps. Specifically, we prove Proposition \ref{lem1} when the function $\widetilde{c}_n:(s,r)\to-c_n(s,r)/\sqrt{m(n)}$ is replaced by the function $f_n$. Then, the properties introduced in the first two steps are used to show the equivalence. To this end, define the random function $Q_{n}$ on the set $\mathcal{D}=\lbrace (s,r,z)\in\mathbb{R}^3:~(s,r)\in\mathcal{S},~\abs{z}\leq r \rbrace$, as follows
\begin{align}
Q_{n}(s,r,z) &=   - \sqrt{\frac{\norm{\vh}^2_2}{m(n)}-\left(\frac{\widehat{\veta}^T\vh}{\sqrt{m(n)}}\right)^2} \sqrt{r^2-z^2}\nonumber\\
&+\frac{\vh^T \widehat{\veta}}{\sqrt{m(n)}}z+f_n(s,r), 
\label{eq:op1}
\end{align}
where $\widehat{\veta} = \widetilde{\veta}/\norm{\widetilde{\veta}}_2$ and define the deterministic function $Q$ on the set $\mathcal{D}$ as follows
\begin{align}
Q(s,r,z) &= -\frac{1}{\sqrt{\alpha}} \sqrt{r^2-z^2}+\sqrt{c_d(s,r)}.
\end{align}
Fix $(s,r,z)$ in the set $\mathcal{D}$. Given that $\vh$ is independent of $\widehat{\veta}$, we have $\vh^T \widehat{\veta}/\sqrt{m(n)} \xrightarrow[]{n\to\infty} 0$. Furthermore, using the WLLN, we have $\norm{\vh}_2^2/{m(n)} \xrightarrow[]{n\to\infty} \frac{1}{{\alpha}}$. Therefore, based on the first step, the function $Q_n$ converges pointwise to the function $Q$. 

Define $\sup_{\mathcal{D}} f(s,r,z)$ as $\sup_{(s,r,z)\in\mathcal{D}} f(s,r,z)$. Consider the following three functions 
\begin{equation}
\begin{cases}\notag
h_{1,n}:(s,r,z)\to Q_n(s,r,z) - Q(s,r,z)\\
h_{2,n}:(s,r,z)\to \hspace{-1mm} \Big( \sqrt{\frac{\norm{\vh}^2_2}{m(n)} -\hspace{-1mm}\Big(\frac{\widehat{\veta}^T\vh}{\sqrt{m(n)}}\Big)^2}\hspace{-1mm}-\frac{1}{\sqrt{\alpha}} \Big) \sqrt{r^2-z^2}\\
h_{3,n}:(s,r)\to f_n(s,r)-\sqrt{c_d(s,r)}.
\end{cases}
\end{equation}
Therefore, we have
\begin{align}
\sup_{\mathcal{D}} \abs{ h_{1,n}(s,r,z) } &\leq \sup_{\mathcal{D}} \abs{ \frac{\vh^T \widehat{\veta}}{\sqrt{m(n)}} z}+\sup_{\mathcal{D}} \abs{  h_{2,n}(s,r,z) }\nonumber\\
&+\sup_{\mathcal{D}} \abs{  h_{3,n}(s,r)},
\end{align}
which leads to the following inequality
\begin{align}\notag%\label{eq:uconv_sp}
&\mathbb{P}\left( \sup_{\mathcal{D}} \abs{ h_{1,n}(s,r,z) } > \epsilon \right) \leq \mathbb{P}\Big( \sup_{\mathcal{D}} \abs{ \frac{\vh^T \widehat{\veta}}{\sqrt{m(n)}} z} > \frac{\epsilon}{3}\Big)\nonumber\\
&+\mathbb{P}\left( \sup_{\mathcal{D}} \abs{  h_{2,n}(s,r,z) } > \frac{\epsilon}{3}\right)+\mathbb{P}\left( \sup_{\mathcal{D}} \abs{  h_{3,n}(s,r) } > \frac{\epsilon}{3}\right).\notag
\end{align}
Since $\vh^T \widetilde{\veta}/\sqrt{m(n)} \xrightarrow[]{n\to\infty} 0$ and $\sup_{\mathcal{D}} \abs{ z}$ is positive and finite, we obtain $$\mathbb{P}\left( \sup_{\mathcal{D}} \abs{ \frac{\vh^T \widehat{\veta}}{\sqrt{m(n)}} z} > \frac{\epsilon}{3}\right)\overset{n\to\infty}{\longrightarrow} 0.$$
Based on the WLLN, we have $$\sqrt{\frac{\norm{\vh}^2_2}{m(n)}-\left(\frac{\widehat{\veta}^T\vh}{\sqrt{m(n)}}\right)^2} \xrightarrow[]{n\to\infty} \frac{1}{\sqrt{\alpha}}.$$ Given that $\sup_{\mathcal{D}} \abs{\sqrt{r^2-z^2} }$ is positive and finite, we obtain $$\mathbb{P}\left( \sup_{\mathcal{D}} \abs{  h_{2,n}(s,r,z) } > \frac{\epsilon}{3}\right) \xrightarrow[]{n\to\infty} 0.$$
Based on the WLLN, we have 
\begin{align}\label{eq:conv}
f_n(s,r)^2 & \xrightarrow[]{n\to\infty} c_d(s,r),
\end{align}
for any fixed $s$ and $r$ in the set $\mathcal{S}$. Assume that $g$ and $q$ are i.i.d. Gaussian random variables. The function $(s,r) \to \min(\abs{q} - \abs{ r g + s q},0)^2$ is bounded in the set $\mathcal{{S}}$, i.e.
\begin{align}\label{eq:bound}
\hspace{-1.5mm}{\min \left(\abs{q} - \abs{ r g + s q},0 \right)^2 } \leq \Big( \abs{{q}} + {B} \abs{ g } + B \abs{ q } \Big)^2.
\end{align}
Note that the right hand side of \eref{bound} has a finite expectation and the function $(s,r,q,g)\to \min(\abs{q} - \abs{ r g + s q},0)^2$ is continuous in the variables $s$, $r$, $q$ and $g$. Hence, it is a measurable function in the variables $q$ and $g$. Moreover, the set $\mathcal{S}$ is compact. Based on \cite[lemma 2.4]{whitney:94}, we conclude that 
\begin{align}\label{eq:uconv}
f_n(s,r)^2 \overset{u.p}{\longrightarrow} c_d(s,r),
\end{align}
where $\overset{u.p}{\longrightarrow}$ denotes the uniform convergence in probability. Based on the fact that $\abs{ \sqrt{x}-\sqrt{y}} \leq \sqrt{\abs{x-y}}$ for any $x \geq 0$, $y\geq 0$, we have 
\begin{align}\label{eq:uconv2}
\abs{ f_n(s,r)-\sqrt{c_d(s,r)} } \leq \sqrt{ \abs{ f_n(s,r)^2-{c_d(s,r)} }  }.
\end{align}
Therefore, for any fixed $\epsilon > 0$, we obtain the following inequality
\begin{align}\label{eq:uconv3}
&\mathbb{P}\left( \sup_{\mathcal{D}} \abs{ f_n(s,r)-\sqrt{c_d(s,r)} } > \epsilon \right) \leq\nonumber\\
&~~~~~\mathbb{P}\left( \sup_{\mathcal{D}} \abs{ f_n(s,r)^2-{c_d(s,r)} } > \epsilon^2 \right).
\end{align}
Based on \eref{uconv} and \eref{uconv3}, the function $f_n$ converges uniformly in probability to the function $(s,r)\to \sqrt{c_d(s,r)}$ which means that 
$$\mathbb{P}\left( \sup_{\mathcal{D}} \abs{  h_{3,n}(s,r) } > \frac{\epsilon}{3}\right) \xrightarrow[]{n\to\infty} 0.$$ 
Hence, the function $Q_n$ converges uniformly in probability to the function $Q$ in the set $\mathcal{D}$. Note that the set $\mathcal{D}=\lbrace (s,r,z)\in\mathbb{R}^3:~(s,r)\in\mathcal{S},~\abs{z}\leq r \rbrace$ is compact. Moreover, the functions $Q_n$ and $Q$ are continuous on the set $\mathcal{D}$ and the function $(s,z)\to -\eta_1 s -\norm{\widetilde{\veta}}_2 z$ is uniformly continuous on the set $\mathcal{D}$. Also, the set $\lbrace (s,r,z)\in\mathcal{D}:~Q(s,r,z)\leq 0 \rbrace$ has a nonempty interior. Based on the proof of Proposition \ref{compactness} provided in Appendix \ref{prop3}, the sequence $\lbrace \lambda_n \rbrace_{n\in\mathbb{N}}$ should satisfy $\lambda_n \geq \norm{\vx_{\text{init}}}_2 w_n \sqrt{n}/\zeta_n$ which means that there exists a sequence $\lbrace \lambda_n \rbrace_{n\in\mathbb{N}}$ such that $\widetilde{\lambda}_n \overset{n\to\infty}{\longrightarrow} \infty$. Then, using Lemma \ref{gresult},
%the optimal objective value of the optimization problem \eref{ao55_form} where $c_n$ is replaced by $f_n$ converges in probability to the optimal objective value of the deterministic problem \eref{aod_form2}. Now, based on 
${\bf Step~1}$, ${\bf Step~2}$ and given the continuity
of the function $c_n$, we conclude that the optimal objective value of the optimization problem \eref{ao55_form} converges in probability to the optimal objective value of the deterministic problem \eref{aod_form2}.

Based on Lemmas \ref{uniqz}, \ref{lem3} and \ref{lem4}, the optimization problem \eref{aod_form2} have a unique optimal solution. Denote by $(s^\ast,r^\ast,z^\ast)$ the unique optimal solution of the problem \eref{aod_form2} and $V^\ast$ the corresponding optimal objective value. Fix $\delta > 0$ and define the sets $\widetilde{\mathcal{D}}(\delta)$ and $\widehat{\mathcal{D}}(\delta)$ as follows: $\widetilde{\mathcal{D}}(\delta)=\lbrace (s,r,z)\in\mathcal{D}:~( \, \abs{s-s^\ast}^2 +\abs{r-r^\ast}^2 +\abs{z-z^\ast}^2)^{1/2} \leq \delta \rbrace$ and $\widehat{\mathcal{D}}(\delta)=\mathcal{D}\setminus \lbrace (s,r,z)\in\mathcal{D}:~(\,\abs{s-s^\ast}^2+\abs{r-r^\ast}^2+\abs{z-z^\ast}^2)^{1/2} < \delta \rbrace$. Consider the following optimization problems
\begin{align}\notag
&\widetilde{V}_n(\delta)=\underset{ \substack{ (s,r,z)\in\widetilde{\mathcal{D}}(\delta)} }{\min} \Gamma_n(s,r,z),~ \widetilde{V}(\delta)=\underset{ \substack{ (s,r,z)\in\widetilde{\mathcal{D}}(\delta)\\ Q(s,r,z) \leq 0} }{\min} \Gamma(s,r,z),
\end{align}
where $\Gamma_n$ is the cost function of the problem \eref{ao55_form} and $\Gamma$ is the cost function of the problem \eref{aod_form2}. Moreover, consider the following optimization problems
\begin{align}\notag
&\widehat{V}_n(\delta)=\underset{ \substack{ (s,r,z)\in\widehat{\mathcal{D}}(\delta)} }{\min} \Gamma_n(s,r,z),~ \widehat{V}(\delta)=\underset{ \substack{ (s,r,z)\in\widehat{\mathcal{D}}(\delta)\\ Q(s,r,z) \leq 0} }{\min} \Gamma(s,r,z).
\end{align}
Given that the sets $\widetilde{\mathcal{D}}(\delta)$ and $\widehat{\mathcal{D}}(\delta)$ are compact and using the above analysis, we have $\widetilde{V}_n(\delta) \xrightarrow[]{n\to\infty} \widetilde{V}(\delta)$ and $\widehat{V}_n(\delta) \xrightarrow[]{n\to\infty} \widehat{V}(\delta)$. Furthermore, we have $\widetilde{V}(\delta) < \widehat{V}(\delta)$, then, there exists $\gamma > 0$ such that $\widetilde{V}(\delta)+\gamma < \widehat{V}(\delta)$. Since $\widetilde{V}_n(\delta) \xrightarrow[]{n\to\infty} \widetilde{V}(\delta)$ and $\widehat{V}_n(\delta) \xrightarrow[]{n\to\infty} \widehat{V}(\delta)$, we have 
\begin{align}
\mathbb{P}\left(  \abs{\widetilde{V}_n(\delta)-\widetilde{V}(\delta)} \leq \frac{\gamma}{2},~\abs{\widehat{V}_n(\delta)-\widehat{V}(\delta)} \leq \frac{\gamma}{2}\right)\overset{n\to\infty}{\longrightarrow} 1.
\end{align}
%Since $\widetilde{V}_n(\delta) \xrightarrow[]{n\to\infty} \widetilde{V}(\delta)$, there exists $n_1 \in \mathbb{N}$ such that $\forall n \geq n_1$, we have 
%$$\abs{\widetilde{V}_n(\delta)-\widetilde{V}(\delta)} \leq \frac{\gamma}{2},$$ 
%with probability going to one as $n$ goes to infinity. Since $\widehat{V}_n(\delta) \xrightarrow[]{n\to\infty} \widehat{V}(\delta)$, there exists $n_2 \in \mathbb{N}$ such that $\forall n \geq n_2$, we have 
%$$\abs{\widehat{V}_n(\delta)-\widehat{V}(\delta)} \leq \frac{\gamma}{2},$$
%with probability going to one as $n$ goes to infinity. Therefore, for any $n\geq \max(n_1,n_2)$, we have 
Therefore, we have the following convergence result
\begin{align}
\mathbb{P}\left(  \widetilde{V}_n(\delta)-\frac{\gamma}{2} \leq \widetilde{V}(\delta),~\widehat{V}(\delta) \leq \widehat{V}_n(\delta)+\frac{\gamma}{2} \right)\overset{n\to\infty}{\rightarrow} 1.
\end{align}
Since $\widetilde{V}(\delta)+\gamma < \widehat{V}(\delta)$, we conclude that for any for any $\delta > 0$, we have 
$$\mathbb{P}\left( \widetilde{V}_n(\delta) < \widehat{V}_n(\delta) \right)\overset{n\to\infty}{\longrightarrow} 1.$$ Therefore, we conclude that for any $\delta >0$, we have 
$$
\mathbb{P}\left( \mathcal{V}_n^\ast \subseteq \widetilde{\mathcal{D}}(\delta) \right)\overset{n\to\infty}{\longrightarrow} 1,
$$ 
where $\mathcal{V}_n^\ast$ denotes the set of optimal solutions of the optimization problem \eref{ao55_form}. Given the uniqueness of the optimal solution of the problem \eref{aod_form2}, we obtain 
\begin{align}
\mathbb{P}\Big( \sup_{ \vx\in\mathcal{V}_n^\ast } \inf_{ \vy\in\mathcal{V}^\ast } \norm{\vx-\vy}_2 \leq  \delta \Big)\overset{n\to\infty}{\longrightarrow} 1,\forall \delta >0,
\end{align}
where $\mathcal{V}^\ast$ denotes the set of optimal solutions of the optimization problem \eref{aod_form2}. This then gives us the statement of Proposition \ref{lem1}.
%where $s_n$, $r_n$ and $z_n$ are the optimal solutions of the optimization problem \eref{ao55_form}. This means that $s_n \xrightarrow[]{n\to\infty} s^\ast$, $r_n \xrightarrow[]{n\to\infty} r^\ast$ and $z_n \xrightarrow[]{n\to\infty} z^\ast$.

%$$
%\widetilde{V}_n(\delta)-\frac{\gamma}{2}+\gamma \leq \widetilde{V}(\delta)+\gamma < \widehat{V}(\delta) \leq \widehat{V}_n(\delta)+\frac{\gamma}{2},
%$$ 
%with probability going to one as $n$ goes to infinity. This means that for any $\delta > 0$, we have 

\subsection{Proof of Lemma \ref{boundp}}
\label{pboundp}
Fix the oversampling ratio such that $\alpha > 2$. The objective is to show that $\exists~T > 0$ such that $$\mathbb{P} \left\lbrace \norm{\mathcal{K}_n}_2 < T \right\rbrace \overset{n\to\infty}{\longrightarrow} 1,$$ where $T$ is a finite constant independent of $n$. To this end, consider the following optimization problem 
\begin{equation}\label{eq:normax}
\begin{aligned}
T_n&=\underset{{\vx}\in\mathcal{S}_{\tau}}{\max}~~~ \norm{\vx}_2^2\\	
&~~~~~~~~\text{s.t.}~~~~  \abs{\va_i^T \vx} \leq y_i, \text{ for }  1 \le i \le m,
\end{aligned}
\end{equation}
where the set $\mathcal{S}_{\tau}=\lbrace \vx\in\mathbb{R}^n:~\norm{\vx}_2^2\leq \tau \rbrace$, and where $\tau$ is a sufficiently large constant independent of $n$. Based on Section \ref{pr_th1} and Appendix \ref{plem:fb}, one can show that the optimal solution of the optimization problem \eref{normax} satisfies the following
\begin{equation}
\mathbb{P}\left( T_n < T \right)   \overset{n\to\infty}{\longrightarrow} 1,
\end{equation} 
where $T$ is a finite constant independent of $n$ and it satisfies $c^\ast < T \leq \tau$ where $c^\ast$ is the optimal objective value of the following problem
\begin{align}\label{eq:dnorm}
&\max_{s^2+r^2\leq \tau} ~~s^2+r^2 ~~\text{s.t.}~~ \sqrt{c_d(s,r)} \leq r/\sqrt{\alpha},~r \geq 0,  
\end{align}
where $s$ and $r$ are defined as follows $r=\norm{\vxt}_2$ where $\vx^T=[s~~\vxt^T]$. Note that $T$ exists since the set $\mathcal{D}_{\text{feas}}$ is bounded (see Lemma \ref{prop}) which means that for sufficiently large $\tau$, we have  $c^\ast < \tau$.
Then, we have $\mathbb{P}\left( \norm{\mathcal{S}_{\tau} \cap \mathcal{K}_n}_2 < T \right)   \overset{n\to\infty}{\longrightarrow} 1$. Given that $T \leq \tau$, we have $\norm{\mathcal{S}_{T} \cap \mathcal{K}_n}_2 \leq \norm{\mathcal{S}_{\tau} \cap \mathcal{K}_n}_2$ which leads to the following $\mathbb{P}\left( \norm{\mathcal{S}_{T} \cap \mathcal{K}_n}_2 < T \right)   \overset{n\to\infty}{\longrightarrow} 1$.  

Since $\norm{\mathcal{S}_{T} \cap \mathcal{K}_n}_2 \leq \norm{\mathcal{K}_n}_2$, we obtain the following inequality $$\mathbb{P}\left( \norm{\mathcal{S}_{T} \cap \mathcal{K}_n}_2 \geq T \right) \leq \mathbb{P}\left( \norm{\mathcal{K}_n}_2 \geq T \right).$$
Note that the feasibility set $\mathcal{K}_n$ is convex. Assume that $\norm{\mathcal{K}_n}_2 \geq T$, then, there exists $\vx \in \mathcal{K}_n$ such that $\norm{\vx}_2 \geq T$. Given that the all zero vector is in the convex set $\mathcal{K}_n$, $\frac{T}{\norm{\vx}_2} \vx \in \mathcal{K}_n$. Therefore, we have $\frac{T}{\norm{\vx}_2} \vx \in \mathcal{S}_{T} \cap \mathcal{K}_n$ which means that $\norm{\mathcal{S}_{T} \cap \mathcal{K}_n}_2 \geq T$. Therefore, $$\mathbb{P}\left( \norm{\mathcal{S}_{T} \cap \mathcal{K}_n}_2 \geq T \right) \geq \mathbb{P}\left( \norm{\mathcal{K}_n}_2 \geq T \right),$$
which means that 
$$\mathbb{P}\left( \norm{\mathcal{S}_{T} \cap \mathcal{K}_n}_2 \geq T \right) = \mathbb{P}\left( \norm{\mathcal{K}_n}_2 \geq T \right).$$
Thus, we obtain $\mathbb{P}\left( \norm{ \mathcal{K}_n}_2 \geq T \right)\overset{n\to\infty}{\longrightarrow} 0$. This completes the proof of Lemma \ref{boundp}.
%%%%%%%%%%%%%%%%%%%%%%%%%%%%%%%%%%%%%%%%%%%%%%%%%%
%%%%%%%%%%%%%%%%%%%%%%%%%%%%%%%%%%%%%%%%%%%%%%%%%%
%%%%%%%%%%%%%%%%%%%%%%%%%%%%%%%%%%%%%%%%%%%%%%%%%%
%%%%%%%%%%%%%%%%%%%%%%%%%%%%%%%%%%%%%%%%%%%%%%%%%%
%%%%%%%%%%%%%%%%%%%%%%%%%%%%%%%%%%%%%%%%%%%%%%%%%%
%%%%%%%%%%%%%%%%%%%%%%%%%%%%%%%%%%%%%%%%%%%%%%%%%%
%%%%%%%%%%%%%%%%%%%%%%%%%%%%%%%%%%%%%%%%%%%%%%%%%%
%%%%%%%%%%%%%%%%%%%%%%%%%%%%%%%%%%%%%%%%%%%%%%%%%%

\section{Deterministic Analysis}

%%%%%%%%%%%%%%%%%%%%%%%%%%%%%%%%%%%%%%%%%%%%%%%%%%
%%%%%%%%%%%%%%%%%%%%%%%%%%%%%%%%%%%%%%%%%%%%%%%%%%
%%%%%%%%%%%%%%%%%%%%%%%%%%%%%%%%%%%%%%%%%%%%%%%%%%
%%%%%%%%%%%%%%%%%%%%%%%%%%%%%%%%%%%%%%%%%%%%%%%%%%
\subsection{Proof of Lemma \ref{prop}}
\label{proppf}
%\begin{itemize}
%\item[(P.1)] {\bf Convexity}
%\end{itemize}
%\noindent :\\
\noindent (P.1) {\bf Convexity}: The deterministic set ${\mathcal D}_{\rm feas}$ is given by
\begin{align}
{\mathcal D}_{\rm feas}=\lbrace (s,r)\in\mathbb{R}^2:~ r\geq 0,~\alpha~c_d(s,r) \leq r^2 \rbrace,
\end{align}
where $\alpha > 1$.
The objective is to show the convexity of the set ${\mathcal D}_{\rm feas}$ with respect to the variables $s$ and $r$. Note that the set ${\mathcal D}_{\rm feas}$ is nonempty. Let $\lambda \in [0,~1]$, $(s_1,r_1) \in {\mathcal D}_{\rm feas}$ and $(s_2,r_2) \in {\mathcal D}_{\rm feas}$. We know that $c_d(s,r)=\mathbb{E}\lbrace \min(\abs{q} - \abs{ r g + s q},0)^2 \rbrace$ for any $s\in\mathbb{R}$ and $r \geq 0$, where $g$ and $q$ are two independent standard normal random variables. Let $g\in \mathbb{R}$, $q \in \mathbb{R}$ and consider the function $c_{\text{rd}}$ defined as follows
$$
c_{\text{rd}}(s,r)=\min(\abs{q} - \abs{ r g + s q},0)^2,
$$
and the function $\widetilde{c}_{\text{rd}}$ defined as follows
$$
\widetilde{c}_{\text{rd}}(s,r)=\min(\abs{q} - \abs{ r g + s q},0).
$$
%First, we show that the function $c_{\text{rd}}$ is convex. 
Since the function $s\to \abs{q} - \abs{ r g + s q}$ is concave in the variables $(s,r)$, we have
\begin{align}\notag
&\abs{q} - \abs{ (\lambda r_1+(1-\lambda) r_2) g + (\lambda s_1+(1-\lambda) s_2) q} \geq \nonumber\\
&\lambda \left\lbrace \abs{q} - \abs{ r_1 g + s_1 q} \right\rbrace + (1-\lambda) \left\lbrace \abs{q} - \abs{ r_2 g + s_2 q} \right\rbrace,\nonumber
\end{align}
which implies that
\begin{align}
&\widetilde{c}_{\text{rd}}(\lambda s_1+(1-\lambda) s_2,\lambda r_1+(1-\lambda) r_2) \geq  \nonumber\\
& \min( \lambda \left\lbrace \abs{q} - \abs{ r_1 g + s_1 q} \right\rbrace + (1-\lambda) \left\lbrace \abs{q} - \abs{ r_2 g + s_2 q} \right\rbrace,0)\nonumber\\
&\geq \lambda \widetilde{c}_{\text{rd}}(s_1,r_1)+(1-\lambda) \widetilde{c}_{\text{rd}}(s_2,r_2).\nonumber
\end{align}
Therefore, we obtain the following inequality
\begin{align}
&c_{\text{rd}}(\lambda s_1+(1-\lambda) s_2,\lambda r_1+(1-\lambda) r_2) \leq \nonumber\\ 
&( \lambda \widetilde{c}_{\text{rd}}(s_1,r_1)+(1-\lambda) \widetilde{c}_{\text{rd}}(s_2,r_2) )^2,
\end{align}
which implies that
\begin{align}
&c_{{d}}(\lambda s_1+(1-\lambda) s_2,\lambda r_1+(1-\lambda) r_2) \leq \nonumber\\ 
&\mathbb{E}\left[ ( \lambda \widetilde{c}_{\text{rd}}(s_1,r_1)+(1-\lambda) \widetilde{c}_{\text{rd}}(s_2,r_2) )^2 \right].
\end{align}
\vsp
\textbf{\emph{Property 1}}: Consider two random variables $X$ and $Y$. %Based on the Cauchy Schwarz inequality, we have
%\begin{align}
%\mathbb{E}(XY) \leq \sqrt{\mathbb{E}(X^2)\times \mathbb{E}(Y^2)}.
%\end{align}
We have the following inequality
\begin{equation}
\mathbb{E}\left[ \left( a X + b Y \right)^2 \right] \leq \left( a \sqrt{\mathbb{E}(X^2)}+ b \sqrt{\mathbb{E}(Y^2)} \right)^2,
\end{equation}
for any $a \geq 0$ and $b \geq 0$.
\vsp

\noindent Based on the above property, we get
\begin{align}
&\alpha c_{{d}}(\lambda s_1+(1-\lambda) s_2,\lambda r_1+(1-\lambda) r_2) \leq \nonumber\\ 
& \left[ \lambda \sqrt{\alpha c_{{d}}(s_1,r_1)} + (1-\lambda) \sqrt{\alpha c_{{d}}(s_2,r_2)}  \right]^2.
\end{align}
Given that $(s_1,r_1) \in {\mathcal D}_{\rm feas}$ and $(s_2,r_2) \in {\mathcal D}_{\rm feas}$, we have $\sqrt{\alpha c_{{d}}(s_1,r_1)} \leq r_1$ and $\sqrt{\alpha c_{{d}}(s_1,r_1)} \leq r_2$. This leads to the following inequality
\begin{align}
&\alpha c_{{d}}(\lambda s_1+(1-\lambda) s_2,\lambda r_1+(1-\lambda) r_2) \leq \nonumber\\ 
& \left[ \lambda r_1 + (1-\lambda) r_2 \right]^2.
\end{align}
Therefore, we have $(\lambda s_1+(1-\lambda) s_2,\lambda r_1+(1-\lambda) r_2) \in {\mathcal D}_{\rm feas}$ which implies the convexity of the set ${\mathcal D}_{\rm feas}$.  This completes the proof of property P.1 in Lemma \ref{prop}.

%\begin{itemize}
%\item[(P.2)] {\bf Boundedness}:
%\end{itemize}
\noindent (P.2) {\bf Boundedness}: Given the expression of the function $c_d$ in \eref{cdref}, note that $[-1,~1]\times \lbrace 0 \rbrace \subset {\mathcal D}_{\rm feas}$. The set $\mathcal{D}_{\text{feas}}$ is symmetric in the variable $s$. Hence, it is sufficient to prove that the set $\mathcal{D}_{\text{feas}}^{+}=\lbrace (s,r)\in\mathbb{R}^2:~ s\geq 0,~r\geq 0,~c_d(s,r) \leq r^2/\alpha \rbrace$ is compact for $\alpha>1$ and there exists $z >0$ such that the set $\mathcal{D}_{\text{feas}}^{+}\subseteq [0,~1]\times [0,~z]$ for $\alpha \geq 2$. 
Note that the function $c_d$ is continuous and we have
\begin{align}\label{eq:inflim}
\begin{cases}
\lim\limits_{r\to +\infty} \frac{c_d(s,r)}{r^2}=1,~\forall s\in\mathbb{R}\\
\lim\limits_{s\to+\infty} c_d(s,r)= +\infty,~\forall r > 0.
\end{cases}
\end{align}
Next, we assume that $s \geq 0$ and we distinguish between two different cases:\\
\textbf{Case 1}: If $r=0$, the function $c_d$ can be rewritten as follows
\begin{equation}\label{eq:eq1lem9}
c_d(s,r)=\begin{cases}
(1-s)^2 & \text{if}~~ s > 1\\
0 & \text{otherwise}.
\end{cases}
\end{equation}
Equation \eref{eq1lem9} shows that
\begin{equation}\label{eq:eq2lem9}
\hspace{-4mm}\begin{cases}
\lbrace (s,r)\in\mathbb{R}^2:~ r=0,~0\leq s \leq 1 \rbrace \subset \mathcal{D}_{\text{feas}}^{+}\\
(s,r)\in \lbrace (s,r)\in\mathbb{R}^2: r=0, s > 1 \rbrace \Rightarrow (s,r)\notin \mathcal{D}_{\text{feas}}^{+}.
\end{cases}
\end{equation}
\textbf{ Case 2}: In what follows, we assume that $r > 0$. Note that 
\begin{equation}
\frac{c_d(1,r)}{r^2}=\frac{1}{\pi}\Big[ \pi+\frac{4}{r^2} \atan\left( {r}/{2} \right)-\atan\left( {2}/{r} \right) - \frac{2}{r} \Big]. \nonumber
\end{equation}
Using a Taylor expansion of the function $r \to \atan(r/2)$ in the neighborhood of zero, one can show that 
\begin{equation}\label{eq:eq3lem9}
\frac{c_d(1,r)}{r^2} \xrightarrow[]{r\to0} \frac{1}{2}.
\end{equation}
Next, we show the following property:
\vsp

\noindent \textbf{\emph{Property 2}}: The function $(s,r) \to \frac{c_d(s,r)}{r^2}$ is strictly increasing in the variable $r > 0$, for fixed $s=1$. Moreover, it is  strictly increasing in the variable $s \geq 1$, for any fixed $r > 0$.
\vsp 

First, consider the function $f_1:r\to \frac{c_d(1,r)}{r^2}$ defined for $r>0$ and for fixed $s=1$. Note that the function $f_1$ is differentiable with first derivative given by 
\begin{align}
f^{\prime}_1(r)&=\frac{1}{\pi}\Big[ -\frac{8}{r^3}\atan\left( {r}/{2} \right)+\frac{8}{r^2} \frac{1}{4+r^2}+\frac{2}{r^2}+\frac{2}{r^2+4}  \Big]\nonumber\\
&=\frac{1}{\pi r^3} \Big[ -8\atan\left( {r}/{2} \right)+4r  \Big].
\end{align}
Note that the function $h:r \to -8\atan\left( \frac{r}{2} \right)+4r$ is differentiable with derivative $h^{\prime}(r)=-16/(r^2+4)+4$ which is strictly positive for any $r>0$. Hence, the function $h$ is strictly increasing and $\lim_{r\to 0} h(r)=0$. This means that $h(r) > 0$, for any $r >0$ which implies that the derivative of the function $f_1$ is strictly positive for any $r > 0$. Therefore, the function $f_1$ is strictly increasing in the variable $r > 0$, for fixed $s=1$. 

Second, consider the function $f_r:s\to \frac{c_d(s,r)}{r^2}$ defined for $s \geq 1$ and for fixed $r>0$. Note that the function $f_r$ is differentiable with first derivative given by
\begin{align}
f^{\prime}_r(s)&=\frac{2s}{r^2}+\frac{2}{\pi r^2}\Big[ (1-s)\atan \frac{1-s}{r} -(1+s)\atan \frac{1+s}{r}  \Big]\nonumber\\
&=\frac{2s}{r^2}+\frac{2}{\pi r^2}\left[ g(1-s)-g(-1-s) \right],
\end{align}
where the function $g:x\to x \atan(x/r)$. The function $g$ is twice differentiable with first derivative given by $$g^{\prime}(x)=\atan(x/r)+rx/(r^2+x^2).$$ Then, we can see that the function $g$ is nonincreasing in the variable $x \leq 0$. This means that $g(1-s)-g(-1-s) \leq 0$, for any $s \geq 1$. Furthermore, the second derivative of the function $g$ can be expressed as $g^{\prime\prime}(x)=2r^3/(r^2+x^2)^2$ which means that the function $g^{\prime}$ is strictly increasing in the variable $x \geq 0$. Hence, the function $s\to g(1-s)-g(-1-s)$ is strictly decreasing in the variable $s \geq 1$ and we also have the following
\begin{equation}
\begin{cases}
\lim\limits_{s\to 1} g(1-s)-g(-1-s)= -2 \atan(2/r)\\
\lim\limits_{s\to \infty} g(1-s)-g(-1-s)= -\pi.
\end{cases}\nonumber
\end{equation}
Therefore, we have $g(1-s)-g(-1-s) > -\pi$ which means that $f^{\prime}_r(s) > 0$ for $s > 1$. Thus, the function $f_r$ is strictly increasing in the variable $s\geq 1$, for any fixed $r > 0$. This completes the proof of the above property.

Based on the continuity of the function $c_d$, \eref{inflim} and {\bf Case 1}, the set $\mathcal{D}_{\text{feas}}^{+}$ is compact for any fixed $\alpha > 1$.

Next, assume that $\alpha \geq 2$. Based on property 2 and \eref{eq3lem9}, we have the following 
$$
(s,r)\in\lbrace (s,r)\in\mathbb{R}^2:~ r>0,~ s > 1 \rbrace \Rightarrow (s,r)\notin\mathcal{D}_{\text{feas}}^{+},
$$ 
for any $\alpha \geq 2$. Combining this result with \eref{eq2lem9}, we conclude that the set $\mathcal{D}_{\text{feas}}$ is a subset of $[-1,~1]\times \mathbb{R}$. Given the continuity of $c_d$, \eref{eq2lem9} and \eref{inflim}, there exists $z >0$ such that $\mathcal{D}_{\text{feas}}\subseteq [-1,~1]\times [0,~z]$. Given \eref{eq3lem9} and property 2, the intersection between the set $\lbrace (s,r)\in\mathbb{R}^2:~s=1 \rbrace$ and $\mathcal{D}_{\text{feas}}$ is only the target signal vector, i.e. $\lbrace (1,0) \rbrace$. This completes the proof of property P.2 in Lemma \ref{prop}.

%\begin{itemize}
%\item[(P.3)] {\bf Boundedness II}:
%\end{itemize}
%Fix $1 < \alpha < 2$. Similar to the previous part, we have 
%\begin{equation}\label{eq:rzeroo}
%\begin{cases}
%\lbrace (s,r)\in\mathbb{R}^2~\mathrm{s.t.}~ r=0,~0\leq s \leq 1 \rbrace \subset \mathcal{D}_{\text{feas}}^{+}\\
%\lbrace (s,r)\in\mathbb{R}^2~\mathrm{s.t.}~ r=0,~ s > 1 \rbrace \not\subset \mathcal{D}_{\text{feas}}^{+}.
%\end{cases}
%\end{equation}
%when $r=0$. Note that the function $c_d$ is continuous and $\lim_{s\to\infty} c_d(s,r)= +\infty$ for any $r\neq 0$. Based on property 2, \eref{inflim}, \eref{rzeroo} and \eref{eq3lem9}, we conclude that there exists $a >0$ and $b>1$ such that $\mathcal{D}_{\text{feas}} \subseteq [-b,~b]\times\mathbb{R}$ and $(b,a)\in\mathcal{D}_{\text{feas}}$. Given the continuity of $c_d$ and \eref{inflim}, there exists $z >0$ such that $\mathcal{D}_{\text{feas}}\subseteq [-b,~b]\times [0,~z]$.

%\begin{itemize}
%\item[(P.3)]  {\bf Boundary}:
%\end{itemize}
\noindent (P.3) {\bf Boundary}: Assume that the oversampling ratio $\alpha \geq 2$. Then, there exists $z >0$ such that $\mathcal{D}_{\text{feas}}\subseteq [-1,~1]\times [0,~z]$. For fixed $s \in[-1,~1]$, the maximum radius of the set $\mathcal{D}_{\text{feas}}$ is the solution of the following problem
\begin{align}\label{eq:maxrad}
{r}^{\ast}(s)&=\max_{(s,r)\in\mathcal{S}} ~~r ~~\text{s.t.}~~ {c_d(s,r)} \leq r^2/{\alpha}.  
\end{align} 
Note that the function $r \to c_d(s,r)/r^2$ is continuous for $r >0$ and  $c_d(s,0)=0$. First, if ${r}^{\ast}(s)=0$, then the result is true. Now, assume that the solution ${r}^{\ast}(s)>0$ and suppose by contradiction that the solution of the above problem satisfies
$$
{c_d(s,{r}^{\ast}(s))} < {{r}^{\ast}(s)}^2/{\alpha}.
$$
First, note that the function $r \to c_d(s,r)/r^2$ is continuous in $(0,~\infty)$. Based on Lemma \ref{prop}, the set $\mathcal{D}_{\text{feas}}$ is convex which implies that the feasibility set of the problem \eref{maxrad} is convex. Based on the proof of P.2, we have
\begin{align}
\lim\limits_{r\to +\infty} \frac{c_d(s,r)}{r^2}=1,~\forall s\in\mathbb{R}.
\end{align}
Now, since $\alpha \geq 2$ and based on the above properties, there exists $\widehat{r}(s) > {r}^{\ast}(s)$ such that 
$$
{c_d(s,\widehat{r}(s))} \leq {\widehat{r}(s)}^2/{\alpha},
$$
which leads to a contradiction.
Now, assume that $s=0$ and note that 
$$
\lim\limits_{r\to 0} \frac{c_d(0,r)}{r^2}=0.
$$
Based on the above properties, we conclude that ${r}^{\ast}(0) > 0$.
This completes the proof of property P.3 in Lemma \ref{prop}.

%\begin{itemize}
%\item[(P.4)] {\bf Slope}:
%\end{itemize}
\noindent (P.4) {\bf Slope}: Based on the previous point, the boundary of the set $\mathcal{D}_{\text{feas}}$ is the set of $(s,r)\in\mathbb{R}^2$ such that $r\geq 0$ and
%\begin{align}
$c_d(s,r)={r^2}/{\alpha}$.
%\end{align}
Next, we study the slope of this boundary curve at $s=1$. To this end, assume that $s(\delta)=1-\delta$ and $r=r(\delta)$ such that $(s(\delta),r_\delta)\in \text{bd}(\mathcal{D}_{\text{feas}})$. Then, $r(\delta)$ must satisfy the following equality
\begin{align}
\frac{r_\delta^2}{\alpha}=c_d(s(\delta),r_\delta)&=\frac{1}{\pi}\Big[   (\delta^2+r_\delta^2) \atan \frac{r_\delta}{\delta}\nonumber\\
&+((2-\delta)^2+r_\delta^2) \atan \frac{r_\delta}{2-\delta}-2r_\delta\Big].
\end{align}
We write $r_\delta=c_1 \delta +c_2 \delta^2+o(\delta^2)$. Then, we get the following
\begin{align}
&c_1^2 \delta ^2 + o(\delta^2)=\frac{\alpha}{\pi} \Big[ (1+c_1^2) \delta^2 \atan(c_1) \nonumber\\
&+\left( 4-4\delta+(1+c_1^2) \delta^2 \right) \left(  c_1\delta/2 +(c_1/4+c_2/2)\delta^2\right) \nonumber\\
&-2(c_1\delta+c_2\delta^2)  \Big].
\end{align}
Dividing by $\delta^2$ and letting $\delta$ go to zero, the slope of the boundary curve should satisfy the following equality
\begin{align}
c_1^2=\frac{\alpha}{\pi} \left[ (1+c_1^2)\atan(c_1)-c_1 \right].
\end{align}
This completes the proof of property P.4 in Lemma \ref{prop}.

%\begin{itemize}
%\item[(P.5)] {\bf Perturbation}:
%\end{itemize}
\noindent (P.5) {\bf Perturbation}: Assume that the oversampling ratio $\alpha > 1$. Note that the set ${\mathcal D}_{\rm feas}^{\epsilon}$ is given by
\begin{align}
{\mathcal D}_{\rm feas}^{\epsilon}=\lbrace (s,r)\in\mathbb{R}^2:~ r\geq 0,~\alpha~c_d(s,r) \leq r^2+\epsilon \rbrace,
\end{align}
and ${\mathcal D}_{\rm feas}^{0}={\mathcal D}_{\rm feas}$. Based on the proof of P.2, the set ${\mathcal D}_{\rm feas}^{\epsilon}$ is compact for any $\epsilon >0$. Also, note that $\mathcal{D}_{\text{feas}}^{\epsilon_1} \subseteq \mathcal{D}_{\text{feas}}^{\epsilon_2}$ for any $0<\epsilon_1 \leq \epsilon_2$. Now, let $\lbrace \epsilon_k \rbrace_{k\in\mathbb{N}}$ be a decreasing sequence of positive numbers such that $\lim_{k\to \infty} \epsilon_k=0$. Based on \cite[Exercise 4.3]{var_ana}, $\lim_{k \to \infty} {\mathcal D}^{\epsilon_k}_{\rm feas}$ exists and we have
\begin{align}
\lim_{k \to \infty} {\mathcal D}^{\epsilon_k}_{\rm feas}=\bigcap_{k\geq 0} {\mathcal D}^{\epsilon_k}_{\rm feas}.
\end{align}
Next, the objective is to show that $\lim_{k \to \infty} {\mathcal D}^{\epsilon_k}_{\rm feas}={\mathcal D}_{\rm feas}$. Note that  ${\mathcal D}_{\rm feas} \subseteq {\mathcal D}^{\epsilon_k}_{\rm feas}$ for any $k \geq 0$. It follows that ${\mathcal D}_{\rm feas} \subseteq \lim_{k \to \infty} {\mathcal D}^{\epsilon_k}_{\rm feas}$. Assume that $(s,r) \in \lim_{k \to \infty} {\mathcal D}^{\epsilon_k}_{\rm feas}$. Then, there exists a decreasing sequence of positive numbers $\lbrace \epsilon_k \rbrace_{k\in\mathbb{N}}$ and a sequence $\lbrace (s_k,r_k) \rbrace_{k\in\mathbb{N}}$ such that $(s_k,r_k) \in  {\mathcal D}^{\epsilon_k}_{\rm feas}$ for any $k \geq 0$ and
\begin{align}
\epsilon_k  \overset{k\to\infty}{\longrightarrow} 0~\text{and}~(s_k,r_k) \overset{k\to\infty}{\longrightarrow} (s,r).
\end{align}
Hence, we have
\begin{align}
r_k\geq 0,~\alpha~c_d(s_k,r_k) \leq r_k^2+\epsilon_k.
\end{align}
Note that the function $c_d$ is continuous. Letting $k$ go to $\infty$, we obtain $(s,r) \in {\mathcal D}_{\rm feas}$ which implies that $  \lim_{k \to \infty} {\mathcal D}^\epsilon_{\rm feas} \subseteq {\mathcal D}_{\rm feas}$. This completes the proof of property P.5 in Lemma \ref{prop}.

\subsection{Proof of Proposition \ref{gos}}
\label{gosp}
%Based on Theorem \ref{lem:fb}, the projected feasibility set ${\mathcal S}_{\rm feas}$ of the PhaseMax method is a subset of the following deterministic set 
%\begin{align}
%{\mathcal D}_{\rm feas}=\lbrace (s,r)\in\mathbb{R}^2:~r\geq 0,~ \alpha~c_d(s,r) \leq r^2 \rbrace,
%\end{align}
%in the high dimensional limit. Therefore, to show Proposition \ref{gos}, 
We start by providing a sufficient condition under which the intersection between the unit circle $\mathcal{C}_{\text{unit}}=\lbrace (s,r)\in\mathbb{R}^2:~s^2+r^2=1\rbrace$ and the deterministic set ${\mathcal D}_{\rm feas}$ defined as 
\begin{align}
{\mathcal D}_{\rm feas}=\lbrace (s,r)\in\mathbb{R}^2:~r\geq 0,~ \alpha~c_d(s,r) \leq r^2 \rbrace,
\end{align}
is only the target signal vectors $(1,0)$ and $(-1,0)$, i.e. $\mathcal{C}_{\text{unit}} \cap {\mathcal D}_{\rm feas} = \lbrace (1,0),(-1,0) \rbrace$. This is equivalent to showing that the function $f$ defined in the set $(-1,~1)$ as follows
\begin{align}
f(s)&:=(1-s^2)-\alpha c_d(s,\sqrt{1-s^2}),%\nonumber\\
%&=1-s^2-\frac{2\alpha}{\pi} \Big[ (1-s) \atan\left(  \sqrt{\frac{1+s}{1-s}} \right), \nonumber\\
%&+ (1+s) \atan\left(  \sqrt{\frac{1-s}{1+s}} \right) - \sqrt{1-s^2} \Big],
\end{align}
is strictly negative. Given the symmetry of the function $f$, it is sufficient to show that $f$ is strictly negative in the set $[0,~1)$. The function $f$ is twice differentiable in the set $(0,~1)$. It can be checked that the derivative of the function $f$ has at most two zeros at $s=0$ and $\widehat{s}\in[0,~1)$. Note that $f(0)=1-\alpha\left( 1-\frac{2}{\pi}\right)$, $\lim_{s\to1} f(s)=0$, $f^{\prime}(0)=0$ and $f^{\prime}(1)=-2+\alpha$. This implies that the function $f$ is strictly negative in the set $(-1,~1)$ if and only if 
\begin{align}\label{eq:ncgos}
f(0)=1-\alpha\Big( 1- {2}/{\pi} \Big) < 0.
\end{align} 
This means that $\mathcal{C}_{\text{unit}} \cap {\mathcal D}_{\rm feas} = \lbrace (1,0),(-1,0) \rbrace$ for any $\alpha > \pi/(\pi-2)$. Next, assume that the oversampling ratio satisfies $\alpha > \pi/(\pi-2)$. But, from Lemma \ref{prop}, selecting the oversampling ratio  in this way ensures that for any $\epsilon>0$ there exists $\delta > 0$ such that
\begin{align}
\mathcal{D}_{\text{feas}}^{\epsilon} \cap \mathcal{C}_{\text{unit}} \subseteq \mathcal{B}_1^\delta \cup \mathcal{B}_2^\delta,
\end{align}
where $\mathcal{B}_1^\delta$ and $\mathcal{B}_2^\delta$ are two balls of radius $\delta$ and center the target signal vectors $\vxi$ and $-\vxi$, respectively. Define the sequence $\lbrace \delta_{\epsilon_k} \rbrace_{k\in\mathbb{N}}$ as follows
\begin{align}\label{eq:depslon3}
\delta_{\epsilon_k} :=\inf \lbrace \delta>0:~ \mathcal{D}_{\text{feas}}^{\epsilon_k} \cap \mathcal{C}_{\text{unit}} \subseteq \mathcal{B}_1^\delta \cup \mathcal{B}_2^\delta \rbrace.
\end{align}
where $\lbrace \epsilon_k \rbrace_{k\in\mathbb{N}}$ is a decreasing sequence of positive numbers with $\lim_{k\to\infty} \epsilon_k=0$.
Based on the proof of Proposition \ref{suffcondd}, it can be checked that $\lim_{k\to\infty} \delta_{\epsilon_k}=0$. Based on Theorem \ref{lem:fb}, it holds that
$$
\lim_{n\rightarrow\infty} \Pro\Big( \mathcal{S}_{\text{feas}}  \subseteq \mathcal{D}_{\text{feas}}^{\epsilon}  \Big) = 1, \forall \epsilon>0,
$$
which means that for any $\epsilon > 0$, we have
$$
\lim_{n\rightarrow\infty} \Pro\Big( \mathcal{S}_{\text{feas}} \cap    \mathcal{C}_{\text{unit}}  \subseteq \mathcal{D}_{\text{feas}}^{\epsilon} \cap   \mathcal{C}_{\text{unit}} \Big) = 1.
$$
Therefore, we conclude that for any decreasing sequence of positive numbers $\lbrace \epsilon_k \rbrace_{k\in\mathbb{N}}$ with $\lim_{k\to\infty} \epsilon_k=0$, there exists a sequence of positive numbers $\lbrace \delta_k \rbrace_{k\in\mathbb{N}}$ such that
\begin{align}
\lim_{n\rightarrow\infty} \Pro\Big( \min\lbrace \norm{\widehat{\vx}-\vxi}_2, \norm{\widehat{\vx}+\vxi}_2 \rbrace \leq \delta_k,~ \forall \widehat{\vx}\in \mathcal{S}_{\text{lamp}} \Big) = 1.
\end{align}
for any $k \geq 0$, where $\lim_{k \to \infty} \delta_{\epsilon_k}=0$ and $\mathcal{S}_{\text{lamp}}$ denotes the set of optimal solutions of the PhaseLamp problem \eref{qb_form}.
%Given the following properties $\mathcal{D}_{\text{feas}}^{\epsilon_1} \subseteq \mathcal{D}_{\text{feas}}^{\epsilon_2}$ for $\epsilon_1 \leq \epsilon_2$ and $\mathcal{D}_{\text{feas}}^{0}=\mathcal{D}_{\text{feas}}$, 
This implies that the set $\mathcal{S}_{\text{lamp}}$ converges to the set $\lbrace \vxi,-\vxi \rbrace$ in the sense that $\sup_{\widehat{\vx}\in \mathcal{S}_{\text{lamp}}} (\min\lbrace \norm{\widehat{\vx}-\vxi}_2, \norm{\widehat{\vx}+\vxi}_2 \rbrace)$ converges to zero in probability. 
This completes the proof of Proposition \ref{gos}. 
%\begin{figure}[h!]
%    \centering
%    \includegraphics[width=0.325\textwidth]{figs/gos.eps}
%    \caption{Illustration of the sufficient condition given in \eref{ncgos} for two different values of $\alpha$.}
%    \label{fig:gos}
%\end{figure}
%\fref{gos} provides a simulation example to illustrate the necessary and sufficient condition given in \eref{ncgos}. It is clear that when $\alpha$ does not satisfy \eref{ncgos} ($\alpha=2.4$), the target signal vector is not the global optimal solution. However, when $\alpha$ satisfies \eref{ncgos} ($\alpha=2.9$), the target signal vector is indeed the unique global optimal solution.
%%%%%%%%%%%%%%%%%%%%%%%%%%%%%%%%%%%%%%%%%%%%%%%%%%
%%%%%%%%%%%%%%%%%%%%%%%%%%%%%%%%%%%%%%%%%%%%%%%%%%
%%%%%%%%%%%%%%%%%%%%%%%%%%%%%%%%%%%%%%%%%%%%%%%%%%
%%%%%%%%%%%%%%%%%%%%%%%%%%%%%%%%%%%%%%%%%%%%%%%%%%
%%%%%%%%%%%%%%%%%%%%%%%%%%%%%%%%%%%%%%%%%%%%%%%%%%
%%%%%%%%%%%%%%%%%%%%%%%%%%%%%%%%%%%%%%%%%%%%%%%%%%
%%%%%%%%%%%%%%%%%%%%%%%%%%%%%%%%%%%%%%%%%%%%%%%%%%
%%%%%%%%%%%%%%%%%%%%%%%%%%%%%%%%%%%%%%%%%%%%%%%%%%
\subsection{Proof of Lemma \ref{uniqz}}
\label{puniqz}
The optimization problem \eref{aod_form2} is equivalent to the following problem
\begin{align}\label{eq:aod_form22}
&\underset{ \substack{ (s,r) \in \mathcal{S}\\ \abs{z} \leq r } }{\max}~\eta_1 s + \norm{\widetilde{\veta}}_2 z ~~\text{s.t.}~~ {\alpha~c_d(s,r)} \leq r^2- z^2. 
\end{align}
It can be noticed that the optimization problem \eref{aod_form22} is feasible only when $r^2-\alpha~c_d(s,r) \geq 0$. Based on \eref{cdexp}, $c_d$ is a nonnegative function. This means that the optimization problem \eref{aod_form22} admits a unique solution in the variable $z$ which is given by 
$$
z^\ast=\sqrt{r^2-\alpha~c_d(s,r)}.
$$
Therefore, the optimization problem \eref{aod_form22} is equivalent to the following problem
\begin{align}\label{eq:aod_form23}
&\underset{ \substack{ (s,r) \in \mathcal{S} } }{\max}~\eta_1 s + \norm{\widetilde{\veta}}_2 \sqrt{r^2-\alpha~c_d(s,r)} ~~\text{s.t.}~~ {\alpha~c_d(s,r)} \leq r^2.  
\end{align}
This completes the proof of Lemma \ref{uniqz}.
\subsection{Proof of Lemma \ref{lem3}}
\label{pr_lem3}
Fix $s$ such that $|s|\leq 1$, fix the oversampling ratio such that $\alpha >2$ and consider the following change of variable $r=\sqrt{t}$. Then, the optimization problem \eref{aod2_form} can be equivalently formulated as follows
\begin{align}\label{eq:tra_opt}
&\underset{  t\geq 0}{\max}~ t-\alpha\,c_d(s,\sqrt{t}), 
\end{align}
where the function $c_d$ is defined in \eref{cdref}.
Due to the symmetry of the cost function, we assume that $s$ is in the set $[0,~1]$. Consider the following function $f_s:t\to -\alpha\,c_d(s,\sqrt{t})$ defined for $t \geq 0$. 
The function $f_s$ can be written for $t>0$ as follows 
\begin{align}
&f_s(t)
%-\frac{\alpha}{\pi} \Bigg[  ( (1-s)^2+t )\left({\pi}/{2}-\text{atan}\left( {(1-s)}/{\sqrt{t}} \right) \right)\nonumber\\
%&+( (1+s)^2+t )\left( {\pi}/{2}-\text{atan}\left( {(1+s)}/{\sqrt{t}} \right) \right)-2\sqrt{t} \Bigg]\nonumber\\
=-\alpha t-\alpha(1+s^2)+\frac{\alpha}{\pi} ( (1-s)^2+t ) \text{atan}\left( \frac{(1-s)}{\sqrt{t}} \right) \nonumber\\
&+ \frac{\alpha}{\pi} ( (1+s)^2+t ) \text{atan}\left( \frac{(1+s)}{\sqrt{t}} \right)+\frac{2\alpha\sqrt{t}}{\pi}.\nonumber
\end{align}
Note that the function $f_s$ is twice differentiable for $t >0$.
The first derivative of the function $f_s$ for $t > 0$ can be expressed as follows
\begin{align}
f_s^{\prime}(t)&=-\alpha+\frac{\alpha}{\pi} \text{atan}\left( {(1-s)}/{\sqrt{t}} \right) +\frac{\alpha}{\pi} \text{atan}\left( {(1+s)}/{\sqrt{t}} \right).
\end{align}
Moreover, the second derivative of the function $f_s$ can be expressed as follows
\begin{align}
f_s^{\prime\prime}(t)&=-\frac{\alpha (1-s)}{2\pi\sqrt{t}[t+(1-s)^2]}-\frac{\alpha(1+s)}{2\pi\sqrt{t}[t+(1+s)^2]}.
\end{align}
By performing a Taylor expansion of $f_s$ at $t=0$, it can be checked that $f_s^{\prime}(0)=0$ when $0\leq s < 1$ and $f_s^{\prime}(0)=-\alpha/2$ when $s=1$.
 Moreover, by performing a Taylor expansion of $f_s^{\prime}$ at $t=0$, we get
%\begin{equation}\notag
%\begin{cases}
%f_s(t)=-\frac{\alpha}{\pi}\left[ \frac{4t^{3/2}}{3(1-s^2)}%+o(t^{3/2}) \right] & \text{if}~~0\leq s < 1 \\
%f_s(t)=-\alpha\left[ \frac{t}{2}+\frac{1}{3\pi} t^{3/2}+o(t^{3/2})%\right] & \text{if}~~s = 1.
%\end{cases}
%\end{equation}
\begin{equation}\notag
\begin{cases}
f_s^{\prime}(t)=-\frac{\alpha}{\pi} \frac{2 \sqrt{t}}{1-s^2}+o(t) & \text{if}~~0\leq s < 1 \\
f_s^{\prime}(t)=-\frac{\alpha}{2} \left[ 1+\frac{\sqrt{t}}{\pi} \right] +o(t) & \text{if}~~s = 1.
\end{cases}
\end{equation}
Note that the function  $f_s^{\prime\prime}$ satisfies $f_s^{\prime\prime}(t) \leq 0$ for any $t \geq 0$ and any fixed $0 \leq s \leq 1$. Therefore, the function $f_s$ is concave for any fixed $0 \leq s \leq 1$. This means that the cost function $t\to t-\alpha\,c_d(s,\sqrt{t})$  is concave for any fixed $0 \leq s \leq 1$.
Now, we distinguish between two different cases:\\
\textbf{Case 1}: Assume that $s=1$. Note that the cost function in \eref{tra_opt} evaluated at $0$ is $0$ and the derivative of the cost function in \eref{tra_opt} at $t=0$ is $1-\frac{\alpha}{2}<0$. Given the concavity of the cost function in \eref{tra_opt}, we conclude that $\sqrt{t^\ast}=0$ is the unique global optimal solution of the optimization problem \eref{tra_opt}. \\
\textbf{Case 2}:  Assume that $0 \leq s <1$. Let $t > 0$, setting the derivative of the cost function in \eref{tra_opt} to zero, we get
\begin{align}\label{eq:stat_t}
1-\alpha+\frac{\alpha}{\pi} \text{atan}\left( \frac{1-s}{\sqrt{t}} \right) +\frac{\alpha}{\pi} \text{atan}\left( \frac{1+s}{\sqrt{t}} \right)=0.
\end{align}
Note that the solutions of the above equation represent the global optimal solutions of the problem \eref{tra_opt}.
Since $\alpha > 2$, equation \eref{stat_t} can be rewritten as follows
\begin{align}
 \text{atan}\left( \frac{2\sqrt{t}}{t+s^2-1} \right)+\pi =\frac{\pi(\alpha-1)}{\alpha},
\end{align}
which leads to the following unique solution of equation \eref{stat_t}
\begin{equation}
\sqrt{t^\ast}=\sqrt{{1}/{\tan\left( {\pi}/{\alpha} \right)^2}+1-s^2}-{1}/{\tan\left( {\pi}/{\alpha} \right)}.
\end{equation}
Therefore, for any $0 \leq s <1$, the optimal solution of the optimization problem \eref{aod2_form} can be expressed as in \eref{vstar}.

%\begin{equation}\label{eq:os}
%r_\alpha(s)=\sqrt{{1}/{\tan\left( {\pi}/{\alpha} \right)^2}+1-s^2}-{1}/{\tan\left( {\pi}/{\alpha} \right)}.
%\end{equation}
Based on the above two cases, we conclude that the optimization problem \eref{aod2_form} admits a unique optimal solution as given in \eref{vstar} for any fixed $\abs{s}\leq 1$ and any $\alpha > 2$. %This then gives us the statement of Lemma \ref{lem3}.

%%%%%%%%%%%%%%%%%%%%%%%%%%%%%%%%%%%%%%%%%%%%%%%%%%
%%%%%%%%%%%%%%%%%%%%%%%%%%%%%%%%%%%%%%%%%%%%%%%%%%
%%%%%%%%%%%%%%%%%%%%%%%%%%%%%%%%%%%%%%%%%%%%%%%%%%
%%%%%%%%%%%%%%%%%%%%%%%%%%%%%%%%%%%%%%%%%%%%%%%%%%
\subsection{Proof of Lemma \ref{lem4}}
\label{pr_lem4}
\noindent Assume that $\alpha > 2$. Based on the assumption that the initial guess vector $\vx_{\text{init}}$ has a positive cosine with the target signal vector $\vxi$, the optimization problem \eref{aod4_form0} can be equivalently formulated as follows
\begin{align}\label{eq:aod4_form_m}
&{\max_{ 0 \leq s\leq 1}}~\eta_1 s + \norm{\widetilde{\veta}}_2 \sqrt{(r_\alpha(s))^2-\alpha~c_d(r_\alpha(s),s)  }.
%&~~~\text{s.t.}~~ \alpha~c_d(s,r^\ast(s)) \leq (r^\ast(s))^2 \nonumber 
\end{align}
The main objective is to show that the cost function of the optimization problem \eqref{eq:aod4_form_m} is strictly concave. To this end, define the function $\widehat{f}:s\to \sqrt{f(s)}$ where the function $f$ is defined as follows
\begin{align}\label{eq:f_fun}
f(s)&:={(r_\alpha(s))^2-\alpha~c_d(r_\alpha(s),s)  } &\nonumber\\
&= \Big[-1-s^2+\frac{2\alpha}{\pi}\sqrt{t^\ast}\nonumber\\
&~~~~~~+\frac{2\alpha s}{\pi}\atan  \frac{s}{\sqrt{t^\ast} + 1/\tan(\pi/\alpha)}  \Big],
\end{align}
where $0 \leq s\leq 1$ and $$\sqrt{t^\ast}=\sqrt{{1}/{\tan\left( {\pi}/{\alpha} \right)^2}+1-s^2}-{1}/{\tan\left( {\pi}/{\alpha} \right)}.$$
Note that the function $\widehat{f}$ is strictly positive when $0 \leq s < 1$. For $0 \leq s < 1$, the derivative of the function $\widehat{f}$ is given by
\begin{equation}
\widehat{f}^{\prime}(s)=\frac{ -2s + \frac{2\alpha}{\pi} \atan  \frac{s}{\sqrt{t^\ast} + 1/\tan(\pi/\alpha)}   }{ 2 \widehat{f}(s) }.
\end{equation}
Now, consider the following function
\begin{equation}
g(s)=-2s + \frac{2\alpha}{\pi} \atan  \frac{s}{\sqrt{t^\ast} + 1/\tan(\pi/\alpha)} .
\end{equation}
For $0 \leq s < 1$, the second derivative of the function $\widehat{f}$ can be expressed as follows
\begin{equation}
\widehat{f}^{\prime\prime}(s)=\frac{2 g^{\prime}(s)\widehat{f}(s)^2-g(s)^2}{4\widehat{f}(s)^2\widehat{f}(s)}.
\end{equation}
Hence, the sign of $\widehat{f}^{\prime\prime}(s)$ only depends on the sign of the function $h:s\to 2 g^{\prime}(s)\widehat{f}(s)^2-g(s)^2$ defined in $[0,~1)$.
%and is given by
%\begin{align}
%h(s)&=4-\frac{8\alpha}{\pi} (\sqrt{t^\ast} - 1/\tan(\pi/\alpha))-\frac{4\alpha}{\pi} \frac{1}{\sqrt{t^\ast}}-\frac{4\alpha}{\pi} \frac{s^2}{\sqrt{t^\ast}}\nonumber\\
%&+2\left( \frac{2\alpha}{\pi} \right)^2 \left( \frac{\sqrt{t^\ast} - 1/\tan(\pi/\alpha)}{\sqrt{t^\ast}} \right)\nonumber\\
%&+2\left( \frac{2\alpha}{\pi} \right)^2 \frac{s}{\sqrt{t^\ast}} \atan\left( \frac{s}{\sqrt{t^\ast} - 1/\tan(\pi/\alpha)} \right)\nonumber\\
%&-2\left( \frac{2\alpha}{\pi} \right)^2  \atan\left( \frac{s}{\sqrt{t^\ast} - 1/\tan(\pi/\alpha)} \right)^2
%\end{align}
It can be checked that the derivative of the function $h$ can be expressed as follows
\begin{equation}
h^{\prime}(s)= \frac{4\alpha s\widehat{f}(s)^2}{\pi\sqrt{\frac{1}{\tan\left( {\pi}/{\alpha} \right)^2}+1-s^2} \Big(\frac{1}{\tan\left( {\pi}/{\alpha} \right)^2}+1-s^2 \Big)},
\end{equation}
which means that $h^{\prime}(s) > 0$ for any $0 < s < 1$. Therefore, the function $h$ is strictly increasing in the set $(0,~1)$ and we also have $\lim_{s\to 1}h(s)=0$. Therefore, $h$ is a strictly negative function in the set $[0,~1)$ which means that the function $\widehat{f}$ is a strictly concave function in $[0,~1)$. Furthermore, the function $\widehat{f}$ is continuous in $[0,~1]$, $\widehat{f}(s) > 0$ in $0 \leq s < 1$ and $\widehat{f}(1)=0$. This implies that the function $\widehat{f}$ is strictly concave in $[0,~1]$. Since the cost function of the optimization problem \eqref{eq:aod4_form0} is the positive weighted sum of a linear function and the function $\widehat{f}$, it is a strictly concave function in $[0,~1]$. 
%%%%%%%%%%%%%%%%%%%%%%%%%%%%%%%%%%%%%%%%%%%%%%%%%%
%%%%%%%%%%%%%%%%%%%%%%%%%%%%%%%%%%%%%%%%%%%%%%%%%%
%%%%%%%%%%%%%%%%%%%%%%%%%%%%%%%%%%%%%%%%%%%%%%%%%%
%%%%%%%%%%%%%%%%%%%%%%%%%%%%%%%%%%%%%%%%%%%%%%%%%%
\subsection{Proof of Lemma \ref{Lamp_01}}
\label{pLamp_01}
To prove Lemma \ref{Lamp_01}, it suffices to show that the function $f$ defined as follows
\begin{align}
&f(s):= \pi c_d(s,\sqrt{s-s^2})-\frac{\pi}{\alpha} (s-s^2),%\nonumber\\
%&=(1-s)\atan \sqrt{\frac{s}{1-s}} +(1+3s)\atan \sqrt{\frac{s(1-s)}{1+s}} \nonumber\\
%& -2\sqrt{s(1-s)} -\frac{\pi}{\alpha} s(1-s),
\end{align} 
has a unique zero in the set $(0,~1)$ and there exists $s \in(0,~1)$ such that $f(s)>0$. Note that the function $f$ is four times differentiable. Moreover, the first derivative of the function $f$ can be expressed as follows
\begin{align}
f^{\prime}(s)&=-\atan\left( \sqrt{\frac{s}{1-s}} \right)+3\atan\left( \sqrt{\frac{s(1-s)}{1+s}} \right)\nonumber\\
&  -\frac{\pi}{\alpha} (1-2s).
\end{align}
Additionally, the second derivative of the function $f$ can be expressed as follows
\begin{align}
f^{\prime\prime}(s)&= \frac{1-6s}{\sqrt{s(1-s)}(1+3s)}+\frac{2\pi}{\alpha}.
\end{align}
It can be checked that the function $f^{\prime\prime}$ is strictly decreasing in the set $(0,~1)$ by computing the third derivative of the function $f$. Furthermore, we have $\lim_{s\to 0 } f^{\prime\prime}(s)=+\infty$ and $\lim_{s\to 1 } f^{\prime\prime}(s)=-\infty$ which means that the function $f^{\prime\prime}$ is strictly decreasing and has exactly one zero at $\widetilde{s}$ in the set $(0,~1)$. Note that $\lim_{s\to 0 } f(s)=\lim_{s\to 1 } f(s)=0$, $\lim_{s\to 0 } f^{\prime}(s)=-\pi/\alpha$, $\lim_{s\to 1 } f^{\prime}(s)=\pi/\alpha-\pi/2$ and the function $f$ is continuous. Hence, the function $f^{\prime}$ has exactly two zeros $\widehat{s}_1$ and $\widehat{s}_2$ in the set $(0,~1)$ and it is strictly increasing in the set $(0,~\widetilde{s})$ then strictly decreasing $(\widetilde{s},~1)$. Therefore, the function $f$ is strictly decreasing in the set $(0,~\widehat{s}_1)$, strictly increasing in the set $(\widehat{s}_1,~\widehat{s}_2)$ and strictly decreasing in the set $(\widehat{s}_2,~1)$. Since $\lim_{s\to 0 } f(s)=\lim_{s\to 1 } f(s)=0$, the function $f$ has exactly one zero in the set $(0,~1)$ and there exists $s \in(0,~1)$ such that $f(s)>0$. This implies that the boundary of the set $\mathcal{D}_{\text{opt}}$ is not a subset of the feasibility set $\mathcal{D}_{\text{feas}}$. This completes the proof of Lemma \ref{Lamp_01}.
%%%%%%%%%%%%%%%%%%%%%%%%%%%%%%%%%%%%%%%%%%%%%%%%%%
%%%%%%%%%%%%%%%%%%%%%%%%%%%%%%%%%%%%%%%%%%%%%%%%%%
%%%%%%%%%%%%%%%%%%%%%%%%%%%%%%%%%%%%%%%%%%%%%%%%%%
%%%%%%%%%%%%%%%%%%%%%%%%%%%%%%%%%%%%%%%%%%%%%%%%%%
%%%%%%%%%%%%%%%%%%%%%%%%%%%%%%%%%%%%%%%%%%%%%%%%%%
%%%%%%%%%%%%%%%%%%%%%%%%%%%%%%%%%%%%%%%%%%%%%%%%%%
%%%%%%%%%%%%%%%%%%%%%%%%%%%%%%%%%%%%%%%%%%%%%%%%%%
%%%%%%%%%%%%%%%%%%%%%%%%%%%%%%%%%%%%%%%%%%%%%%%%%%
\subsection{Proof of Lemma \ref{Lamp_02}}
\label{pLamp_02}
To prove Lemma \ref{Lamp_02}, it suffices to show that the function $f$ defined as follows 
\begin{align}
f(s)=& (1+{c}^2) \atan(c) +\left( \frac{(1+s)^2}{(1-s)^2}+{c}^2 \right) \times\nonumber\\&\atan\left( \frac{c(1-s)}{1+s} \right)
-\frac{2c}{1-s} -\frac{\pi {c}^2}{\alpha},
\end{align}
has at most one zero in the set $(0,~1)$ and $f(0)>0$ when a zero exists, where $c>0$. Note that the function $f$ is twice differentiable in the set $(0,~1)$ where the first derivative is given by
\begin{align}
f^{\prime}(s)=& \frac{s(s+1)}{(1-s)^3} \left[ \atan\left( \frac{c(1-s)}{1+s} \right) - \frac{c(1-s)}{1+s} \right].
\end{align}
It can be noticed that the function $f^{\prime}$ is strictly negative in the set $(0,~1)$ which means that the function $f$ is strictly decreasing in the set $(0,~1)$. Additionally, we have
\begin{equation}
\begin{cases}
\lim\limits_{s\to 0}f(s)=2(1+c^2)\atan(c)-2c-\frac{\pi c^2}{\alpha}=f_0(c)\\
\lim\limits_{s\to 1}f(s)=(1+c^2)\atan(c)-c-\frac{\pi c^2}{\alpha}=f_1(c).
\end{cases}
\end{equation}
Hence, if $c$ is in the set $\lbrace x\in\mathbb{R}:~x>0,~f_0(x) \leq 0 \rbrace$ or $c$ is in the set $\lbrace x\in\mathbb{R}:~x>0,~f_1(x) \geq 0 \rbrace$, then, the function $f$ has no zeros in the set $(0,~1)$. If $c$ is in the set $\mathcal{C}=\lbrace x\in\mathbb{R}:~x>0,~f_0(x) > 0,~f_1(x) < 0 \rbrace$, then, the function $f$ has exactly one zero in the set $(0,~1)$. Note that it can be checked that the set $\mathcal{C}$ is nonempty. Assume that $c\in\mathcal{C}$, then, $f(0) > 0$. This completes the proof of Lemma \ref{Lamp_02}.

%%%%%%%%%%%%%%%%%%%%%%%%%%%%%%%%%%%%%%%%%%%%%%%%%%
%%%%%%%%%%%%%%%%%%%%%%%%%%%%%%%%%%%%%%%%%%%%%%%%%%
%%%%%%%%%%%%%%%%%%%%%%%%%%%%%%%%%%%%%%%%%%%%%%%%%%
%%%%%%%%%%%%%%%%%%%%%%%%%%%%%%%%%%%%%%%%%%%%%%%%%%
%%%%%%%%%%%%%%%%%%%%%%%%%%%%%%%%%%%%%%%%%%%%%%%%%%
%%%%%%%%%%%%%%%%%%%%%%%%%%%%%%%%%%%%%%%%%%%%%%%%%%
%%%%%%%%%%%%%%%%%%%%%%%%%%%%%%%%%%%%%%%%%%%%%%%%%%
%%%%%%%%%%%%%%%%%%%%%%%%%%%%%%%%%%%%%%%%%%%%%%%%%%
\subsection{Proof of Lemma \ref{lemm9}}
\label{plemm9}
Select the input cosine similarity of PhaseMax $\rho_{\text{init}}$ such that $\rho_{\text{init}} > \widehat{\rho}_{s}(\alpha)$. We use induction to prove Lemma \ref{lemm9}. We know that the result is true for $k=0$. Now, assume that the optimal solution of PhaseLamp at iteration $k$ satisfies $\rho^k_{\text{init}} > \widehat{\rho}_{s}(\alpha)$, where $k \geq 0$. Next, we show that the optimal solution of PhaseLamp at iteration $k+1$ also satisfies the inequality. Given that the function $x \to \frac{x}{\sqrt{1-x^2}}$ is strictly increasing in $[0,1]$, we have 
$$
\frac{\rho^k_{\text{init}}}{\sqrt{1-{\rho^k_{\text{init}}}^2}} > \frac{\widehat{\rho}_{s}(\alpha)}{\sqrt{1-\widehat{\rho}_{s}(\alpha)^2}}.
$$
Based on Section \ref{suffcond}, the optimal solution of PhaseLamp at iteration $k+1$ belongs to the following set
\begin{align}\notag
{\mathcal{D}}_{\text{fp}}(\rho_{\text{init}})=\left\lbrace (s,r)\in\mathbb{R}^2:~r\geq 0,~\chi_k (1-s) \leq r \right\rbrace,
\end{align}
where $\chi_k={\rho^k_{\text{init}}}/{\sqrt{1-{\rho^k_{\text{init}}}^2}}$. Based on Lemma \ref{prop}, the set $\mathcal{D}_{\text{feas}}$ is a subset of $[-1,~1]\times [0,~\infty]$ for $\alpha>2$. This means that $\widehat{\vx}_{k+1}$ satisfies
$$
\frac{\widehat{\rho}_{s}(\alpha)}{\sqrt{1-\widehat{\rho}_{s}(\alpha)^2}} (1-s) \leq r,
$$
where $\widehat{\vx}_{k+1}=[s~~\widetilde{\vx}_{k+1}^T]^T$ and $r=\norm{\widetilde{\vx}_{k+1}}_2$.  

If $r=0$, it is obvious that $s$ should be $1$. In this case, $\rho_{\text{init}}^{k+1}=1$ which means that the optimal solution of PhaseLamp at iteration $k+1$ also satisfies the inequality.

Now, assume that $r\neq 0$. Therefore, we have
$$
\widehat{\rho}_{s}(\alpha) \leq \frac{r}{\sqrt{r^2+(1-s)^2}}.
$$
Using a geometric argument, one can show that 
$$
\frac{r}{\sqrt{r^2+(1-s)^2}} < \frac{s}{\sqrt{r^2+s^2}},
$$
which means that
$$
\widehat{\rho}_{s}(\alpha) < \frac{s}{\sqrt{r^2+s^2}}=\rho_{\text{init}}^{k+1}.
$$
This means that the optimal solution of PhaseLamp at iteration ${k+1}$ satisfies the statement of Lemma \ref{lemm9}. This completes the proof of Lemma \ref{lemm9}.
%%%%%%%%%%%%%%%%%%%%%%%%%%%%%%%%%%%%%%%%%%%%%%%%%%
%%%%%%%%%%%%%%%%%%%%%%%%%%%%%%%%%%%%%%%%%%%%%%%%%%
%%%%%%%%%%%%%%%%%%%%%%%%%%%%%%%%%%%%%%%%%%%%%%%%%%
%%%%%%%%%%%%%%%%%%%%%%%%%%%%%%%%%%%%%%%%%%%%%%%%%%
%%%%%%%%%%%%%%%%%%%%%%%%%%%%%%%%%%%%%%%%%%%%%%%%%%
%%%%%%%%%%%%%%%%%%%%%%%%%%%%%%%%%%%%%%%%%%%%%%%%%%
%%%%%%%%%%%%%%%%%%%%%%%%%%%%%%%%%%%%%%%%%%%%%%%%%%
%%%%%%%%%%%%%%%%%%%%%%%%%%%%%%%%%%%%%%%%%%%%%%%%%%

\section{Additional Technical Lemmas}

%%%%%%%%%%%%%%%%%%%%%%%%%%%%%%%%%%%%%%%%%%%%%%%%%%
%%%%%%%%%%%%%%%%%%%%%%%%%%%%%%%%%%%%%%%%%%%%%%%%%%
%%%%%%%%%%%%%%%%%%%%%%%%%%%%%%%%%%%%%%%%%%%%%%%%%%
%%%%%%%%%%%%%%%%%%%%%%%%%%%%%%%%%%%%%%%%%%%%%%%%%%
\begin{lemma}\label{gresult}
Consider the following optimization problem
\begin{align}\label{eq:gret}
\underset{ \vx\in\mathcal{S}_{\vx}}{\min}~f(\vx)+\lambda_n \rho(c_n(\vx)),
\end{align}
where $\lbrace \lambda_n \rbrace_{n\in\mathbb{N}}$ is a sequence of positive finite numbers, $\mathcal{S}_{\vx}$ is a compact set in $\mathbb{R}^d$ for fixed $d$, the function $f$ is uniformly continuous, the function $c_n$ is continuous on the set $\mathcal{S}_{\vx}$ and $c_n$ is a random function. Assume that $c_n$ converges uniformly in probability to a deterministic function $c$. Moreover, assume that the function $c$ is continuous and the set $\lbrace \vx \in \mathcal{S}_{\vx}:~c(\vx) \leq 0 \rbrace$ has a nonempty interior. 
Then, if $\lambda_n \overset{n\to\infty}{\longrightarrow} \infty$, the optimal objective of the problem \eref{gret} converges in probability to  the optimal objective value of the following problem
\begin{align}\label{eq:gret2}
&\underset{ \vx\in\mathcal{S}_{\vx}}{\min}~f(\vx) \\
&~~\text{s.t.}~ c(\vx) \leq 0.\nonumber
\end{align}
\end{lemma}
\begin{IEEEproof}
Consider a sequence of positive numbers $\lbrace \lambda_n \rbrace_{n\in\mathbb{N}}$ such that $\lambda_n \overset{n\to\infty}{\longrightarrow} \infty$. We further assume that the set $\mathcal{R}=\lbrace \vx \in \mathcal{S}_{\vx}:~c(\vx) \leq 0 \rbrace$ has a nonempty interior which means that there exists $\widehat{\delta} > 0$ such that for any $0<\delta \leq \widehat{\delta}$, the set $\mathcal{R}_{-\delta}=\lbrace \vx \in \mathcal{S}_{\vx}:~c(\vx) \leq -\delta \rbrace$ is not empty. Fix $\epsilon>0$, the function $f$ is uniformly continuous on the feasibility set $\mathcal{S}_{\vx}$, then, there exists $\gamma > 0$ such that $\forall~\vx, \vy \in \mathcal{S}_{\vx}$ with $\norm{\vx-\vy}\leq \gamma$ implies that $\abs{f(\vx)-f(\vy)}\leq \epsilon$. Now, consider the following set $\mathcal{R}_{+\delta}=\lbrace \vx \in \mathcal{S}_{\vx}:~c(\vx) \leq \delta \rbrace$ which is nonempty. 

Fix $\delta > 0$. First, assume that the set $\widetilde{\mathcal{S}}_{\vx}=\mathcal{S}_{\vx} \setminus \mathcal{R}_{+\delta}$ is not empty. Note that $c(\vx) > 0$ for any $\vx \in \widetilde{\mathcal{S}}_{\vx}=\mathcal{S}_{\vx} \setminus \mathcal{R}_{+\delta}$ and the set $\mathcal{S}_{\vx}$ is compact. This implies that there exists $\zeta_{+} > 0$ such that $$\inf_{\vx \in \widetilde{\mathcal{S}}_{\vx}} c(\vx) = \zeta_{+} > 0. $$
Given that the random function $c_n$ converges uniformly to the function $c$ and the fact that 
$$
\abs{\inf_{\vx \in \widetilde{\mathcal{S}}_{\vx}} c_n(\vx)-\inf_{\vx \in \widetilde{\mathcal{S}}_{\vx}} c(\vx)} \leq \sup_{\vx \in \widetilde{\mathcal{S}}_{\vx}} \abs{ c_n(\vx)-c(\vx) },
$$
we get the following convergence result
\begin{align}\label{eq:ineq01}
\mathbb{P}\Big(  \inf_{\vx \in \widetilde{\mathcal{S}}_{\vx}} c_n(\vx) > \zeta_{+}/2\Big) \overset{n\to\infty}{\longrightarrow} 1.
\end{align}
Second, assume that the set $\widetilde{\mathcal{S}}_{\vx}=\mathcal{S}_{\vx} \setminus \mathcal{R}_{+\delta}$ is empty. Note that the convergence result in \eref{ineq01} still hold. Hence, we conclude that for any $\delta >0$, we have
\begin{align}\label{eq:ineq1}
\mathbb{P}\Big(  \inf_{\vx \in \widetilde{\mathcal{S}}_{\vx}} c_n(\vx) > \zeta_{+}/2\Big) \overset{n\to\infty}{\longrightarrow} 1.
\end{align}
Similarly, there exists  $\zeta_{-} < 0$ such that $\max_{\vx \in \widehat{\mathcal{S}}_{\vx}} c(\vx) = \zeta_{-} < 0$ and 
\begin{align}\label{eq:ineq2}
\mathbb{P}\Big(  \max_{\vx \in \widehat{\mathcal{S}}_{\vx}} c_n(\vx) < \zeta_{-}/2\Big) \overset{n\to\infty}{\longrightarrow} 1,
\end{align}
for any $0<\delta \leq \widehat{\delta}$, where the set $\widehat{\mathcal{S}}_{\vx}= \mathcal{R}_{-\delta}$. Based on \eref{ineq1} and given that the sequence $\lbrace \lambda_n \rbrace_{n\in\mathbb{N}}$ diverges, we have the following equality for any $\delta >0$
\begin{align}
\min_{\vx\in\mathcal{S}_{\vx}} f(\vx)+\lambda_n \rho(c_n(\vx))=\min_{\vx\in\mathcal{R}_{+\delta}} f(\vx)+\lambda_n \rho(c_n(\vx)),
\end{align}
with probability going to one as $n$ goes to infinity. This implies that for any $\delta >0$
\begin{align}
\min_{\vx\in\mathcal{S}_{\vx}} f(\vx)+\lambda_n \rho(c_n(\vx))\geq \min_{\vx\in\mathcal{R}_{+\delta}} f(\vx), \notag
\end{align}
with probability going to one as $n$ goes to infinity. Given the uniform continuity of the function $f$ and the continuity of $c$ on the compact set $\mathcal{S}_{\vx}$, there exists $\delta(\epsilon) >0$ such that
$$
\min_{\vx\in\mathcal{R}_{+\delta(\epsilon)}} f(\vx)\geq \min_{\vx\in\mathcal{R}} f(\vx)-\epsilon,
$$ 
which means that
\begin{align}\label{eq:ineq3}
\min_{\vx\in\mathcal{S}_{\vx}} f(\vx)+\lambda_n \rho(c_n(\vx))\geq \min_{\vx\in\mathcal{R}} f(\vx)-\epsilon, 
\end{align}
with probability going to one as $n$ goes to infinity.
%where the last inequality follows due the uniform continuity of the function $f$ on the set $\mathcal{S}_{\vx}$. 
Based on \eref{ineq2}, we have the following equality for any $0<\delta \leq \widehat{\delta}$
\begin{align}
\min_{\vx\in\mathcal{R}_{-\delta}} f(\vx)+\lambda_n \rho(c_n(\vx))=\min_{\vx\in\mathcal{R}_{-\delta}} f(\vx),
\end{align}
with probability going to one as $n$ goes to infinity. This implies that for any $0<\delta \leq \widehat{\delta}$
\begin{align}\label{eq:ineq4}
\min_{\vx\in\mathcal{S}_{\vx}} f(\vx)+\lambda_n \rho(c_n(\vx))&\leq\min_{\vx\in\mathcal{R}_{-\delta}} f(\vx),
\end{align}
with probability going to one as $n$ goes to infinity. Given the uniform continuity of the function $f$ and the continuity of $c$ on the compact set $\mathcal{S}_{\vx}$, there exists $0< \delta(\epsilon) \leq \widehat{\delta}$ such that
$$
\min_{\vx\in\mathcal{R}_{-\delta(\epsilon)}} f(\vx) \leq \min_{\vx\in\mathcal{R}} f(\vx) + \epsilon,
$$ 
which means that
\begin{align}\label{eq:ineq5}
\min_{\vx\in\mathcal{S}_{\vx}} f(\vx)+\lambda_n \rho(c_n(\vx))&\leq \min_{\vx\in\mathcal{R}} f(\vx) + \epsilon,
\end{align}
with probability going to one as $n$ goes to infinity.

%where the last inequality follows due the uniform continuity of the function $f$ on the set $\mathcal{S}_{\vx}$.  
\noindent Now, based on \eref{ineq3} and \eref{ineq5}, we conclude that 
\begin{align}\label{eq:convgres}
\abs{ \min_{\vx\in\mathcal{S}_{\vx}} f(\vx)+\lambda_n \rho(c_n(\vx))- \min_{\vx\in\mathcal{R}} f(\vx)} \leq \epsilon,
\end{align}
with probability going to one as $n$ goes to infinity. Since $\epsilon$ is an arbitrary positive scalar, \eref{convgres} implies that $\min_{\vx\in\mathcal{S}_{\vx}} f(\vx)+\lambda_n \rho(c_n(\vx))$ converges in probability to  $\min_{\vx\in\mathcal{R}} f(\vx)$.
\end{IEEEproof}
%Next, we use Lemma \ref{gresult} to show that analyzing the optimization problem \eref{ao4_form} is equivalent to analyzing a deterministic problem in the high dimensional limit
%%%%%%%%%%%%%%%%%%%%%%%%%%%%%%%%%%%%%%%%%%%%%%%%%%
%%%%%%%%%%%%%%%%%%%%%%%%%%%%%%%%%%%%%%%%%%%%%%%%%%
%%%%%%%%%%%%%%%%%%%%%%%%%%%%%%%%%%%%%%%%%%%%%%%%%%
%%%%%%%%%%%%%%%%%%%%%%%%%%%%%%%%%%%%%%%%%%%%%%%%%%


%\bibliographystyle{apacite}
\bibliographystyle{IEEEtran}
\bibliography{reference,refs}

\begin{thebibliography}{10}
\providecommand{\url}[1]{#1}
\csname url@samestyle\endcsname
\providecommand{\newblock}{\relax}
\providecommand{\bibinfo}[2]{#2}
\providecommand{\BIBentrySTDinterwordspacing}{\spaceskip=0pt\relax}
\providecommand{\BIBentryALTinterwordstretchfactor}{4}
\providecommand{\BIBentryALTinterwordspacing}{\spaceskip=\fontdimen2\font plus
\BIBentryALTinterwordstretchfactor\fontdimen3\font minus
  \fontdimen4\font\relax}
\providecommand{\BIBforeignlanguage}[2]{{%
\expandafter\ifx\csname l@#1\endcsname\relax
\typeout{** WARNING: IEEEtran.bst: No hyphenation pattern has been}%
\typeout{** loaded for the language `#1'. Using the pattern for}%
\typeout{** the default language instead.}%
\else
\language=\csname l@#1\endcsname
\fi
#2}}
\providecommand{\BIBdecl}{\relax}
\BIBdecl

\bibitem{Lampouss17}
O.~Dhifallah, C.~Thrampoulidis, and Y.~M. Lu, ``Phase retrieval via linear
  programming: Fundamental limits and algorithmic improvements,'' in \emph{2017
  55th Annual Allerton Conference on Communication, Control, and Computing
  (Allerton)}, Oct 2017, pp. 1071--1077.

\bibitem{Candes:2013xy}
E.~J. Candes, T.~Strohmer, and V.~Voroninski, ``Phaselift: {Exact} and stable
  signal recovery from magnitude measurements via convex programming,''
  \emph{Communications on Pure and Applied Mathematics}, vol.~66, no.~8, pp.
  1241--1274, 2013.

\bibitem{jaganathan2015phase}
K.~Jaganathan, Y.~C. Eldar, and B.~Hassibi, ``Phase retrieval: An overview of
  recent developments,'' \emph{arXiv preprint arXiv:1510.07713}, 2015.

\bibitem{Waldspurger:2015rz}
I.~Waldspurger, A.~d'Aspremont, and S.~Mallat, ``Phase recovery, maxcut and
  complex semidefinite programming,'' \emph{Mathematical Programming}, vol.
  149, no. 1-2, pp. 47--81, 2015.

\bibitem{Netrapalli:2013qv}
P.~Netrapalli, P.~Jain, and S.~Sanghavi, ``Phase retrieval using alternating
  minimization,'' in \emph{Advances in {Neural} {Information} {Processing}
  {Systems}}, 2013, pp. 2796--2804.

\bibitem{Candes:2015fv}
E.~J. Candes, X.~Li, and M.~Soltanolkotabi, ``Phase retrieval via {Wirtinger}
  flow: {Theory} and algorithms,'' \emph{Information Theory, IEEE Transactions
  on}, vol.~61, no.~4, pp. 1985--2007, 2015.

\bibitem{WangGY:2016}
G.~Wang, G.~B. Giannakis, and Y.~C. Eldar, ``Solving {Systems} of {Random}
  {Quadratic} {Equations} via {Truncated} {Amplitude} {Flow},''
  \emph{arXiv:1605.08285}, May 2016.

\bibitem{She15}
Y.~Shechtman, Y.~C. Eldar, O.~Cohen, H.~N. Chapman, J.~Miao, and M.~Segev,
  ``Phase retrieval with application to optical imaging: A contemporary
  overview,'' \emph{IEEE Signal Processing Magazine}, vol.~32, no.~3, pp.
  87--109, May 2015.

\bibitem{balan2009painless}
R.~Balan, B.~G. Bodmann, P.~G. Casazza, and D.~Edidin, ``Painless
  reconstruction from magnitudes of frame coefficients,'' \emph{Journal of
  Fourier Analysis and Applications}, vol.~15, no.~4, pp. 488--501, 2009.

\bibitem{Chen:2015eu}
Y.~Chen and E.~J. Candes, ``Solving {Random} {Quadratic} {Systems} of
  {Equations} {Is} {Nearly} as {Easy} as {Solving} {Linear} {Systems},''
  \emph{arXiv:1505.05114}, 2015.

\bibitem{LuL:17}
\BIBentryALTinterwordspacing
Y.~M. Lu and G.~Li, ``Phase transitions of spectral initialization for
  high-dimensional nonconvex estimation,'' \emph{arXiv:1702.06435 [cs.IT]},
  2017. [Online]. Available: \url{https://arxiv.org/abs/1702.06435}
\BIBentrySTDinterwordspacing

\bibitem{phmax2}
\BIBentryALTinterwordspacing
S.~Bahmani and J.~Romberg, ``{Phase Retrieval Meets Statistical Learning
  Theory: {A} Flexible Convex Relaxation},'' \emph{CoRR}, vol. abs/1610.04210,
  2016. [Online]. Available: \url{http://arxiv.org/abs/1610.04210}
\BIBentrySTDinterwordspacing

\bibitem{phmax}
\BIBentryALTinterwordspacing
T.~Goldstein and C.~Studer, ``{{PhaseMax}: Convex Phase Retrieval via Basis
  Pursuit},'' \emph{CoRR}, vol. abs/1610.07531, 2016. [Online]. Available:
  \url{http://arxiv.org/abs/1610.07531}
\BIBentrySTDinterwordspacing

\bibitem{Fienup:82}
J.~R. Fienup, ``Phase retrieval algorithms: a comparison,'' \emph{Applied
  Optics}, vol.~21, no.~15, pp. 2758--2769, 1982.

\bibitem{Hand:2016cs}
\BIBentryALTinterwordspacing
P.~Hand and V.~Voroninski, ``An {Elementary} {Proof} of {Convex} {Phase}
  {Retrieval} in the {Natural} {Parameter} {Space} via the {Linear} {Program}
  {PhaseMax},'' \emph{arXiv:1611.03935 [cs, math]}, Nov. 2016, arXiv:
  1611.03935. [Online]. Available: \url{http://arxiv.org/abs/1611.03935}
\BIBentrySTDinterwordspacing

\bibitem{Ouss:17}
\BIBentryALTinterwordspacing
O.~{Dhifallah} and Y.~M. {Lu}, ``{Fundamental Limits of PhaseMax for Phase
  Retrieval: A Replica Analysis},'' \emph{Proc. International Workshop on
  Computational Advances in Multi-Sensor Adaptive Processing (CAMSAP)}, 2017.
  [Online]. Available: \url{https://arxiv.org/abs/1708.03355}
\BIBentrySTDinterwordspacing

\bibitem{hand2016elementary}
P.~Hand and V.~Voroninski, ``An elementary proof of convex phase retrieval in
  the natural parameter space via the linear program phasemax,''
  \emph{arXiv:1611.03935}, 2016.

\bibitem{chris:151}
\BIBentryALTinterwordspacing
C.~Thrampoulidis, E.~Abbasi, and B.~Hassibi, ``Precise error analysis of
  regularized m-estimators in high-dimensions,'' \emph{CoRR}, vol.
  abs/1601.06233, 2016. [Online]. Available:
  \url{http://arxiv.org/abs/1601.06233}
\BIBentrySTDinterwordspacing

\bibitem{chris:152}
C.~Thrampoulidis, S.~Oymak, and B.~Hassibi, ``Regularized linear regression: A
  precise analysis of the estimation error,'' in \emph{Proceedings of The 28th
  Conference on Learning Theory}, vol.~40.\hskip 1em plus 0.5em minus
  0.4em\relax Paris, France: PMLR, 03--06 Jul 2015, pp. 1683--1709.

\bibitem{Gordon:85}
Y.~Gordon, ``Some inequalities for {Gaussian} processes and applications,''
  \emph{Israel Journal of Mathematics}, vol.~50, no.~4, pp. 265--289, Dec 1985.

\bibitem{Sto:13}
\BIBentryALTinterwordspacing
M.~Stojnic, ``A framework to characterize performance of {LASSO} algorithms,''
  \emph{CoRR}, vol. abs/1303.7291, 2013. [Online]. Available:
  \url{http://arxiv.org/abs/1303.7291}
\BIBentrySTDinterwordspacing

\bibitem{thrampoulidis2015asymptotically}
C.~Thrampoulidis, A.~Panahi, and B.~Hassibi, ``Asymptotically exact error
  analysis for the generalized equation-lasso,'' in \emph{Information Theory
  (ISIT), 2015 IEEE International Symposium on}.\hskip 1em plus 0.5em minus
  0.4em\relax IEEE, 2015, pp. 2021--2025.

\bibitem{thrampoulidis2015lasso}
C.~Thrampoulidis, E.~Abbasi, and B.~Hassibi, ``Lasso with non-linear
  measurements is equivalent to one with linear measurements,'' in
  \emph{Advances in Neural Information Processing Systems}, 2015, pp.
  3420--3428.

\bibitem{Phmaxcomp}
\BIBentryALTinterwordspacing
S.~Fariborz, A.~Ehsan, and H.~Babak, ``A precise analysis of phasemax in phase
  retrieval,'' \emph{CoRR}, vol. abs/1801.06609, 2018. [Online]. Available:
  \url{http://arxiv.org/abs/1801.06609}
\BIBentrySTDinterwordspacing

\bibitem{fasta}
\BIBentryALTinterwordspacing
T.~Goldstein, C.~Studer, and R.~Baraniuk, ``A field guide to forward-backward
  splitting with a {FASTA} implementation,'' \emph{arXiv eprint}, vol.
  abs/1411.3406, 2014. [Online]. Available:
  \url{http://arxiv.org/abs/1411.3406}
\BIBentrySTDinterwordspacing

\bibitem{Lange:16}
K.~Lange, \emph{{MM} Optimization Algorithms}.\hskip 1em plus 0.5em minus
  0.4em\relax {SIAM}, 2016.

\bibitem{Gert:09}
G.~R. Lanckriet and B.~K. Sriperumbudur, ``On the convergence of the
  concave-convex procedure,'' in \emph{Advances in Neural Information
  Processing Systems 22}.\hskip 1em plus 0.5em minus 0.4em\relax Curran
  Associates, Inc., 2009, pp. 1759--1767.

\bibitem{SIopt17}
\BIBentryALTinterwordspacing
M.~Mondelli and A.~Montanari, ``Fundamental limits of weak recovery with
  applications to phase retrieval,'' \emph{CoRR}, vol. abs/1708.05932, 2017.
  [Online]. Available: \url{https://arxiv.org/abs/1708.05932}
\BIBentrySTDinterwordspacing

\bibitem{phlift13}
E.~J. Candès, T.~Strohmer, and V.~Voroninski, ``Phaselift: Exact and stable
  signal recovery from magnitude measurements via convex programming,''
  \emph{Communications on Pure and Applied Mathematics}, vol.~66, no.~8, pp.
  1241--1274, 2013.

\bibitem{phasepack17}
R.~Chandra, Z.~Zhong, J.~Hontz, V.~McCulloch, C.~Studer, and T.~Goldstein,
  ``Phasepack: A phase retrieval library,'' \emph{Asilomar Conference on
  Signals, Systems, and Computers}, 2017.

\bibitem{Gor88}
Y.~Gordon, ``On {Milman}'s inequality and random subspaces which escape through
  a mesh in $\mathbb{R}^n$,'' in \emph{Geometric Aspects of Functional
  Analysis}.\hskip 1em plus 0.5em minus 0.4em\relax Berlin, Heidelberg:
  Springer Berlin Heidelberg, 1988, pp. 84--106.

\bibitem{rockafellar70}
\BIBentryALTinterwordspacing
R.~Rockafellar, \emph{Convex Analysis}, ser. Princeton landmarks in mathematics
  and physics.\hskip 1em plus 0.5em minus 0.4em\relax Princeton University
  Press, 1970. [Online]. Available:
  \url{https://books.google.com/books?id=1TiOka9bx3sC}
\BIBentrySTDinterwordspacing

\bibitem{whitney:94}
W.~K. Newey and D.~McFadden, ``{Large sample estimation and hypothesis
  testing},'' in \emph{{Handbook of Econometrics}}, ser. Handbook of
  Econometrics, R.~F. Engle and D.~McFadden, Eds.\hskip 1em plus 0.5em minus
  0.4em\relax Elsevier, 1994, vol.~4, ch.~36, pp. 2111--2245.

\bibitem{var_ana}
R.~T. Rockafellar and R.~J.-B. Wets, \emph{Variational Analysis}.\hskip 1em
  plus 0.5em minus 0.4em\relax Springer, 1998.

\end{thebibliography}

\end{document}